%% file: mnras_QS_QUENCH_MANGA_ARXIV_VERSION.tex
\documentclass[fleqn,usenatbib]{mnras}

\usepackage{newtxtext,newtxmath}
\usepackage[T1]{fontenc}
\usepackage{ae,aecompl}

\usepackage{graphicx}	% Including figure files
\usepackage{amsmath}	% Advanced maths commands
\usepackage{amssymb}	% Extra maths symbols

\usepackage{float}

\usepackage{pdflscape}
\usepackage{xcolor}
\usepackage[normalem]{ulem}
\usepackage[export]{adjustbox}

\usepackage{xspace}
\usepackage{subfig}
\usepackage{caption}

\usepackage{soul} % overstraking text: \st{} 

\defcitealias{Citro2017}{\scshape C17}
\defcitealias{Quai2018}{\scshape Q18}

\newcommand{\OH}{[\ion{O}{iii}]/H$\alpha$\xspace}
\newcommand{\NO}{[\ion{N}{ii}]/[\ion{O}{ii}]\xspace}
\newcommand{\OHb}{[\ion{O}{iii}]/H$\beta$\xspace}
\newcommand{\NS}{[\ion{N}{ii}]/[\ion{S}{ii}]\xspace}
\newcommand{\OO}{[\ion{O}{iii}]/[\ion{O}{ii}]\xspace}
\newcommand{\NeO}{[\ion{Ne}{iii}]/[\ion{O}{ii}]\xspace}

\newcommand{\OIII}{[\ion{O}{iii}]\xspace}
\newcommand{\halpha}{H$\alpha$\xspace}
\newcommand{\hbeta}{H$\beta$\xspace}
\newcommand{\OII}{[\ion{O}{ii}]\xspace}
\newcommand{\NII}{[\ion{N}{ii}]\xspace}
\newcommand{\SII}{[\ion{S}{ii}]\xspace}
\newcommand{\NeIII}{[\ion{Ne}{iii}]\xspace}
\newcommand{\SIII}{[\ion{S}{iii}]\xspace}

\newcommand{\und}{QG\xspace}
\newcommand{\undw}{QG\xspace}
\newcommand{\tsf}{SF\xspace}

\newcommand{\mass}{log(M/M$_\odot$)\xspace}
\newcommand{\smass}{M$_\odot$\xspace}

\title{Spatially resolved signature of quenching in star-forming galaxies}

\author[S. Quai et al.]{
Salvatore Quai,$^{1,2}$\thanks{E-mail: salvatore.quai@unibo.it}
Lucia Pozzetti,$^{2}$
Michele Moresco,$^{1,2}$ 
Annalisa Citro,$^{3}$
Andrea Cimatti,$^{1,4}$
\newauthor Jarle Brinchmann$^{5,6}$,
Madusha L. P. Gunawardhana$^{6}$
and Mieke Paalvast$^{6}$
\\
$^{1}$Dipartimento di Fisica e Astronomia, Universit\`a di Bologna, Via Gobetti 93/2, I-40129, Bologna, Italy\\
$^{2}$INAF - Osservatorio di Astrofisica e Scienza dello Spazio di Bologna, Via Gobetti 93/3, I-40129, Bologna, Italy\\
$^{3}$ The Leonard E. Parker Center for Gravitation, Cosmology and Astrophysics, Department of Physics, University of Wisconsin-Milwaukee, 3135 N Maryland Avenue, Milwaukee, WI 53211, USA\\
$^{4}$INAF - Osservatorio Astrofisico di Arcetri, Largo E. Fermi 5, I-50125, Firenze, Italy\\
$^{5}$Instituto de Astrof\`{i}sica e Ci\^{e}ncias do Espa\c{c}o, Universidade do Porto, CAUP, Rua das Estrelas, PT4150-762 Porto, Portugal\\
$^{6}$Leiden Observatory, Leiden University, PO Box 9513, 2300 RA, Leiden, The Netherlands\\}
\date{Accepted XXX. Received YYY; in original form ZZZ}

\pubyear{2019}

\begin{document}
\label{firstpage}
\pagerange{\pageref{firstpage}--\pageref{lastpage}}
\maketitle

% Abstract of the paper
\begin{abstract}
Understanding when,  how and where star formation ceased (quenching) within galaxies is still a critical subject in galaxy evolution studies.
%It is still critical in galaxy evolution to understand when, how and where (within galaxies) the star formation ceased (quenching). 
Taking advantage of the new methodology developed by \citet{Quai2018} to select recently quenched galaxies, we explored the spatial information provided by IFU data to get critical insights on this process. 
In particular, we analyse $10$ SDSS-IV MaNGA galaxies that show regions with low \OH compatible with a recent quenching of the star formation.
We compare the properties of these $10$ galaxies with those of a control sample of $8$ MaNGA galaxies with ongoing star formation in the same stellar mass, redshift and gas-phase metallicity range. 
The quenching regions found are located between $0.5$ and $1.1$ effective radii from the centre. This result is supported by the analysis of the average radial profile of the ionisation parameter, which reaches a minimum at the same radii, while the one of the star-forming sample shows an almost flat trend. These quenching regions occupy a total area between $\sim$ 15\% and 45\% of our galaxies. Moreover, the average radial profile of the star formation rate surface density of our sample is lower and flatter than that of the control sample, at any radii, suggesting a systematic suppression of the star formation in the inner part of our galaxies. %sharp decline in their star-formation rate. 
Finally, the radial profile of gas-phase metallicity of the two samples have a similar slope and normalisation.
Our results cannot be ascribed to a difference in the intrinsic properties of the analysed galaxies, suggesting a quenching scenario more complicated than a simple inside-out quenching.
\end{abstract}

% Select between one and six entries from the list of approved keywords.
% Don't make up new ones.
\begin{keywords}
galaxies: general -- galaxies: evolution -- galaxies: ISM
\end{keywords}

%%----------------------------------------------------------------------------------%%
\section{Introduction}
\label{ch:MaNGA}

Galaxies have pronounced bimodal distributions of their main properties \citep[e.g.][]{Strateva2001, Kauffmann2003, Blanton2003, Hogg2003, Balogh2004, Baldry2004,  Bell2012}.
At higher redshifts, this bimodality has been confirmed up to z $\sim2$ \citep[e.g.][]{Willmer2006, Cucciati2006, Cirasuolo2007, Cassata2008, Kriek2008, Williams2009, Brammer2009, Muzzin2013}.
%Thus, they are classified in two distinct categories: late-type that actively forming stars (typically blue) and early-type that stopped their star-formation and are passively evolving now (typically red).
Moreover, there are strong evidences of a continuous growth, both in number density and stellar mass, of the red and passively evolving early-type population from z$\sim 1 - 2$ to the present  \citep[e.g.][]{Bell2004, Blanton2006, Bundy2006, Faber2007, Mortlock2011, Ilbert2013, Moustakas2013}, suggesting that a large fraction of late-type galaxies transforms into early-type ones, as a consequence of the suppression of the star formation, together with a change in morphologies \citep[e.g.][]{Pozzetti2010, Peng2010}. 
These transitional scenarios is thought to be dependent on the environment where galaxies are located \citep[e.g.][]{Goto2003, Balogh2004, Bolzonella2010, Peng2010}.
However, understanding when and how the star formation ceases (the so-called star formation \emph{quenching}) and where it starts and propagates within star-forming galaxies is still one of the key open questions of galaxy evolution. 

The formation and evolution of disc galaxies in a hierarchical Universe \citep{Fall1980} leads to a scenario in which the outskirts of disc galaxies should form later than the inner part, by acquiring gas at higher angular momenta from the surrounding corona \citep[the so-called \emph{inside-out} growth][]{Larson1976}.
Inside-out growth is also predicted by hydrodynamical simulations \citep[e.g.][]{Pichon2011, Stewart2013} and it is supported by chemical evolution models \citep[e.g.][]{Boissier1999, Chiappini2001}. 
This scenario is in agreement with numbers of observational evidences \citep[e.g.][]{Prantzos2000, Gogarten2010, Spindler2018}.
Indeed, a natural consequence of inside-out growth is that central regions of galactic discs are, on average, older and more metal-rich than the outskirts \citep[e.g.][]{Zaritsky1994, RosalesOrtega2011,SanchezBlazquez2014, GonzalezDelgado2014, GonzalezDelgado2015, GonzalezDelgado2016, Goddard2017a, Goddard2017b}. 
This almost ubiquitous behaviour can be explained by a common evolution of gas, chemical history and stars \citep{Ho2015}, bearing in mind that without a continuous replenishing of fresh gas, galaxies would have fuel to sustain at most $\sim1$ Gyr of star formation \citep{Tacconi2013}. In other words, the star-forming galaxies need for a systematic supply of new gas and, together with evidence that inside-out growth is still active in outer part of most local star-forming galaxies \citep[e.g.][]{Wang2011, MunozMateos2011, Pezzulli2015}, it suggests that galactic halos are still providing high angular momentum gas to assemble the out-skirts of galaxies.
Moreover, starting from the evidence that hot coronae must rotate more slowly than the disc (i.e. pressure gradients provide support against gravity) \citet{Pezzulli2016} discussed that a misalignment between disc and halo velocity implies a systematic radial gas flow towards the inner parts of galaxies. Taking into account this effect and disentangling it from the contribution of inside-out growth in their models, these flows show a strong impact on the structural and chemical evolution of galaxies, naturally creating strong steep abundance gradient.
%%%

In this scenario, which mechanism drives the quenching of the star formation and how it can prevent further inflow of gas? 
Observational evidences of a systematic suppression of the star formation in the inner part of galaxies below the star-forming main sequence has been interpreted as an inside-out quenching \citep[e.g.][]{Tacchella2015, Belfiore2018, Ellison2018, Morselli2018, Lin2019}. Being linked to AGN activities and gas outflows, they have suggested the negative AGN feedback as the mechanism that can trigger the interruption of the star formation from the centre and then, towards the outskirts.  
However, \citet{Tacchella2016} and recently \citet{Matthee2019} and \citet{Wang2019} argued that the evidence of symmetry around the star-forming main sequence in the SFR - stellar mass diagram suggests an evolution of galaxies through phases of elevation and suppression of the star formation, without the need for a permanent quenching. This phenomenon is more clear in the inner part of galaxies because of the higher star formation efficiency, since higher gas fractions and shorter depletion times implicate shorter reaction time to the change in the reservoir of gas.

Therefore, identifying actual quenching galaxies that are leaving the blue cloud to reach the red sequence is still challenging. Having intermediate colours between blue late-type and red early-type galaxies, the so-called `green valley` galaxies \citep{Wyder2007, Martin2007, Salim2007, Schiminovich2007, Mendel2013, Salim2014} have been considered as promising candidate for the transiting population. 
\citet{Schawinski2014}, instead argued that green valley galaxies are actually separable into two populations of galaxies that share the same intermediate colours: (i) the green tail of the blue late-type galaxies with low specific star-formation rate but no sign of rapid transition towards early-type (quenching timescale of several Gyr) and (ii) a population of migrating early-type galaxies which are evolving (with timescale $\sim1$ Gyr) to red and passive galaxies, as a result of major mergers of late-type galaxies. 
\citet{Belfiore2017, Belfiore2018} recently found that the ionised optical spectra of most green valley galaxies are dominated by central low-ionization emission-lines (cLIER) due to old post-AGB stars radiation. 
The uniformity of old stellar populations suggests that green valley galaxies can be a  `quasi-static` population subjected to a slow-quenching. 
However, to account for the rate of growth of the red population and for the exiguity of transiting galaxies, there should be found galaxy populations in which star formation quenches on short timescales \citep[e.g.][]{Tinker2010, Salim2014}.
Several hypothesis have been proposed to settle this puzzle. 
Some typical examples of galaxies quickly transforming into passively evolving galaxies are (i) galaxies which show both disturbed morphologies and intermediate colours \citep[e.g.][]{Schweizer1992, Tal2009} or (ii) strong morphological disturbances due to recent mergers  \citep{Hibbard1996, Rothberg2004, Carpineti2012}, (iii) young elliptical galaxies \citep{Sanders1988, Genzel2001, Dasyra2006} that are often characterized by low-level of recent star-formation \citep{Kaviraj2010} represent examples of galaxies that are quickly transforming into passively evolving galaxies. 
Studies regarding the so-called `post-starburst` systems attempted to link the evolution of transient population with the properties of local early-type galaxies. This population shows strong Balmer absorption lines (H$\delta$ with equivalent width $>5$ \AA, in particular), typical of stellar populations dominated by A type stars with ages between 300 Myr and 1 Gyr after the interruption of the star-formation \citep[e.g.][]{Couch1987}. Some of them have spectra compatible with passive evolution and no sign of emission lines \citep[e.g.][]{Quintero2004, Poggianti2004, Balogh2011, Muzzin2012, Mok2013, Wu2014} while others show emission lines (i.e. usually strong [\ion{O}{II}]$\lambda$3726-29 emission) and are often called `strong-H$\delta$' galaxies \citep[e.g.][]{LeBorgne2006, Wild2009,  Wild2016}.
The properties of these galaxies are interpreted as sign of a recent fast-quenching \citep{Dressler1983, Zabludoff1996, Quintero2004, Poggianti2008, Wild2009}. 
As a matter of fact, all these previous studies focused on galaxies observed $0.3-1$ Gyr after the quenching phase.  
This delay, therefore, prevents to clearly unveil which processes drive the radical change in galaxy properties.

If, on one hand, stellar mass and metallicity are tracers of secular evolution of galaxies, on the other hand it is well known that the ionisation parameter (hereafter U) provides powerful constraints on the recent activity within galaxies \citep[e.g.][]{Dopita2000, Kewley2001, Dopita2006, Levesque2010, Kewley2013, Kashino2016}. 
Variations in the UV radiation strongly affect the galactic spectra. For example, the NUV continuum light, which it is primarily produced in the photosphere of long-lived stars more massive than $3$M$_\odot$, can trace the star formation on a timescale of $\sim100$Myr.
The Balmer lines, instead, are generated from the recombination of Hydrogen ionised by photons with energy higher than $912$ \AA\, and only stars more massive than late-B stars irradiate a sufficient amount of UV flux to do this task. 
Thus, the \halpha luminosity can trace SFR over the lifetime of these stars of tens of Myr.
Spectral lines such as \OIII\!$\lambda$5007 and \NeIII\!$\lambda$3869 can instead be produced only by the even more energetic photons coming from the short-lived, super massive O and early B stars. Therefore, these spectral lines are expected to disappear from galaxy spectra on timescales of $10$-$80$ Myr, which correspond to the lifetime of the most massive stars, once that the SF stops. 
\citet[][hereafter \citetalias{Citro2017}]{Citro2017} and \citet[][hereafter \citetalias{Quai2018}]{Quai2018} developed an innovative approach which aims at finding galaxies immediately after the quenching.
The method is bases on the use of ratios between high-ionization potential lines (which can be produced only by very high energetic photons) such as \OIII and \NeIII , and low-ionization potential lines (which require lower energy photons) such as \halpha, \hbeta, \OII. 
\citetalias{Citro2017} proved that \OH ratio is a very sensitive tracer of the ongoing quenching as it drops by a factor $\sim 10$ within $\sim 10$ Myr from the quenching assuming a sharp interruption of the star formation, and even for a smoother and slower star formation decline (i.e. an exponential declining star formation history with {\emph e}-folding time $\tau = 200$ Myr) the \OH decreases by a factor $\sim 2$ within $\sim 80$ Myr from the quenching.
The \OH ratio is affected by a significant degeneracy between ionization and metallicity (herafter Z), in the sense that [O III]$\lambda$5007 emission can be depressed also by high metallicity (U-Z degeneracy, hereafter).
In \citetalias[][]{Quai2018}, we found that the U-Z degeneracy can be mitigated by using  couples of emission line ratios orthogonally dependent on ionisation (i.e. \OH) and metallicity \citep[e.g. \NO is a good tracer of gas-phase metallicity, as discussed in][]{Kewley2002, Nagao2006}. 
In Q18 we used the \OH vs. \NO diagnostic diagram in the SDSS to identify a sample of candidates quenching galaxies (QGs), i.e. in the early phase of quenching star formation, as a population well segregated from the global sample of galaxies with ongoing star-formation, showing \OH ratios, at fixed \NO, so low that they cannot be explained by metallicity effects.
%From a sample of about $174.000$ SDSS-DR8 star-forming galaxies at $0.04 \leq$ z $< 0.21$, they analysed a sample of $\sim 26.000$ galaxies with undetected \OIII in which they identified about $300$ quenching galaxy candidates as those showing \OH ratios, at fixed \NO, so low that they cannot be explained by metallicity effects (see also \autoref{fig:QGs_plane}).
%They represent a population of galaxies well segregated from the global sample of galaxies with ongoing star-formation.

Since the advent of integral field unit (IFU) spectroscopy era, galaxies can be studied with enough spatial resolution to allow analysis of physical properties even at galactocentric distances larger than $2$ effective radii. 
In this paper, we extend the method devised in \citetalias{Quai2018} to select quenching galaxies in the SDSS main sample to IFU data from the SDSS-IV MaNGA survey \citep{Bundy2015, Blanton2017}. Our aim is to search for regions where quenching had started and, therefore, to derive spatial information on the quenching process within galaxies. 
This paper is intended to be a pilot study, where we analyse the more promising galaxies starting from the sample of SDSS QGs previously analyzed by \citetalias{Quai2018}. We already planned to extend this study to the whole MaNGA population of star forming objects to derive the total fraction of galaxies partially quenching and their properties.

We structure this paper as follows: in \autoref{sec:sample_MaNGA} we briefly recall the method introduced in \citetalias{Quai2018} and we describe our MaNGA sample. We use \autoref{sec:cases} to focus on two cases illustrating the detailed procedure and analysis done and then, in \autoref{sec:global_trends} we present the general properties of the entire sample. Finally, in \autoref{sec:Summary} we discuss our results and we provide our concluding remarks.

%%%%%%%%%%%%%%%%%%%%%%%%%%%%%%%%%%%%%%%%%%%%%%%%%%%%%%%%%%%%%%%%%%%%%%%%%%%%%%%%%%%%%%%%%%%%%%%%
\section{Method and sample }
\label{sec:sample_MaNGA}

%\begin{figure*}[t]
%  \begin{center}
%    \showthe\linewidth % Use this to determine the width of the figure.
%    \showthe\columnwidth
%	\showthe\textheight
%	\showthe\textwidth	
%    \includegraphics[width=\columnwidth]{plot_QGs_INTRO_v13.png}
%    \caption{\label{fig:sin_cos} Plot of the sine and cosine functions.}
%  \end{center}
%\end{figure*}
%\end{document}

\begin{figure}
\centering
\includegraphics[width=1\linewidth]{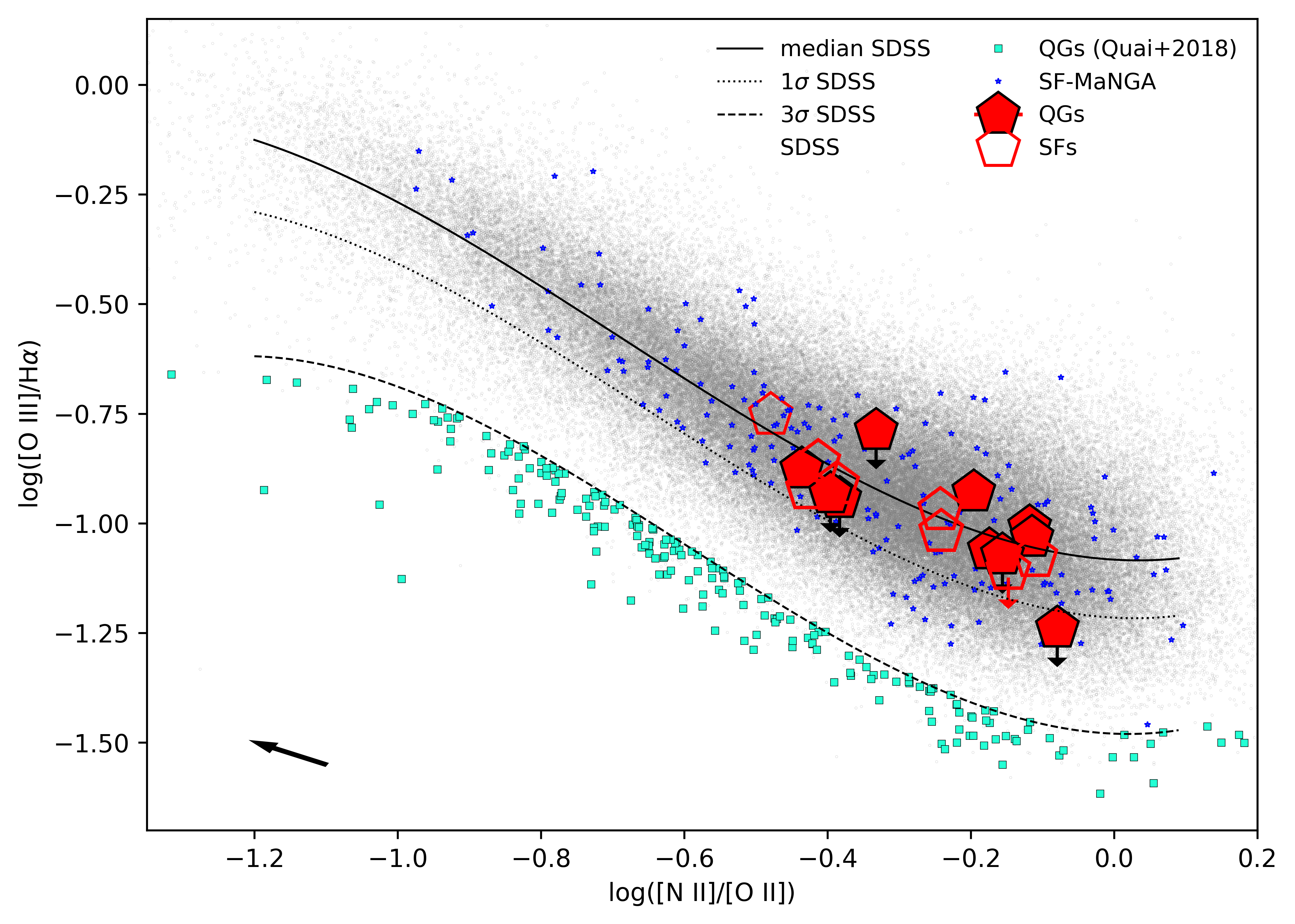} 
\caption{The diagnostic  \OH vs \NO diagram. 
The black curves represent median, $1 \sigma$ and $3\times1\sigma$ limits of the SDSS star-forming galaxies sample \citepalias[see][]{Quai2018}, which are represented by grey dots. The black arrow in the bottom-left corner represents the direction of the dust vectors for the \citet{Calzetti2000} extinction law, for an E(B-V)$=0.3$. The cyan squared dots below the $3\times1\sigma$ limits represent the SDSS quenching candidates selected in \citetalias{Quai2018}. The blue dots represent the SDSS galaxies that have a match in MaNGA-DR14.   The red pentagons represent the SDSS position of the MaNGA galaxies analysed in this paper: full symbols for the galaxies with quenching regions (QRG) and empty symbols for the star forming (SF) galaxies, as defined in \autoref{sec:QGs_and_SFs}. The arrows indicate galaxies with the upper limits in \OH.}
 \label{fig:QGs_plane}
\end{figure}
%%%%%%%%%%%%%%%%%%%%%%%%%%%%%%%%%%%%%%%%%%%%%%%%%%%%%%%%%%%%%%%%
%The our method can easily be extended to IFU data by merely considering the spaxels\footnote{Spatial pixels.} which lie below the $3\times1\sigma$ curve of the \OH vs \NO diagnostic diagram as regions of the galaxy that are experiencing the quenching of the star-formation.
%LP% As shown by \citetalias{Citro2017}, the \OH emission lines ratio rapidly reacts to the star formation quenching. However,  this ratio is affected by a significant degeneracy between ionisation and metallicity, because the  \OIII$\lambda$5007 emission line can be depressed both by a reduction of ionising photons and by high metallicity.
In \citetalias{Quai2018}, from the analysis of a sample of $\sim$174.000 star-forming galaxies at $0.04 <$ z $< 0.21$ extracted from the SDSS-DR8 catalogue, %\citepalias{Quai2018}, 
using the devised \OH vs. \NO diagnostic diagram, we
%LP% devised a method aimed both at avoiding the degeneracy and selecting reliable quenching galaxies candidates. 
%LP% It was found, indeed, that pairs of emission line ratios nearly orthogonally dependent on ionisation (i.e. \OH) and metallicity (i.e. \NO) can strongly mitigate the degeneracy (see Q18 for further details).
%\citepalias[see][ciao]{Quai2018}. 
%LP% Hence, in the \OH vs \NO plane they 
identified about $300$ quenching galaxy candidates satisfying the following criteria:
\begin{enumerate}
\item \OIII weak enough to be undetected inside the SDSS fibre (i.e. S/N(\OIII) $<2$),
\item \OH ratios, at fixed \NO (i.e. fixed gas-phase metallicity), lower than the $3\times1\sigma$ value of the SDSS star-forming distribution (see \autoref{fig:QGs_plane}). They represent a population of galaxies well segregated from the global sample of galaxies with ongoing star-formation.
%lie below the $3\times1\sigma$ of the SDSS \OH vs \NO diagram (see \autoref{fig:QGs_plane}). In other words, they show \OH ratios, at fixed \NO, which are too low to be explained by metallicity effects. 
\end{enumerate}
In order to derive spatial information on the quenching process within the galaxies, we extend these criteria to MaNGA IFU observations by exploiting the \OH vs \NO diagnostic of spatially resolved galaxies. 
To this aim, we cross-match the $\sim174.000$ galaxies selected in the main SDSS survey \citepalias{Quai2018} with the MaNGA data-release 14 \citep{Abolfathi2018}, finding $208$ matches. 
However, none of $\sim300$ SDSS quenching primary candidates selected in \citetalias{Quai2018} has been observed with MaNGA. 
Nevertheless, we find matches with MaNGA data for $10$ galaxies, among $\sim$ 26000 galaxies with \OIII undetected (S/N(\OIII) $< 2$) within the SDSS fibre, which should represent promising candidates of galaxies which could be in the very first phase of the quenching. 
In fact, in \citetalias{Quai2018} we performed a survival analysis (ASURV, i.e. Kaplan-Meier estimator) of their \OH distribution in slices of \NO, and found that about $50\%$ ($3\%$) of them (that we called [\ion{O}{III}]undet galaxies) are statistically distributed below $1\sigma$ ($3\sigma$) curve, respectively, and therefore candidates quenching galaxies, while the other ones should actually be normal star-forming galaxies with fainter emission lines. 
Among the 10 MANGA matches galaxies we discard MaNGA 1-245686 because it appears almost edge-on  (i.e. a ratio b/a $= 0.2$) and we do no further analyse also MaNGA 1-38802 because it is at a redshift considerably higher (i.e. z $=0.11$) than the other [\ion{O}{III}]undet galaxies in the sample. 
The remaining $8$ [\ion{O}{III}]undet  galaxies are located at redshift between $0.04$ and  $0.06$ and have masses between $10^{9.6}$ and $10^{10.8}$ M$_{\sun}$. 
We decide to include as a control sample $12$ SDSS star-forming galaxies with similar mass and \NO range, whose emission line ratios lie along the median SDSS sequence of star-forming galaxies within the \OH vs \NO diagram, .
The diagnostic diagram for the original Q18 sample and for the MaNGA galaxies considered in this analysis is presented in \autoref{fig:QGs_plane}.
%Our aim is to search for galaxies with regions that are quenching, using the same diagnostic used in SDSS \citepalias{Quai2018}, but applied to each resolved galaxy regions. Therefore, we use these information to classify our targets
Our aim is to search for galaxies with regions which are in the quenching phase, using the same diagnostic used in SDSS \citepalias[see][]{Quai2018}, but applied to each resolved galaxy regions.

%Among them, $10$ galaxies have a central SDSS \OH ratio low enough to be promising candidates of galaxies which could be in the very first phase of the quenching. 
%decide, therefore, to analyse a small sample of $22$ MaNGA galaxies (out to 203 matches).

%\autoref{fig:QGs_plane} shows the position of the objects on the \OH vs \NO plane. 

%%%%%%%%%%%%%%%%%%%%%%%%%%%%%%%%%%%%%%%%%%%%%%%%%%%%%%%%%%%%%%%%
\subsection{From MaNGA to pure-emission cube} 
\label{sec:data_reduction}
Starting from the MaNGA datacubes processed by the data reduction pipeline \citep[DRP,][]{Law2016}, the final emission lines maps are obtained applying the following spectral-fitting procedure, similar to that proposed by \cite{Belfiore2016}:
\begin{enumerate}
    \item \emph{Increasing the signal-to-noise of the continuum.}
    To create a pure-emission datacube, it is necessary to accurately subtract the stellar continuum from the original datacube. 
    At first, the noise is corrected for the effect of the spatially correlated noise between adjacent spaxels, as discussed in \cite{Garcia-Benito2015}.
    Then, in order to increase the signal-to-noise ratio (S/N) of the continuum and at the same time preserve the spatial resolution, spaxels which S/N lower than $10$ in the restframe $4740-4840$~\AA \, range are binned together with a Voronoi tessellation approach\footnote{The Voronoi tessellation routine can be found at  \url{http://www-astro.physics.ox.ac.uk/~mxc/software}.} \citep{Cappellari2003}. 
    Spaxels with undetected continuum (i.e. S~/~N $<2$) are not included in the binning, and they are no further considered in our analysis. 
    The size of the bins is not forced to be larger than the typical MaNGA point spread function (PSF, i.e. $\sim 2.5$ arcsec at FWHM, see \autoref{tab:sample_prop}), therefore, it is possible that adjacent bins are statistically correlated. 

\item \emph{Fitting the continuum.} In the spatially binned spectra the emission-lines and the strong sky-lines (i.e. \ion{O}{I}$\lambda$5577, NaD$\lambda$5890,\ion{O}{I}$\lambda$6300,\ion{O}{I}$\lambda$6364)  are masked within a window of 1400 km s$^{-1}$. Then, the spectral continuum has been fitted choosing among various simple MILES stellar population models\citep{Vazdekis2012} using penalised pixel fitting\footnote{pPXF code can be downloaded from \url{http://www-astro.physics.ox.ac.uk/~mxc/software}.} \citep[pPXF,][]{Cappellari2004} without taking into account dust extinction and using a set of additive polynomials up to the 4th order to correct the continuum shape. 

\item \emph{The pure-emission datacube.} The best-fit continuum of each spatial bin is subtracted from the single original spaxels composing the bins, and the resulting data cube is composed by spaxels of pure-emission spectra.  
\end{enumerate}

\subsection{Emission-lines maps} 
In this section, we describe the routine that we apply to pure-emission datacube to obtain maps of individual emission-lines (i.e. \halpha, \hbeta, \OIII$\lambda$5007, \OII$\lambda$3726-29, \NII$\lambda$6584).

\begin{enumerate}
\item \emph{Increasing the signal-to-noise of nebular lines.}  \halpha fluxes are measured in each spaxel from the pure-emission datacube. 
In order to reach an S/N(\halpha)$>5$ we perform a further Voronoi binning tessellation, not considering spaxels with S/N(\halpha)$<1$, which are no further considered in our analysis.
This procedure allows studying nebular emission properties also in the outskirts of galaxies, at the cost of slightly worsening the spatial resolution. 
We find that no spaxels needs to be binned inside the effective radius of the analysed galaxies since their S/N(\halpha) is always higher than $5$. Therefore, the original central spatial resolution is preserved and dominated by the point spread function (PSF) of MaNGA datacubes, %LP% which has a typical value of $2.5\arcsec$ at FWHM (
i.e. an area covered by almost $20$ spaxels.
%It is important to highlight that the spaxels rejected from the Voronoi binning of the continuum because of their undetected stellar continuum are now re-included in the Voronoi binning of the nebular lines and rejected again only if their S/N(\halpha) $<1$.

\item \emph{Fluxes and errors.} In each spaxel, fluxes are measured by integrating the Gaussian best fit to the lines \halpha, \hbeta, \OIII, \OII (we consider \OII $=$ \OII\!$\lambda$3726 $+$  \OII\!$\lambda$3729), \NII\!$\lambda$6584 (hereafter \NII), and \SII\!$\lambda \lambda$6717-6731. Errors on the fluxes are obtained by the propagation of errors on a Gaussian amplitude and standard deviation. 

In our analysis, we need reliable measures of \NII and \OII; hence the spaxels with S/N $< 2$ in these lines are not considered either. 
Instead, since the fingerprint of the method is the weakness or lack of the \OIII emission,  spaxels with S/N(\OIII) $< 2$ are kept as upper-limit values with \OIII = $2 \times \sigma$\OIII, where $\sigma$\OIII is the error on the \OIII flux. 

 \end{enumerate}

\subsection{The derived quantities from MaNGA data}
%In Appendix A
The maps of \halpha, \hbeta, \OIII, \OII and \NII, which form the starting point of our classification criteria (see \autoref{sec:selection}), are corrected for dust attenuation based on the 
\halpha/\hbeta ratio.
In order to perform a proper correction for dust extinction, spaxels with S/N(\hbeta) $<3$ and S/N(\halpha) $<5$ are no further considered in the analysis. 
For the other spaxels, the colour excess E(B-V) is derived adopting the \cite{Calzetti2000} attenuation law and assuming the Case B recombination and 
a Balmer decrement \halpha/\hbeta $= 2.86$ \citep[typical
of \ion{H}{ii} regions with electron temperatures T$_\text{e} = 10^4$ K and electron 
density n$_\text{e} \sim 10^2-10^4 \,\text{cm}^{-3}$,][]{Osterbrock1989, Dopita2003} .
Negative values  of E(B-V) between about $-0.05$ and $\sim0$ (i.e. inverted Balmer decrement, with $\sim 2.7 \leq $ \halpha/\hbeta $< 2.86$) are found in almost all galaxies in our sample, with percentages between $2\%$ and $14\%$ of the spaxels (but the galaxy 1-352114 shows E(B-V)$<0$ in $\sim52\%$ of its spaxels).
However,  these values are still compatible with case B, though at electron temperatures between $10^4$ and $2\times10^4$ K \citep[i.e. $2.74 \leq $ \halpha/\hbeta $< 2.86$,][]{Hummer1987}.
We assign E(B-V) $= 0$ to these spaxels.
%\autoref{fig:EBV} shows the E(B-V) maps of MaNGA 1-43012 and 1-178443 which represent case studies that we will extensively present in \autoref{sec:cases}.
%%%%%%%%%%%%%%%%%%%%%%%%%%%%%%%%%%%%%%%%%%%%%%%%%%%%%%%%%%%%%%%%
%\begin{figure}
%\includegraphics[width=0.495\linewidth]
%{plot_EBV_OIIIundet_MANGAid_8135-12702_vTHESIS} 
%\includegraphics[width=0.495\linewidth]{plot_EBV_SF_MANGAid_7962-6104_vTHESIS} \\
%    \caption{The E(B-V) maps of MaNGA 1-43012 (\emph{left}) and MaNGA 1-178443  (\emph{right}). Overlapped in magenta are the hexagonal shapes of the MaNGA IFU bundles, while the black circles represent the R$_{50}$. The $2.5\arcsec$ circle in the bottom-right corner of the maps represent the typical PSF (FWHM) of MaNGA data.}    
%    \label{fig:EBV}
%\end{figure}
%%%%%%%%%%%%%%%%%%%%%%%%%%%%%%%%%%%%%%%%%%%%%%%%%%%%%%%%%%%%%%%%
%%%%%%%%%%%%%%%%%%%%%%%%%%%%%%%%%%%%%%%%%%%%%%%%%%%%%%%%%%%%%%%%%%%%%%%%%%%%%%%%%%%%%%%%%%%%%%%%%%%%%%%%%%%%%%%%%%%%%%%%%%%%%%%%

%\begin{landscape}
\begin{figure*}
\begin{minipage}{\linewidth}
\centering
\captionsetup{width=.8\linewidth}
\includegraphics[width=0.171\linewidth]{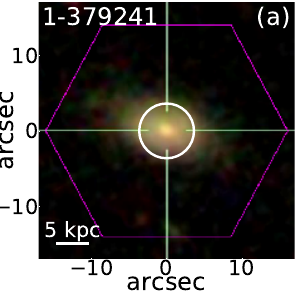}\!
\includegraphics[width=0.211\linewidth]{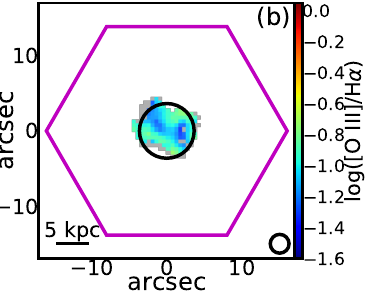}\!
\includegraphics[width=0.211\linewidth]{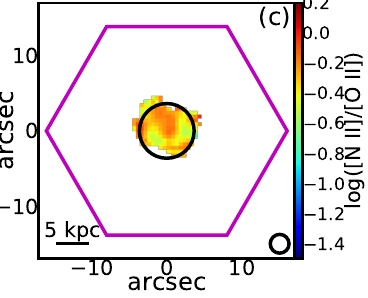}\!
\includegraphics[width=0.22\linewidth]{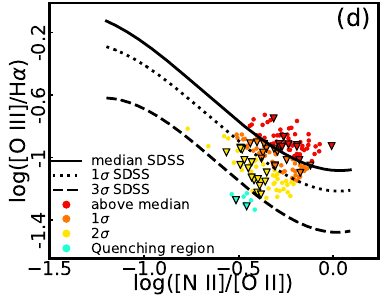} \!
\includegraphics[width=0.171\linewidth]{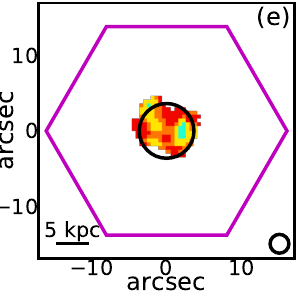} \\

\includegraphics[width=0.171\linewidth]{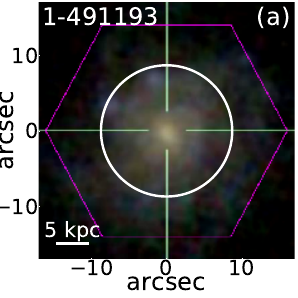}\!
\includegraphics[width=0.211\linewidth]{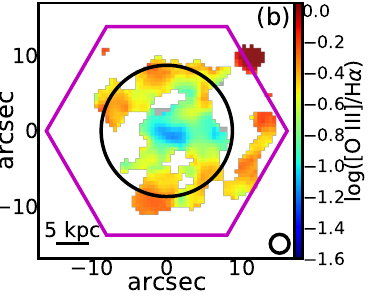}\!
\includegraphics[width=0.211\linewidth]{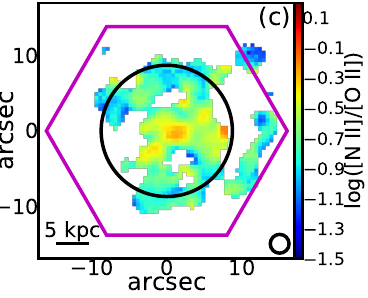}\!
\includegraphics[width=0.22\linewidth]{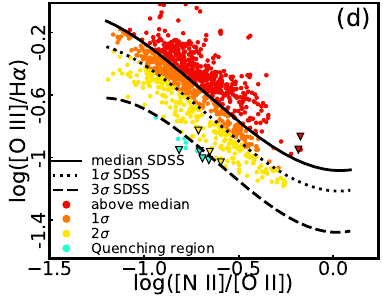} \!
\includegraphics[width=0.171\linewidth]{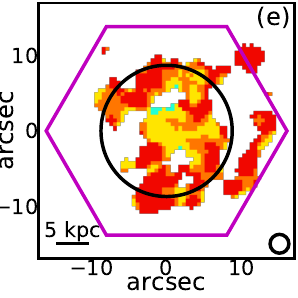} \\

\includegraphics[width=0.171\linewidth]{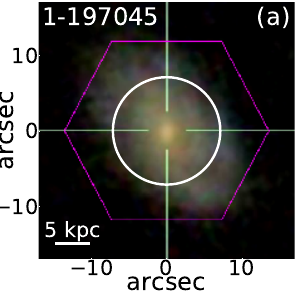}\!
\includegraphics[width=0.211\linewidth]{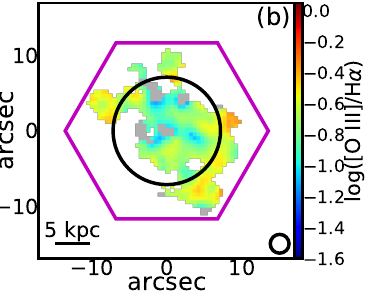}\!
\includegraphics[width=0.211\linewidth]{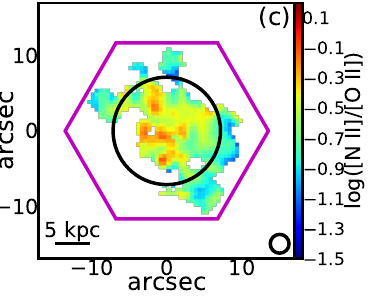}\!
\includegraphics[width=0.22\linewidth]{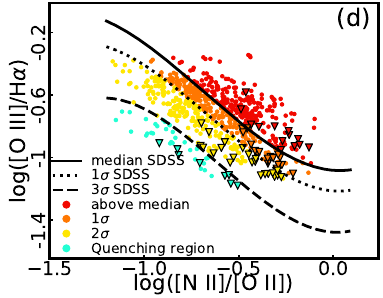} \!
\includegraphics[width=0.171\linewidth]{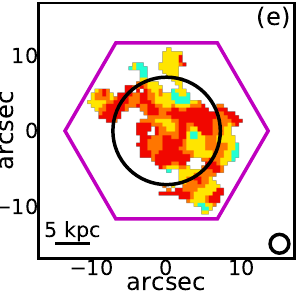} \\

\includegraphics[width=0.171\linewidth]{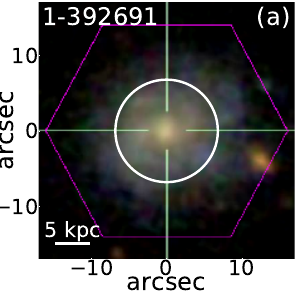}\!
\includegraphics[width=0.211\linewidth]{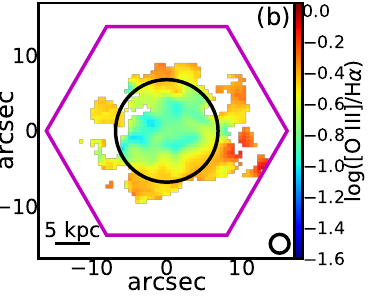}\!
\includegraphics[width=0.211\linewidth]{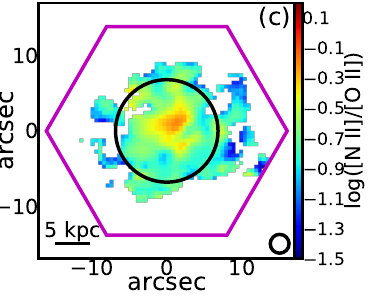}\!
\includegraphics[width=0.22\linewidth]{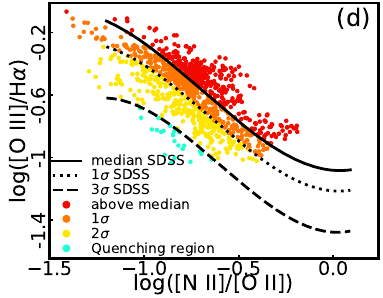} \!
\includegraphics[width=0.171\linewidth]{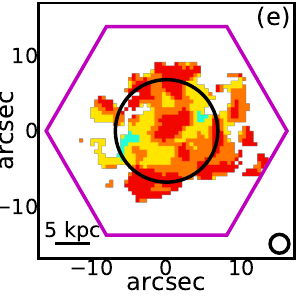} \\
\includegraphics[width=0.171\linewidth]{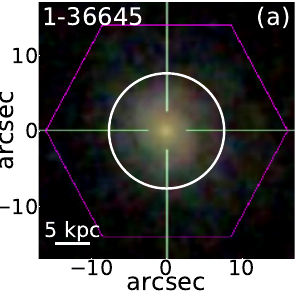}\!
\includegraphics[width=0.211\linewidth]{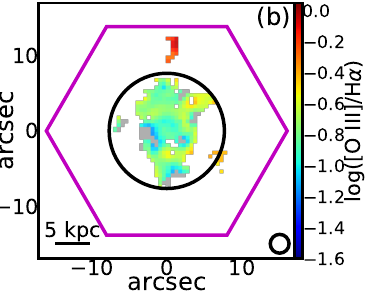}\!
\includegraphics[width=0.211\linewidth]{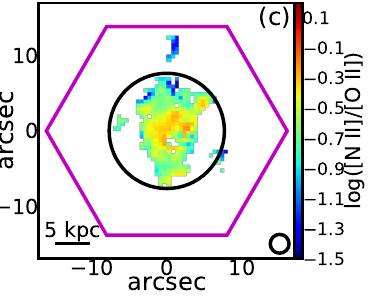}\!
\includegraphics[width=0.22\linewidth]{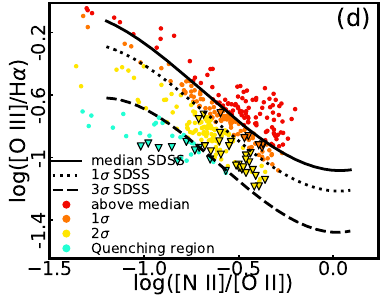} \!
\includegraphics[width=0.171\linewidth]{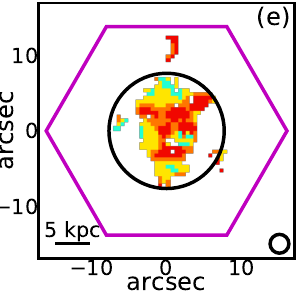}

\includegraphics[width=0.171\linewidth]{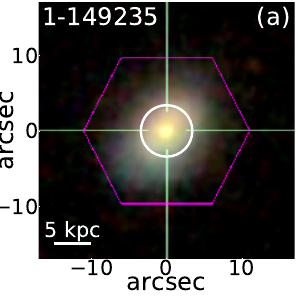}\!
\includegraphics[width=0.211\linewidth]{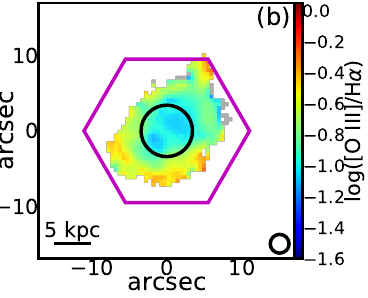}\!
\includegraphics[width=0.211\linewidth]{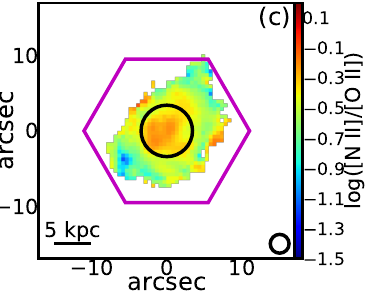}\!
\includegraphics[width=0.22\linewidth]{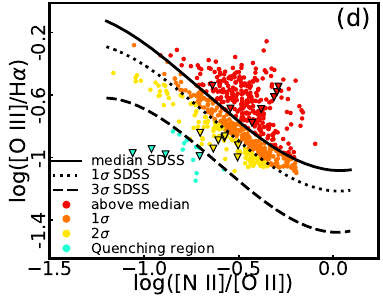} \!
\includegraphics[width=0.171\linewidth]{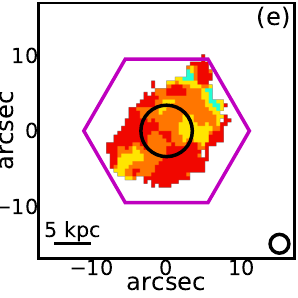} \\

\includegraphics[width=0.171\linewidth]{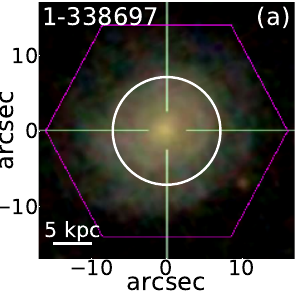}\!
\includegraphics[width=0.211\linewidth]{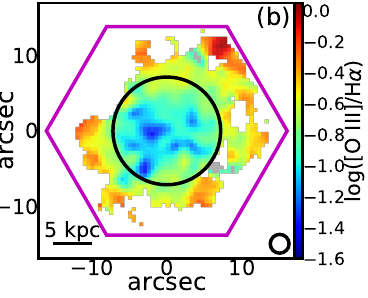}\!
\includegraphics[width=0.211\linewidth]{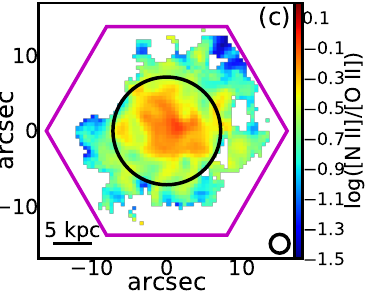}\!
\includegraphics[width=0.22\linewidth]{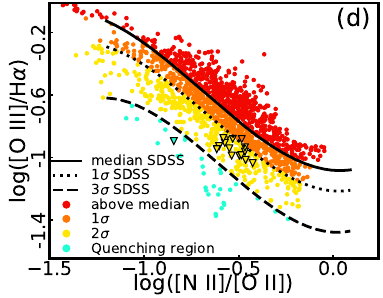} \!
\includegraphics[width=0.171\linewidth]{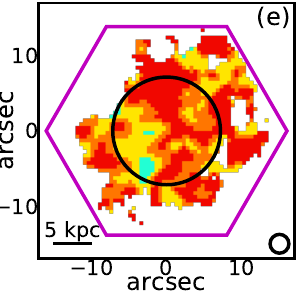} 
    \captionof{figure}{A summary of the $10$ QG galaxies in our sample.  
    (a) The g-r-i images composite from SDSS. Each image covers a region of $17 \times  17$ arcsec$^2$ and in the bottom-left corner of each image is reported the scale of $5$ kpc. (b) The dust-corrected \OH maps. The grey areas show regions with S/N(\OIII) $<2$.  (c) The dust-corrected \NO maps. (d) The \OH vs \NO diagnostic diagram for the quenching. The spaxels are colour-coded according to their position on the plane: red dots for those lying above the median curve, orange dots for them between the median and $1\sigma$ , yellow dots for spaxels which lie between $1\sigma$  and $3\times1\sigma$  and, finally, cyan dots for spaxels below the $3\times1\sigma$  curve that, according with our classification criteria described in the text, represent likely quenching regions. The triangles represent spaxels with an upper limit in \OH (i.e. spaxels with S/N(\OIII) $<2$). (e) The map of the galaxies colour-coded according to the position of spaxels as in (d).
   In (a), (b), (c) and (e) the overlapped-magenta hexagonal shapes the MaNGA IFU bundles, while the %white 
   circle represents the R$_{50}$. Finally, the $2.5\arcsec$ circle in the bottom-right corner of the maps in (b), (c) and (e) represent the typical PSF (FWHM) of MaNGA data. }    
    \label{fig:OIII_info_1}
\end{minipage}
\end{figure*}
%\end{landscape}

%\begin{landscape}
\begin{figure*}\ContinuedFloat
\begin{minipage}{\linewidth}
\centering

\includegraphics[width=0.171\linewidth]{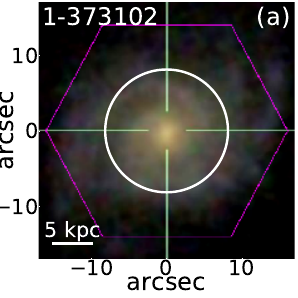}\!
\includegraphics[width=0.211\linewidth]{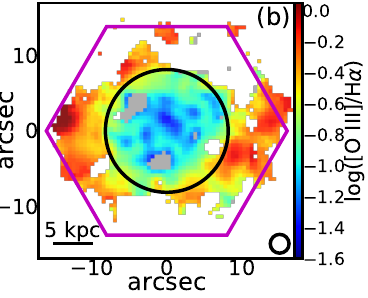}\!
\includegraphics[width=0.211\linewidth]{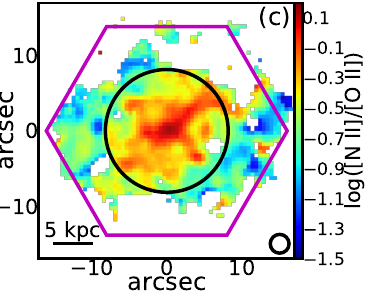}\!
\includegraphics[width=0.22\linewidth]{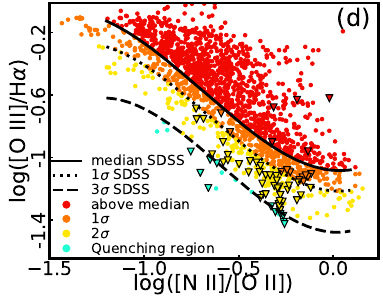} \!
\includegraphics[width=0.171\linewidth]{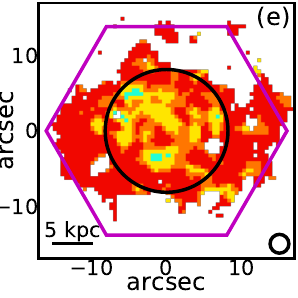} \\

\includegraphics[width=0.171\linewidth]{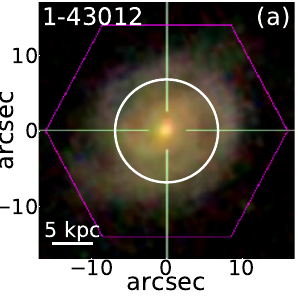}\!
\includegraphics[width=0.211\linewidth]{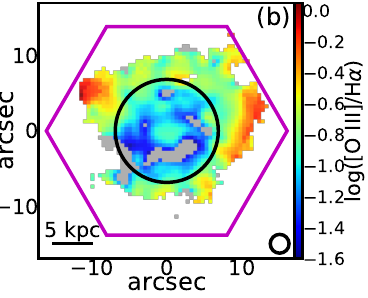}\!
\includegraphics[width=0.211\linewidth]{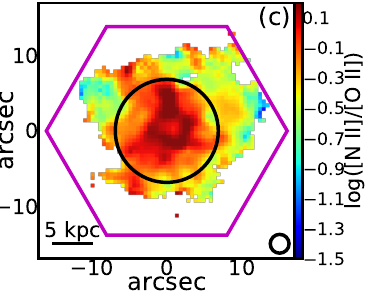}\!
\includegraphics[width=0.22\linewidth]{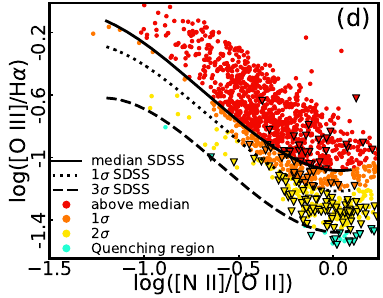} \!
\includegraphics[width=0.171\linewidth]{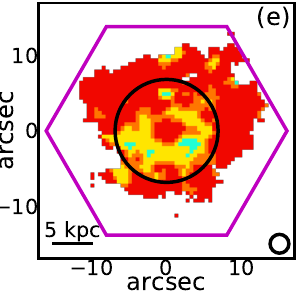} \\

\includegraphics[width=0.171\linewidth]{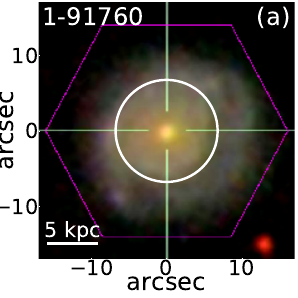}\!
\includegraphics[width=0.211\linewidth]{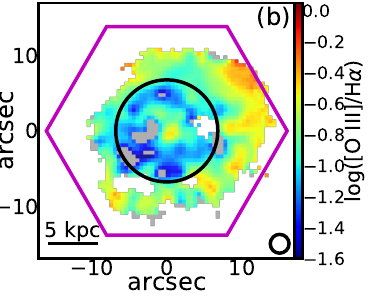}\!
\includegraphics[width=0.211\linewidth]{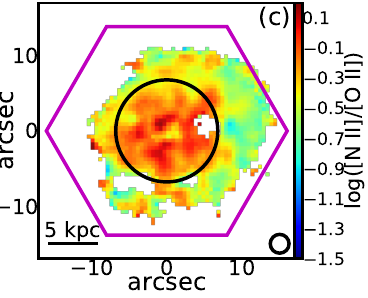}\!
\includegraphics[width=0.22\linewidth]{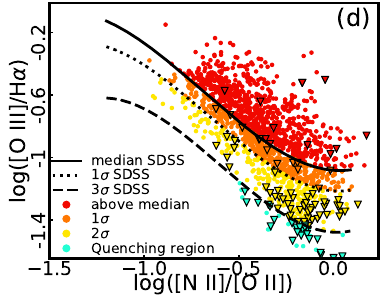} \!
\includegraphics[width=0.171\linewidth]{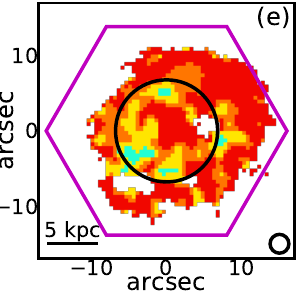} \\

   \captionof{figure}{Continued.}    
    \label{fig:OIII_info_2}
\end{minipage}
\end{figure*}
%\end{landscape}
%%%%%%%%%%%%%%%%%%%%%%%%%%%%%%%%%%%%%%%%%%%%%%%%%%%%%%%%%%%%%%%%%%%%%%%%%%%%%%%%%%%%%%%%%%%%%%%%%%%%%%%%%%%%%%%%%%%%%%%%%%%%%%%%
%As explained in the following, in our analysis we need reliable measures of \NII and \NO; hence the spaxels with S/N $< 2$ in these lines are not considered neither. 
%Instead, since the fingerprint of the method is the weakness or lack of the \OIII emission,  spaxels with S/N(\OIII) $< 2$ are kept as upper-limit values with \OIII = $2 \times \sigma$\OIII, where $\sigma$\OIII is the error on the \OIII flux. 

%%%%%%%%%%%%%%%%%%%%%%%%%%%%%%%%%%%%%%%%%%%%%%%%%%%%%%%%%%%%%%%%%%%%%%%%%%%%%%%%%%%%%%%%%%%%%%%%%%%%%%%%%%%%%%%%%%%%%%%%%%%%%%%%
%\begin{landscape}
\begin{figure*}
\begin{minipage}{\linewidth}
\centering
\captionsetup{width=.8\linewidth}

\includegraphics[width=0.171\linewidth]{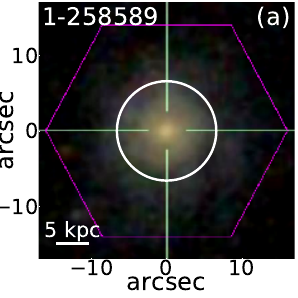}\!
\includegraphics[width=0.211\linewidth]{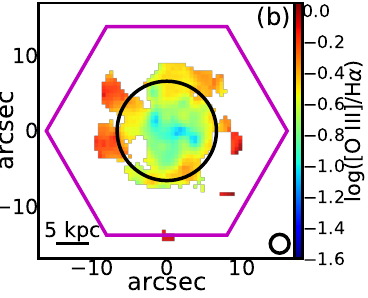}\!
\includegraphics[width=0.211\linewidth]{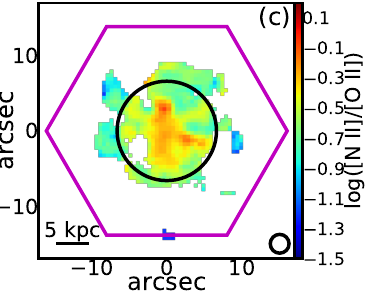}\!
\includegraphics[width=0.22\linewidth]{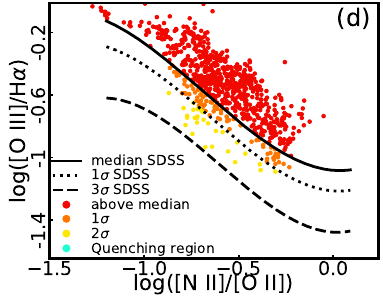} \!
\includegraphics[width=0.171\linewidth]{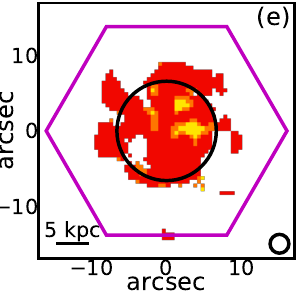} \\

\includegraphics[width=0.171\linewidth]{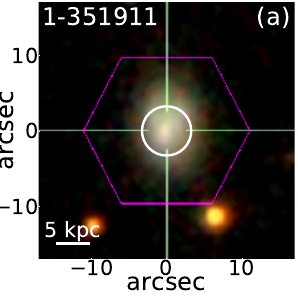}\!
\includegraphics[width=0.211\linewidth]{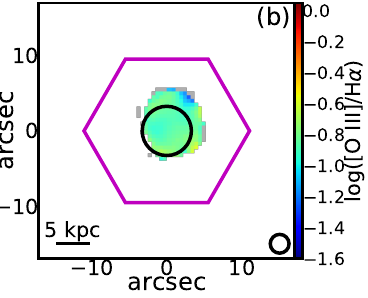}\!
\includegraphics[width=0.211\linewidth]{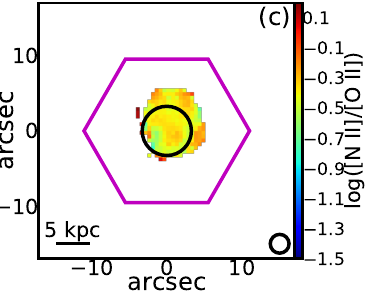}\!
\includegraphics[width=0.22\linewidth]{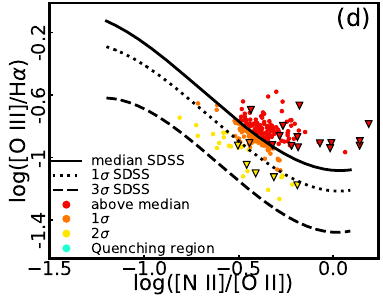} \!
\includegraphics[width=0.171\linewidth]{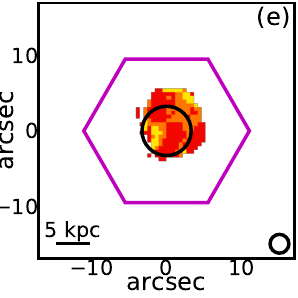} \\

\includegraphics[width=0.171\linewidth]{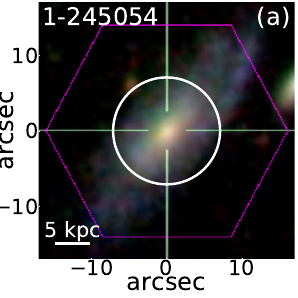}\!
\includegraphics[width=0.211\linewidth]{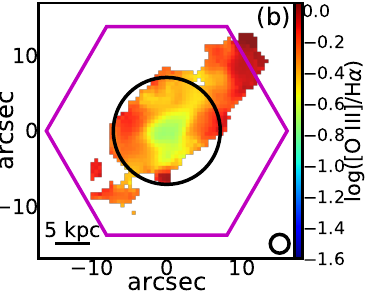}\!
\includegraphics[width=0.211\linewidth]{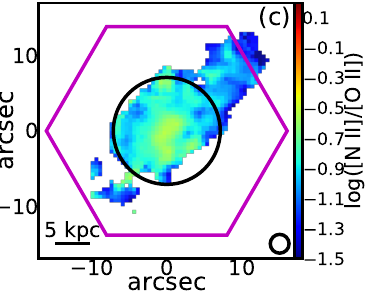}\!
\includegraphics[width=0.22\linewidth]{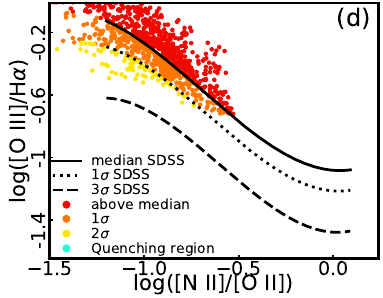} \!
\includegraphics[width=0.171\linewidth]{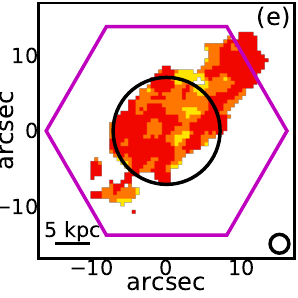} \\ 
\captionof{figure}{A summary of the $8$ SF galaxies in our sample.  
    (a) The g-r-i images composite from SDSS. Each image covers a region of $17 \times  17$ arcsec$^2$ and in the bottom-left corner of each image is reported the scale of $5$ kpc. (b) The dust-corrected \OH maps. The grey areas show regions with S/N(\OIII) $<2$.  (c) The dust-corrected \NO maps. (d) The \OH vs \NO diagnostic diagram for the quenching. The spaxels are colour-coded according to their position on the plane: red dots for those lying above the median curve, orange dots for them between the median and $1\sigma$ , yellow dots for spaxels which lie between $1\sigma$  and $3\times1\sigma$  and, finally, cyan dots for spaxels below the $3\times1\sigma$  curve. The triangles represent spaxels with an upper limit in \OH (i.e. spaxels with S/N(\OIII) $<2$). (e) The map of the galaxies colour-coded according to the position of spaxels as in (d).
   In (a), (b), (c) and (e) the overlapped-magenta hexagonal shapes the MaNGA IFU bundles, while the %white 
   circle represents the R$_{50}$. Finally, the $2.5\arcsec$ circle in the bottom-right corner of the maps in (b), (c) and (e) represent the typical PSF (FWHM) of MaNGA data. }    
    \label{fig:SF_info_1}
\end{minipage}
\end{figure*}
%\end{landscape}
%%%%%%%%%%%%%%%%%%%%%%%%%%%%%%%%%%%%%%%%%%%%%%%%%%%%%%%%%%%%%%%

%\begin{landscape}
\begin{figure*}\ContinuedFloat
\begin{minipage}{\linewidth}
\centering

%%%%%%%%%%%%%%%%%%%%%%%%%%%%%%%%%%%%%
\includegraphics[width=0.171\linewidth]{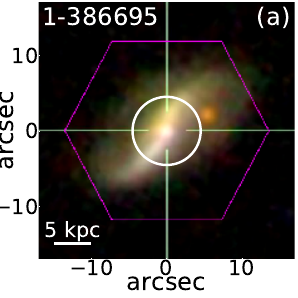}\!
\includegraphics[width=0.211\linewidth]{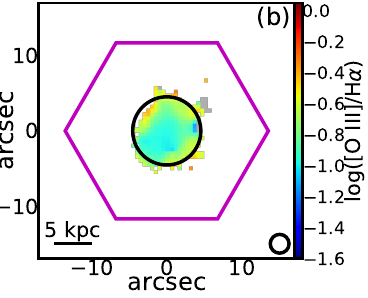}\!
\includegraphics[width=0.211\linewidth]{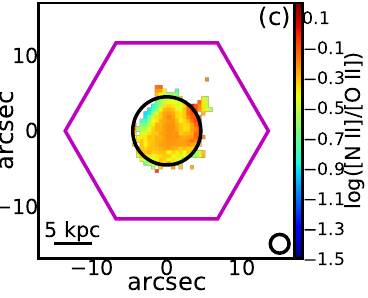}\!
\includegraphics[width=0.22\linewidth]{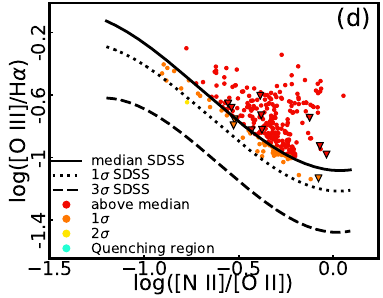} \!
\includegraphics[width=0.171\linewidth]{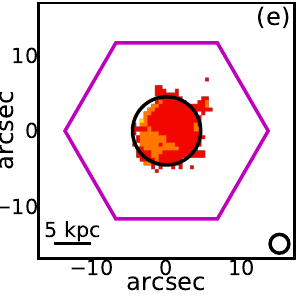} \\

\includegraphics[width=0.171\linewidth]{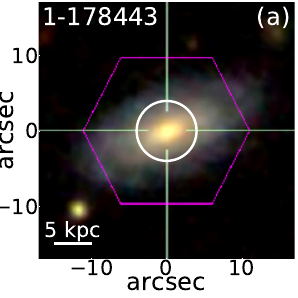}\!
\includegraphics[width=0.211\linewidth]{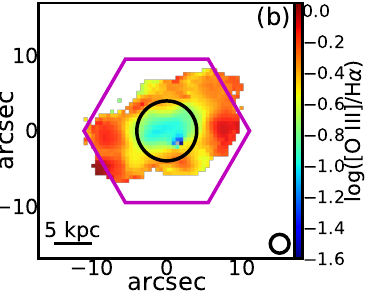}\!
\includegraphics[width=0.211\linewidth]{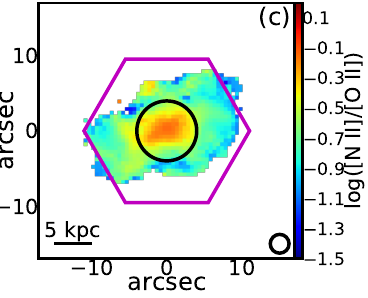}\!
\includegraphics[width=0.22\linewidth]{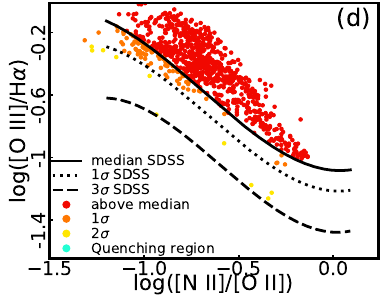} \!
\includegraphics[width=0.171\linewidth]{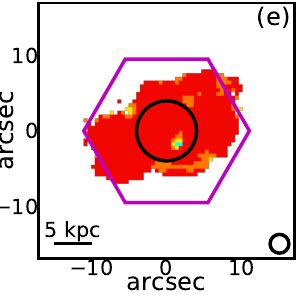} \\

\includegraphics[width=0.171\linewidth]{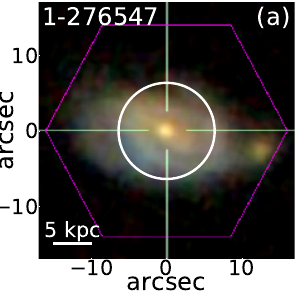}\!
\includegraphics[width=0.211\linewidth]{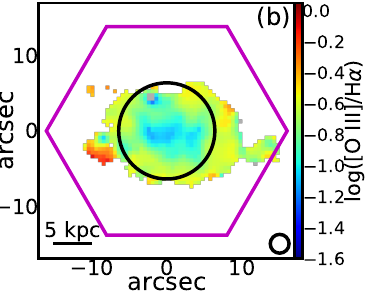}\!
\includegraphics[width=0.211\linewidth]{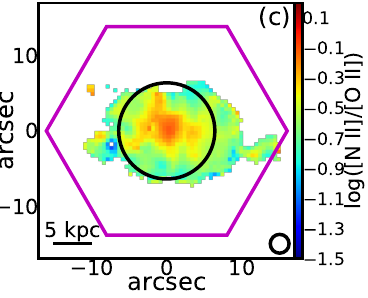}\!
\includegraphics[width=0.22\linewidth]{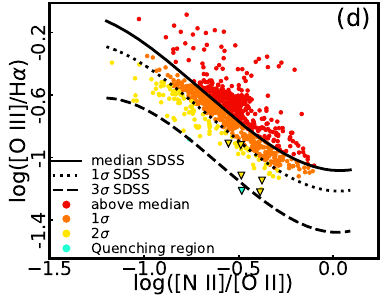} \!
\includegraphics[width=0.171\linewidth]{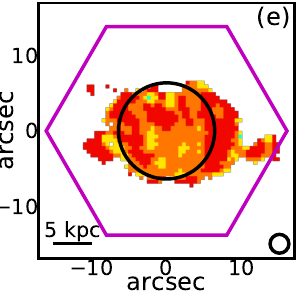} 
 \\
 
 \includegraphics[width=0.171\linewidth]{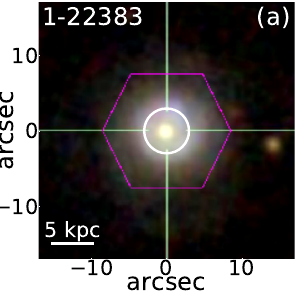}\!
\includegraphics[width=0.211\linewidth]{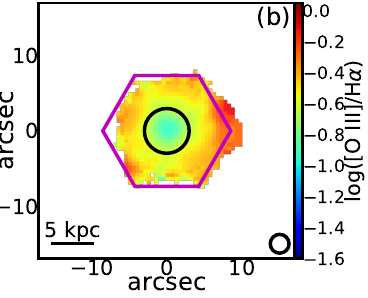}\!
\includegraphics[width=0.211\linewidth]{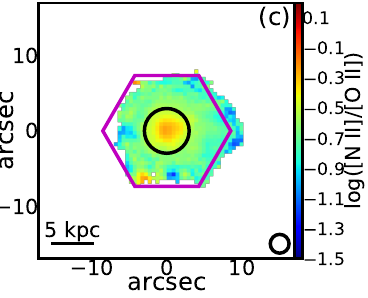}\!
\includegraphics[width=0.22\linewidth]{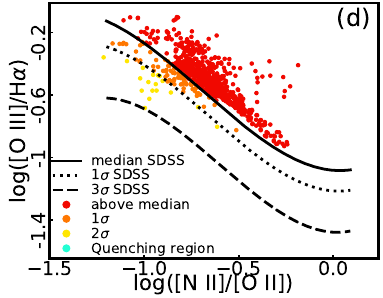} \!
\includegraphics[width=0.171\linewidth]{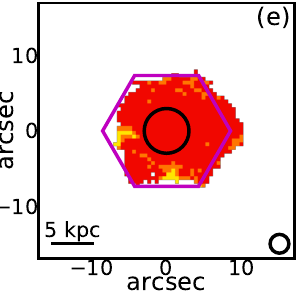} \\

\includegraphics{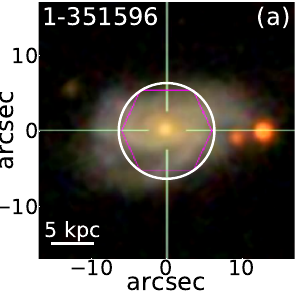}\!
\includegraphics[width=0.211\linewidth]{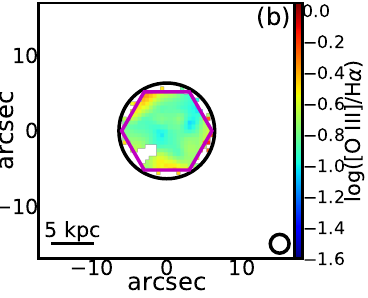}\!
\includegraphics[width=0.211\linewidth]{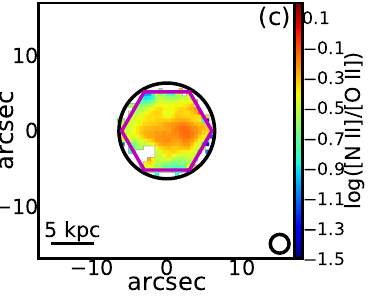}\!
\includegraphics[width=0.22\linewidth]{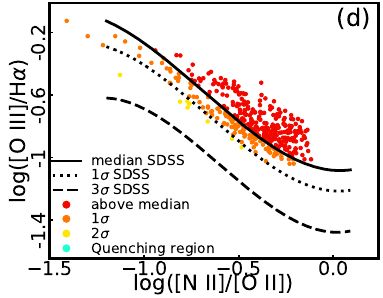} \!
\includegraphics[width=0.171\linewidth]{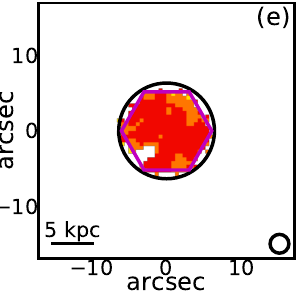} \\
   \captionof{figure}{Continued.}    
    \label{fig:SF_info_2}
\end{minipage}
\end{figure*}
%\end{landscape}

%\end{document}
%%%%%%%%%%%%%%%%%%%%%%%%%%%%%%%%%%%%%%%%%%%%%%%%%%%%%%%%%%%%%%%%%%%%%%%%%%%%%%%%%%%%%%%%%%%%%%%%%%%%%%%%%%%%%%%%%%%%%%%%%%%%%%%%

The dust corrected fluxes are converted to luminosity surface densities (erg s$^{-1}$ kpc$^{-2}$).
Then, the SFR surface density ($\Sigma$SFR) is derived using the dust corrected \halpha luminosity surface density and adopting the 
\citet{Kennicutt1998} conversion factor for \cite{Kroupa2001} initial mass function (IMF): 
 \begin{equation}
 \Sigma\text{SFR =} \Sigma (\text{L(H}\alpha)/10^{41.28}) \;[\text{M}_\odot \text{yr}^{-1} \text{kpc}^{-2}].
 \end{equation}
% to help the comparison between the galaxies. 

In order to obtain estimates of the ionisation parameter log~U and gas-phase metallicity Z from the observables, in the \OH vs \NO plane, we compared the observed values with a grid of theoretical values obtained with photo-ionisation models by \citetalias{Citro2017}.  
To do this, we interpolate the original models with a denser grid in which the theoretical Z spans from $0.004$ to $0.04$ with steps of $0.001$ and log~U from  $-3.6$ to $-2.5$ with steps of $0.01$. 
When a spaxel lies in a region of the diagram which is not covered by the models we assign the value linearly  extrapolated (see \autoref{fig:QGs_wDIST_Models}).  
This assumption has an impact on galactic regions with log(\NO)  higher than about $-0.1$ (e.g. spaxels in the central region of MaNGA 1-43012, see \autoref{fig:OIII_info_1}). 
We stress that these estimates of metallicity are not obtained from a calibration of the \NO or other emission line ratios \citep[e.g.][]{Nagao2006, Curti2017} but they are relative to the outcome of the photo-ionisation models by \citetalias{Citro2017} and they are indicative for separating galaxies with different gas-phase metallicity and should be considered as relative values.

Redshifts, optical colors and effective radii (R$_{50}$, i.e. elliptical Petrosian 50\% light radius in SDSS r-band) are obtained from the NASA Sloan Atlas v1\_0\_1 \citep{Blanton2011}, while NUV band magnitude are taken from the Galaxy Evolution Explorer \citep[GALEX,][]{Martin2005}.
Stellar masses, total star-formation rates (SFR) are taken from the database of the Max Planck Institute for Astrophysics and the John Hopkins University (MPA-JHU measurements\footnote{see \url{http://wwwmpa.mpa- garching.mpg.de/SDSS/}.}) as in \citetalias{Quai2018}.
We use also the SDSS morphological probability distribution of the galaxies provided by \citet{Huertas-Company2011}, which is built by associating a probability to each galaxy to belong to one of four morphological classes (Scd, Sab, S0, E).

%%%%%%%%%%%%%%%%%%%%%%%%%%%%%%%%%%%%%%%%%%%%%%%%%%%%%%%%%%%%%%%%
%-------------------------------------------%
\subsection{The classification scheme}
\label{sec:selection}
In this Subsection, we present the classification scheme applied to the $20$ MaNGA galaxies in our sample.
We stress that none of the \citetalias{Quai2018} best candidates from SDSS are in the MaNGA catalogue. 
Thus, we 
do not expect to find galaxies in an advanced phase of quenching, but more likely galaxies which could have just started it.

In \autoref{fig:OIII_info_1} and \autoref{fig:SF_info_1} we show the key information needed to characterise the sample, along with the g-r-i images from SDSS.  
Starting from the maps of dust corrected \OH (i.e. our observable for the ionisation status) and \NO (i.e. the observable for the metallicity) of each galaxy, we build the spatially resolved \OH vs \NO diagnostic diagram for the quenching.
We classify the spaxels into $4$ groups according to their position on the plane compared to the SDSS distribution: (i) spaxels lying above the median curve of the SDSS, represent galaxy regions whose ionisation status is compatible with ongoing star formation; (ii) spaxels between the median and the $1\sigma$ limit of the SDSS distribution, are regions characterised by slightly lower ionisation, though still compatible with emission due to star formation; (iii) spaxels between $1\sigma$ and $3\times1\sigma$ SDSS limits, are galactic regions in a grey area between star formation and quenching;  (iv) spaxels lying below the $3\times1\sigma$ limit of the SDSS distribution are galaxy regions which are likely experiencing the star formation quenching.

\subsection{The sample of Quenching Galaxy candidates and the sample of Star-Forming galaxies}
\label{sec:QGs_and_SFs}
Once that all the spaxels within each galaxies have been classified, we define as QGs (i.e. Quenching Galaxy candidates), those galaxies which have at least $1.5\%$ of their spaxels below the $3\times1\sigma$ curve, representing a conservative excess of spaxels with respect to those expected below the $3\times1\sigma$ (i.e. $\sim 0.13\%$) for a star-forming galaxy. The QGs will be further analysed as galaxies with regions potentially undergoing the quenching.

In particular, %LP% the already mentioned \autoref{fig:OIII_info_1} refers to the 
we find $10$ QGs galaxies which show such plausible quenching regions (see \autoref{fig:OIII_info_1}). 
On the contrary, the other $10$ galaxies do not show any sign of quenching, with the most of their spaxels lying above and along the median of the  SDSS star-forming galaxies relation, as shown in \autoref{fig:SF_info_1}. Hence, their behaviour in the \OH vs \NO diagram is consistent with that of a typical star-forming galaxy.  
We show in \autoref{fig:BPT} and in the on-line material that also their resolved BPT diagram confirms their star-forming nature. Hence, we can simply call them star-forming galaxies (SFs). In the following, we compare their properties (i.e. parameter of ionisation log~U, gas-phase metallicity Z, star formation rate densities $\Sigma$SFR, etc.) with those of the QGs ones. 

%Hence, we can adopt them as a sample of typical star-forming galaxies. 
%Moreover, in and we compare their properties with those of the QGs. 
%%%%%%%%%%%%%%%%%%%%%%%%%%%%%%%%%%%%%%%%%%%%%%%%%%%%%%%%%%%%%%%%%%%%%%%%%%%%%%%%%%%%%%%%%%%%%%%%%%%%%%%%%%%%%%%%%%%%%%%%%%%%%%%%
%\begin{landscape}
\setlength{\tabcolsep}{0.5em} % for the horizontal padding
\begin{table*}
%\begin{minipage}{\linewidth}
\centering
\captionof{table}{Main properties of our MaNGA QGs and SFs samples. In bold are indicated two galaxies analysed in detail in the text.}
\label{tab:sample_prop}
\begin{tabular}{lcccccccccccccc} 
sample & MaNGA-ID & z & RA & DEC & log(M$_*$/M$_\odot$) & E(B-V) & (NUV-u) & (u-r)& M$g$ & sSFR & R$_{50}$ & n (Sers.) & b/a  & Morph.  \\
& &  &  &  &   && & &[mag] & [yr$^{-1}$] & [arcsec]    & [arcsec] &  \\
 \hline 
QGs			&	1-379241		&	0.0405		&	119.3		& 	52.7		&	 9.74	  	&	0.18		&	1.63		&	1.68		&	-19.6		&	-10.5		&	3.1			&	1.3			&	0.5			&	Sab \\
			&	1-491193		&	0.0405		&	171.5		& 	22.1		&	 9.61		&	0.20		&	0.16		&	1.26		&	-19.3		&	-10.4		&	8.6			&	1.3			&	0.9			&	Scd\\
 			&	1-197045		&	0.0430		&	212.1		&	52.9		& 	 9.96		&	0.28		&	0.05		&	1.36		&	-19.5		&	-10.7		&	6.0			&	0.8			&	0.6			&	Sab\\
 			&	1-392691		&	0.0435		&	156.2		& 	36.0		&  	 9.75		&	0.00		&	0.55		&	1.44		&	-19.8		& 	-9.9 		&	6.4 		& 	1.3 		&   0.8 		& 	Scd\\
 			&	1-36645  		& 	0.0440 		&  	40.5 		& 	-1.0 		&  	 9.65		&	0.26		&	0.72		&	1.17		&	-19.0		&	-9.7  		&	6.6			& 	1.5 		&   0.8 	 	& 	/ \\
 			&	1-149235 		& 	0.0464 		& 	169.3 		& 	51.0 		&  	10.21		&	0.29		&	1.02		&	1.31		&	-20.1		& 	-10.1		&	3.1			& 	1.3 		&  	0.7	   		& 	Sab/Scd \\
 			&	1-338697 		& 	0.0499 		& 	115.0 		& 	43.0 		&  	10.19		&	0.28		&	/			&	1.24		&	-20.2		&	-10.0 		&	6.7			& 	1.0 		&  	0.9 		& 	Scd\\
 			&	1-373102 		& 	0.0511 		&  	223.7 		& 	30.6 		&  	10.17		&	0.26		&	0.46		&	1.24		&	-20.1		&	-9.9  		&	7.8 		& 	1.4 		&  	0.8 		& 	Scd\\
 			&	\bf{1-43012}  	& 	\bf{0.0527} &  	\bf{112.9} 	&  	\bf{38.3} 	&  	\bf{10.48}	&	\bf{0.33}	&	\bf{0.76}	&	\bf{1.56}	& 	\bf{-20.5}	&  	\bf{-10.5}	& 	\bf{6.2}  	&  	\bf{1.1} 	&  	\bf{0.8} 	&  	\bf{Scd}\\
 			&	1-91760  		& 	0.0660 		& 	240.0 		& 	54.8 		&  	10.76		&	0.37		&	/			&	1.40		&	-20.9		&	-10.2 		&	6.4  		& 	0.8 		&   0.9 		& 	Scd\\
\hline				
$<$QGs$>$ 	&  					& 	0.0478  	&  				&  				& 	10.05  		&	0.25		&	0.67		&	1.37		& 	-19.9 		&	-10.2 		& 	6.1 		& 	1.2  		& 	0.8  		&		\\
\hline				
\hline 				
SFs 		& 	1-258589 		&	0.0405 		&	186.7 		&	44.9 		&	 9.72		&  	0.35		&	0.56		&	1.14		&	-19.4  		&	-10.0 		& 	6.7    		& 	1.6   		&   0.9  		& 	/ \\
			&	1-351911 		&	0.0420 		&	122.0 		&	51.8   		&	 9.72		&  	0.31		&	/			&	1.12		&	-19.2  		&	-9.8		& 	2.8     	&   1.1   		&   0.7  		& 	Scd\\
			&	1-245054 		&	0.0428 		&	212.5 		&	53.6 		&	 9.88		&  	0.24		&	0.24		&	1.22		&	-19.7 		&	-9.9		& 	5.5    		&   2.2   		&   0.4  		& 	Sab\\
			&	1-386695 		&	0.0474 		&	138.0 		&	27.9 		&	10.11 		&  	0.22		&	1.08		&	1.27		&	-20.1 		&	-10.1		& 	3.7     	&   1.4    		&   0.3  		& 	Sab\\
			&	\bf{1-178443} 	&	\bf{0.0477} &	\bf{260.8} 	&	\bf{27.6} 	&	\bf{10.35} 	&  	\bf{0.30}	&	\bf{1.0}	&	\bf{1.28}	&	\bf{-20.5}  &	\bf{-9.9}	& 	\bf{3.2}    &  \bf{2.3}  	&  	\bf{0.5}  	&  \bf{Sab}\\
			&	1-276547 		&	0.0487 		&	163.5 		&	44.4 		&	10.20 		&  	0.45		&	0.90		&	1.11		&	-20.6 		&	-9.9		& 	5.2    		&   0.8  		&   0.5  		& 	Scd\\
			&	1-22383   		&	0.0542 		&	253.3 		&	64.5 		&	10.21 		&  	0.15		&	0.57		&	1.08		&	-20.7 		&	-9.6		& 	3.0    		&   1.5   		&   0.9  		& 	/\\
			&	1-351596 		&	0.0554 		&	118.6 		&	49.8 		&	10.41 		&  	0.36		&	0.99		&	1.33		&	-21.0 		&	-10.1		& 	5.3    		&   0.9  		&   0.5  		& 	Sab/Scd\\
\hline									  
$<$SFs$>$ 	&  					& 	0.0473  	&  				&  				& 	10.08  		&  	0.30		&	0.76		&	1.19		& 	-20.2 		&	-9.9 		& 	4.4 		& 	1.5  		& 	0.6  		&	\\
\hline
\end{tabular}		 
%\end{minipage}\hfill
\end{table*}

The main global properties of the QGs and SFs are listed in \autoref{tab:sample_prop}. 
By construction, the two samples have a similar stellar mass and redshift range, with an average (and also median) mass of $10^{10}$ M$_\odot$ and a mean redshift of z $\sim0.048$.
However, we find that two SF galaxies (i.e. 1352114 and 1-197704) have a central \NO $\sim -0.6$, which is $\approx 2$ dex lower than the lowest QGs. Therefore, their gas-phase metallicity is considerably lower than the metallicity range of the QGs sample.
We exclude these two objects, further analysing the remaining $8$ SFs galaxies.
In \autoref{fig:QGs_plane} we report the position in the \OH vs \NO diagram of the SDSS measures of the galaxies in the two samples. 

Both samples show, on average, a typical Sersic profile of disc galaxies (i.e. $<$n$_\text{Sersic}\!> 1.2-1.4$), and they show $<$b/a$>$ (i.e. the ratio between the semi-axis of the galactic plane) higher than $0.5 - 0.6$.
%, though QG 1-245686 and SF 1-386695 are almost edge-on (i.e. b/a $< 0.3$).
%The redshift of QGs appears, on average, a bit higher than the SFs one. However, if we exclude from the average the galaxy QG 1-38802, which is at z $\sim 0.11$,  we obtain a $<$z$_\text{QGs}\!> \sim 0.048$, that is almost equal to $<$z$_\text{SFs}\!>$. 
Instead, we find differences in the specific-SFR (sSFR) and R$_{50}$: the QGs  have, on average, lower sSFR and larger R$_{50}$ than the SFs ones. 
Moreover, we find that QGs have, on average, a slightly redder dust corrected colour (u-r) than SFs (i.e. (u-r) $\sim1.4$ and $\sim 1.2$, respectively).  Instead, QGs show a slightly bluer not dust corrected NUV-u colour than SFs (i.e. $\sim0.7$ and $\sim0.8$, respectively). It is not surprising that at these colours, galaxies of about $10^{10}$ M$_\odot$ lie below the Green Valley \citep[e.g.][]{Schawinski2014}. In fact, the evolution of the colours in quenching galaxies is slower than that of the emission line ratios and it requires timescales larger than $1$ Gyr to reach typical green valley colours \citepalias[e.g.][]{Citro2017}.
Finally, it is interesting to note that stellar masses, colours, SFRs and the other parameters measured in QGs are consistent with those of the quenching candidates derived by \citetalias{Quai2018}.

As mentioned earlier, we expect about $50\%$ of the [\ion{O}{III}]undet SDSS galaxies to be in quenching, and we find that $5$ out of the $8$ analysed [\ion{O}{III}]undet galaxies belong to the QG sample, while the other ones are actually star-forming galaxies. The discrepancy can be ascribed to an increased deepness of MaNGA data with respect to the SDSS ones, resulting 
in a still weak, but measurable \OIII (thanks to the higher S/N).
%\OIII still weak, but measurable with a higher S/N.
Instead, it is interesting that 5 out of the 12 galaxies originally selected as star-forming are instead classified as QG galaxies. 
We will investigate the distribution of the quenching regions within QGs in the following sections. Here we mention that they
%Later, we will investigate the distribution of the quenching regions within QGs, however, we anticipate here that they 
are mainly placed off-centre, which explain why the regions inside the SDSS fibre have been classified as star-forming.

%For continuity with the nomenclature introduced in \autoref{ch:Catching}, we target these objects as Galaxies with Quenching Regions ({\bf QGs}). 
%\end{landscape}

%%%%%%%%%%%%%%%%%%%%%%%%%%%%%%%%%%%%%%%%%%%%%%%%%%%%%%%%%%%%%%%%%

To summarise, according to the distribution of the spaxels on the \OH vs \NO diagnostic diagram for the quenching, we obtain two MaNGA samples:
\begin{itemize}
\item {\bf QGs:} $10$ galaxies that show regions (at least $1.5\%$ of the total galaxy) satisfying our quenching criteria (i.e. lie below the $3\times1\sigma$ of the SDSS star-forming distribution). 
\item {\bf SFs:} $8$ star-forming galaxies which have same redshifts,  stellar masses and gas-phase metallicity range of the QGs.
\end{itemize}
In \autoref{sec:global_trends}, we will extensively analyse the global behaviours of the two samples and we will compare their properties. In the next section, we will focus on the study of two galaxies, one for each sample, with the purpose of providing the details of the analysis that we performed on each galaxies in our sample.

%%%%%%%%%%%%%%%%%%%%%%%%%%%%%%%%%%%%%%%%%%%%%%%%%%%%%%%%%%%%%%%%%
\subsection{The impact of dust extinction on ionisation and metallicity indicators}
As shown in \citetalias{Quai2018}, we can mitigate the U-Z degeneracy using the resolved \OH vs \NO diagram.
The wavelength separation between the lines in the two ratios requires caution because of the not negligible effect of dust extinction. The classical approach relying on the Balmer decrement could be not accurate in recovering the intrinsic emission lines of an object deviating from the average star-forming galaxies. 
%We could use other line ratios less sensitive to this effect. For example, the \OHb ratio would have the same sensitivity to the ionisation parameter of \OH with the advantages to be less affected by dust extinction. To guarantee a high level of precision in the ratio measurement, we should impose an S/N(\hbeta) $\geq$ 5. However, this threshold would introduce a strong bias toward high SFR, to the disadvantage of the quenching galaxies we want to select. 
Other emission line ratios can be used which are less sensitive to this effect. For example, the \OHb ratio would have the same sensitivity to the ionisation parameter of \OH with the ad- vantages to be less affected by dust extinction. However, in order to guarantee a high level of precision in the ratio measurement, we should impose an S/N(\hbeta) $\geq 5$. This threshold would introduce a strong bias toward high SFR, penalizing the statistics  of the quenching  galaxies we are interested in selecting.
Therefore, in order to evaluate the impact of dust extinction, we tested an alternative diagnostic diagram, with \OHb not corrected for dust extinction (in place of dust-corrected \OH) vs dust-corrected \NO. 
%%%%%%%%%%%%%%%%%%%%%%%%%%%%%%%%%%%%%%%%%%%%%%%%%%%%%%%%%%%%%%%%
\begin{figure}
\centering
\includegraphics[width=0.8\linewidth]
{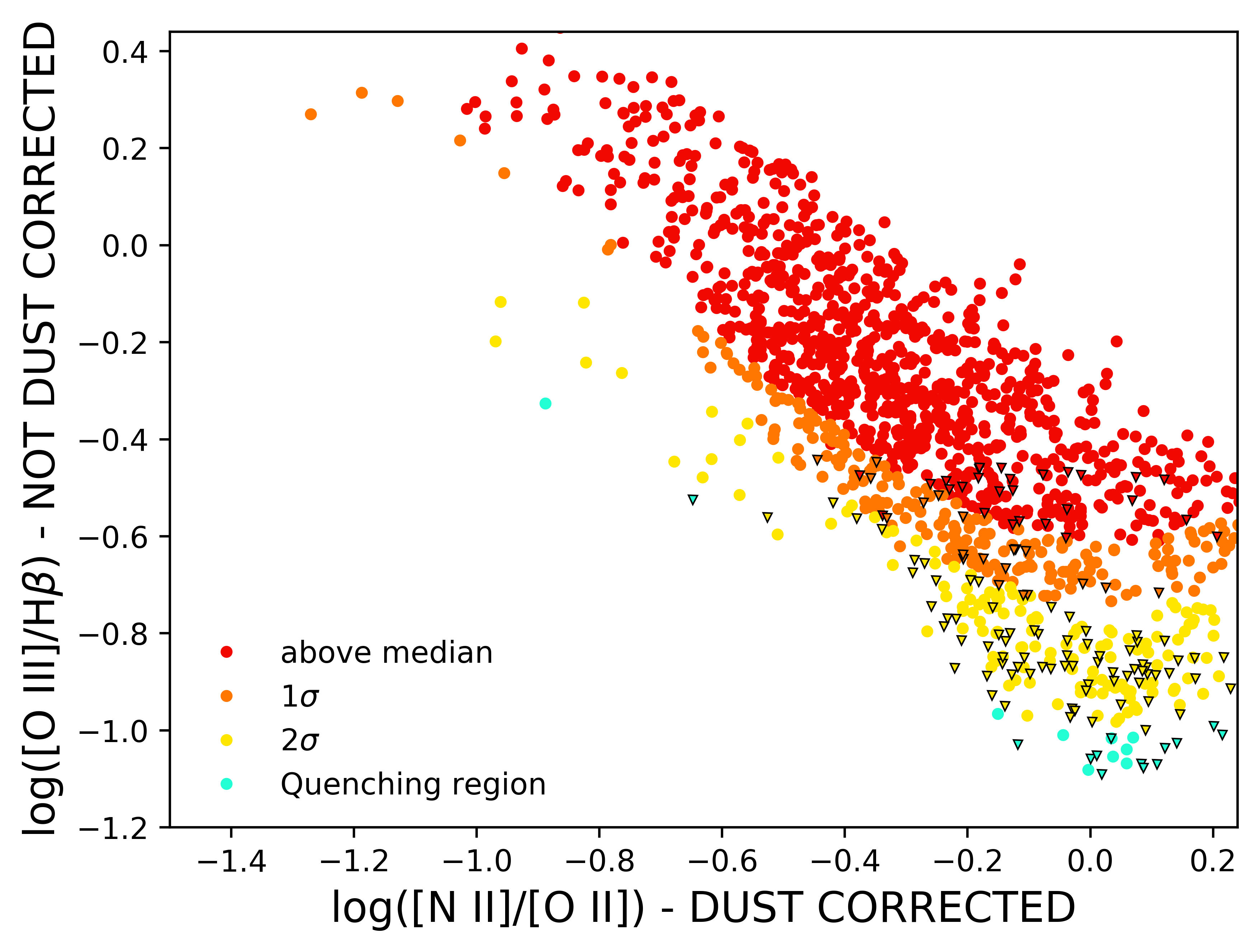} 
    \caption{The resolved \OHb (not corrected for dust extinction) vs \NO (corrected for dust extinction) diagram of \und 1-43012. The dots colour code is based on the position of each spaxel on the dust-corrected \OH vs \NO diagram, and it is the same as in \autoref{fig:QGs_plane}. The cyan is representing quenching regions, followed by the yellow for the galactic regions that lie between $3\times1\sigma$ and $1\sigma$ of the diagram, orange for those between $1\sigma$ and the median and red for regions of pure star-formation that are above the median of the diagram.}    
    \label{fig:hb_no}
\end{figure}
\autoref{fig:hb_no} shows the \OHb vs \NO of QG 1-43012. We find that the spaxels classified as quenching regions according to their position on the \OH vs \NO diagram (i.e. the spaxels lying below the $3\times1\sigma$ of the SDSS relation, see \autoref{fig:OIII_info_1}) remain those showing the lowest \OHb values at fixed \NO. We find the same result also in the other QGs (see the online materials), hence we can state that our classification and results do not depend on the dust correction.

We note also that the \OO ratio is sensitive to the ionisation status, as well. However, this ratio is also known to be rather sensitive to the gas-phase metallicity \citep[e.g.][]{Nagao2006}, and it would be less effective to mitigate the U-Z degeneracy.
Furthermore, the line ratio between \NeIII$\lambda3869$ and \OII (\NeO) is less affected by dust-extinction than \OH and it is sensitive to the ionisation level of a star-forming galaxy. 
However, the \NeIII line is usually faint to be detected at high S/N. Therefore, we are able to measure this line only in the central region of some galaxies in our SF sample.
%, making the interpretation of the result challenging and focused on the regions of the galaxies already covered by the SDSS fibre.
Finally, at larger wavelength, the line ratio [\ion{S}{iii}]/[\ion{S}{ii}] between the  lines \SIII$\lambda\lambda9060-9532$ and the doublet \SII$\lambda\lambda6726-6731$ is another ionisation tracer.
The [\ion{S}{iii}] lines are measurable in MaNGA data up to redshift z$\sim 0.08$, however, at the redshifts of our targets these lines end up in a spectral region dominated by a series of OH skylines, and therefore very difficult to be measured.
%and it results very difficult to interpret their values}. 
%To perform the continuum fit with pPXF, we cut the original spectra at $\lambda 7200$\AA. We try, therefore, to measure the \SIII lines by fitting the stellar continuum with a straight line but we found not reliable, faint and irregular residuals challenging to be interpreted as actual emission lines.

Similar remarks can be done about the metallicity indicator. We could, in principle, use different couples of emission lines closer in wavelength than \NII and \OII, whose ratio is sensitive to the gas-phase metallicity. For example, \NS shows a sensitivity to metallicity similar to the. However, the doublet \SII$\lambda\lambda6717-6731$ is considerably fainter than the \OII$\lambda\lambda3726-3729$ one and the cut in S/N with \NS would end up in excluding wider galactic area than with \NO.
%However, the test of the equivalence between \OHb not corrected for dust extinction and \OH allows to safely use \NO as metallicity indicator.

Finally, we need to assess how much dust-obscuration corrections affect the measurement in the \OH vs \NO diagnostic diagram. In \autoref{fig:QGs_plane} and \autoref{fig:QGs_wDIST_Models} show the direction of dust vectors obtained by assuming the \citet{Calzetti2000} extinction law for an E(B-V)$=0.3$ (the direction is the same for the \citet{Cardelli1989} extinction law). The direction is almost parallel to the median, the $1\sigma$ and $3\sigma$  curves of the distribution and also to the iso-U lines of the \citetalias{Citro2017} models. This test guarantees that different dust laws do not affect our results.

Summarising, we can conclude that the \OH vs \NO diagram is robust against the Balmer decrement approach for correcting dust extinction and that these line ratios are the most suitable for mitigating the U-Z degeneracy.

%In order to obtain estimates of the ionisation parameter log~U and gas-phase metallicity Z from the observables, in the \OH vs \NO plane, we compared the observed values with a grid of theoretical values obtained with photo-ionisation models by \citetalias{Citro2017}.  
%To do this, we interpolate the original models with a denser grid in which the theoretical Z spans from $0.004$ to $0.04$ with steps of $0.001$ and log~U from  $-3.6$ to $-2.5$ with steps of $0.01$. 
%When a spaxel lies in a region of the diagram that is not covered by the models we assign the values of the closest knot on the grid. 
%This assumption has an impact in galactic regions with \NO  higher than $0$, for which the metallicity estimation of Z $ = 0.04$ shall be regarded as a lower limit. 

%%%%%%%%%%%%%%%%%%%%%%%%%%%%%%%%%%%%%%%%%%%%%%%%%%%%%%%%%%%%%%%%%
\section{Results I. A detailed case study}
\label{sec:cases}

We will discuss the general results of the two populations in \autoref{sec:global_trends}, presenting individual details of the objects in our sample in the online materials.
With the purpose of illustrating our research method, we show here the detailed analysis of two objects: QG 1-43012 representing an example of a \und, and SF 1-178443 among the galaxies in the SF sample.
We choose SF 1-178443 because it has mass and redshift similar to those of QG 1-43012. This allows a direct comparison of the two systems, 
especially in terms of the 
%in terms, in particular, of 
ionisation parameter.
\begin{figure*}
%\vspace{\baselineskip}
%\noindent
%\begin{minipage}{\linewidth}
\centering
\includegraphics{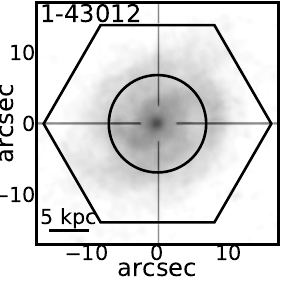}   
\includegraphics{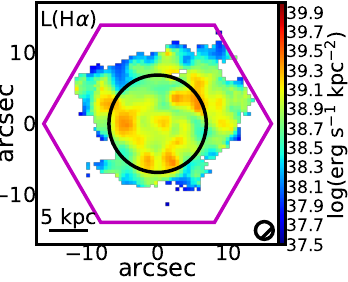}  
\includegraphics{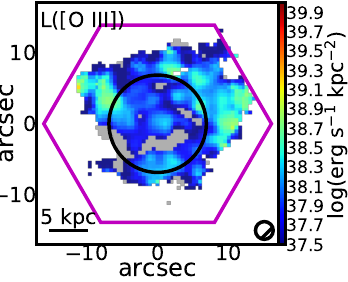}  
\includegraphics{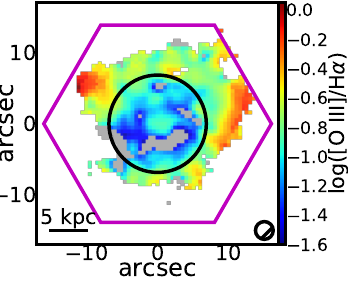} 
\includegraphics{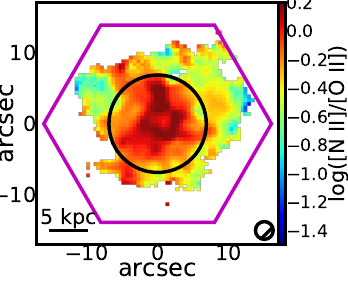}  \\

\includegraphics{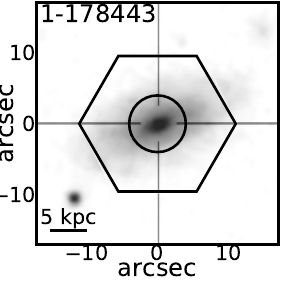}  
\includegraphics{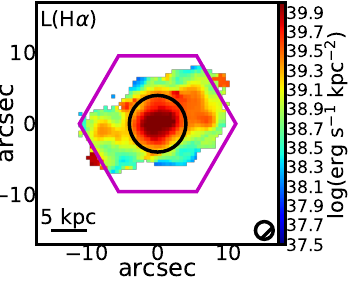}  
\includegraphics{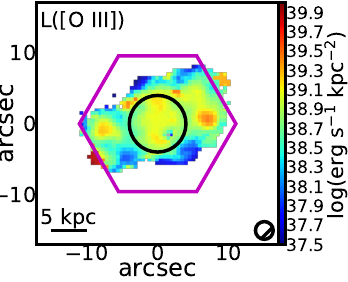}
\includegraphics{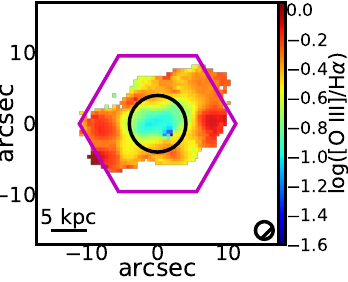} 
\includegraphics{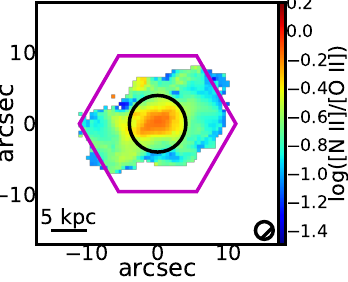} 
    \captionof{figure}{Spatially resolved maps of the two case galaxies. Each rows shows maps of \und 1-43012 (top) and maps of SF 1-178443 (bottom). Each column shows, from left to right (1) r-band images (2) luminosity surface density maps of dust-corrected \halpha, (3) luminosity surface density maps of dust-corrected \OIII, (4) dust corrected \OH maps and (5) dust corrected \NO maps. Spaxels coloured in grey represent regions with S/N(\OIII)$<2$).  Overlapped in magenta are the hexagonal shapes of the MaNGA IFU bundles, while the black circles represent the R$_{50}$. The $2.5\arcsec$ circle in the bottom-right corner of the maps represent the PSF (FWHM) of MaNGA datacubes.
    The galaxies go over the edge of the IFU shape because of the effect of the dithering, resulting in a coverage of a larger area of the sky. 
    }    
    \label{fig:maps_comp}
\end{figure*}
%\end{minipage}
%%%%%%%%%%%%%%%%%%%%%%%%%%%%%%%%%%%%%%%%%%%%%%%%%%%%%%%%%%%%%%%%

%%%%%%%%%%%%%%%%%%%%%%%%%%%%%%%%%%%%%%%%%%%%%%%%%%%%%%%%%%%%%%%%
\subsection{Emission lines maps}

\autoref{fig:maps_comp} shows the r-band image, the \halpha and \OIII luminosity surface density maps and the  \OH and \NO maps for the two galaxies. 
We find some differences, both structural and physical, between the two targets.  
They differ in size, being the SF smaller by a factor of $\sim0.5$ than the \und one (i.e. R$_{50}$ $\sim 4.5$ kpc and $\sim 6$ kpc, respectively) despite they have similar masses (i.e. \mass $=$ 10.48 and 10.35, respectively).
The analysis of \autoref{fig:maps_comp} shows that:
\begin{itemize}
\item QG 1-43012 has some spiral arms in the r-band and. Therefore, according to the morphological probability distribution of the SDSS galaxies provided by \citet{Huertas-Company2011}, it can be classified as a Scd galaxy. Instead, it remains difficult to see any significant spiral arm in the r-band image of SF  1-178443, while it shows a prominent bulge (or pseudo-bulge) and it has been classified as a Sab galaxy. 
\item The \halpha emission is not homogeneously distributed in the QG, showing clumps which reach the maximum intensity of
%The \halpha emission is distributed not homogeneously in the \und.  It shows clumps which reach a maximum intensity of 
$\Sigma$log~L(\halpha) $\sim 39.3$ erg s$^{-1}$ kpc$^{-2}$.
The SF galaxy has a \halpha distribution which is mostly concentrated and homogeneously distributed in the region inside the effective radius, where the emission reaches at values higher than $\Sigma$log~L(\halpha)  $\sim 40$ erg s$^{-1}$ kpc$^{-2}$ and then degrades at lower values toward the outskirts. 

\item The \und has a globally weak emission in \OIII, that rarely exceeds $\Sigma$log~L(\OIII) $\sim38.5$ erg s$^{-1}$ kpc$^{-2}$ and, as a result, the $12.6\%$ of its spaxels have an upper limit in \OIII (i.e. S/N(\OIII)$<2$). 
Instead, the \OIII emission of the SF galaxy follows the pattern of the \halpha although being slightly weaker, as we expected since it arises from stellar ionising sources. 
In this case, only a few spaxels (i.e. $0.4\%$) have S/N(\OIII) $<2$. 

\item The distribution of \OH ratio (i.e. our ionisation level indicator) in the \und\ (see \autoref{fig:maps_comp}) does not show a uniform gradient from the centre towards outer regions, but it reaches a minimum in an irregular annular region between $\sim 2$ and $\sim 5.5$ kpc (i.e. between $\sim0.3$ and $\sim0.9$ R/R$_{50}$) around the centre of the galaxy, then increasing towards more considerable distances. 
Instead, in the case of the SF galaxy, the \OH shows a typical gradient with the \OH raising from the centre towards the outskirts of the galaxy.

\item The distribution of \NO  (i.e. our metallicity indicator) shows an opposite behaviour with respect to \OH, in both \und and SF galaxies, with values increasing towards the inner parts of the galaxies.  
This relation between the \OH and \NO distribution is in part due to the well-known U-Z degeneracy between the ionisation parameter and gas metallicity \citep[see][]{Citro2017, Quai2018}. 

\end{itemize}

% AAA
%%%%%%%%%%%%%%%%%%%%%%%%%%%%%%%%%%%%%%%%%%%%%%%%%%%%%%%%%%%%%%%%
\subsection{The quenching diagnostic diagram}

\subsubsection{The \OH vs \NO diagram}
%%%%%%%%%%%%%%%%%%%%%%%%%%%%%%%%%%%%%%%%%%%%%%%%%%%%%%%%%%%%%%%%
\begin{figure}
\centering
%\includegraphics[width=0.4\linewidth]
%{plotQGsA_weightedDist_withModels_OIIIundet_MANGAid_8135-12702_vTHESIS} 
%\includegraphics[width=0.4\linewidth]
%{plotQGsA_weightedDist_withModels_SF_MANGAid_7962-6104_vTHESIS} \\
\includegraphics{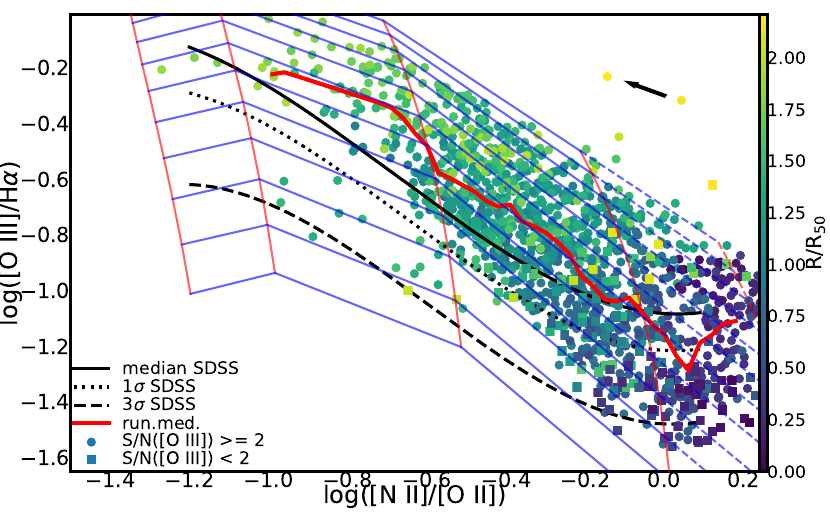} 
\includegraphics{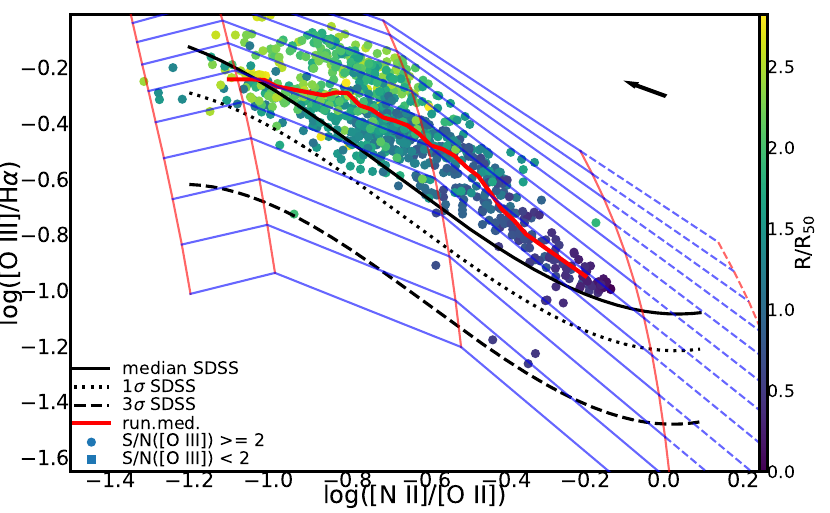}
\\
    \caption{The resolved \OH vs \NO diagram of \und 1-43012 (\emph{left}) and SF 1-178443 (\emph{right}). Each round dot represents a spaxel in which the S/N(\OIII) $\geq 2$, while the square dots represent spaxels in which the S/N(\OIII) $<2$ and their \OH values are upper-limits. The colours of the dots change according to the distance R/R$_{50}$ of the spaxels from the centre of the galaxy. The red curve represents the running median (continue) of the relation. Instead, the black curves (polynomial of degree 4) represent the median (continue),$1\sigma$ (dotted) and $3\times1\sigma$ (dashed) of the distribution of SDSS star-forming galaxies \citepalias[see][]{Quai2018}. Superimposed is reported the grid of photo-ionisation models by \citetalias{Citro2017}, with the red straight lines representing different metallicities (i.e. Z = $\{0.004, 0.008, 0.02, 0.04\}$ from left to right) and the blue straight lines representing different levels of the ionisation parameter U (i.e. from log~U -2.3 in the top to -3.6 in the bottom). The blue and red dashed lines represent the model values linearly extrapolated beyond the coverage of the model grid up to Z$ = 0.054$. The black arrow in the top-right corner represents the direction of the dust vectors for the \citet{Calzetti2000} extinction law, for an E(B-V)$=0.3$.}    
    \label{fig:QGs_wDIST_Models}
\end{figure}
%%%%%%%%%%%%%%%%%%%%%%%%%%%%%%%%%%%%%%%%%%%%%%%%%%%%%%%%%%%%%%%%
In \autoref{fig:QGs_wDIST_Models} we show the \OH vs \NO  diagram  of the two galaxies with the spaxels coloured according to their galactocentric distance and a grid of ionisation models by \citetalias{Citro2017}. 
The U-Z degeneracy is strongly mitigated, with the gas-phase metallicity Z increasing with \NO while the ionisation parameter log~U varying with \OH at fixed metallicity.
%In \autoref{fig:QGs_wDIST_Models} we also show the \OH vs \NO diagrams of the two galaxies, with the spaxels coloured as a function of the distance from the centre. 
Results from \autoref{fig:QGs_wDIST_Models} suggest that a negative gradient of metallicity with radial distances is present in both galaxies.
However, in SF 1-178443 at fixed metallicity, the ionisation parameter does not vary significantly while in \und 1-43012 it shows a large spread revealing differences in the ionising stellar populations in different regions of the galaxy. 
We stress that in this plane, at fixed values of \NO (i.e. fixed metallicity), spaxels lying below the $3\times1\sigma$ limit curve of the relation obtained from the star-forming population of SDSS represent regions compatible with the quenching. 
This region corresponds roughly to a log~U $< -3.4$. 
While in next sections we show in more details the U and Z profiles for our targets, from \autoref{fig:QGs_wDIST_Models} is already evident that the \und, on average, is more metallic than the SF one. 
About $72\%$ of its spaxels have a super-solar metallicity (i.e. Z$>0.02$), against $15.6\%$ of the SF one. 
Moreover, the spaxels of the \und are spread across the entire plane covering the entire scale of ionisation levels, from log~U $-2.4$ to $-3.6$.
%, although the bulk of them (i.e. $\sim 85\%$) lie in the pure star-forming region above the $1\sigma$ curve of the SDSS distribution.
About $1.6\%$ of its spaxels are in the quenching region below the $3\times1\sigma$ curve of the SDSS distribution, and  $14\%$ of the spaxels lie between $1\sigma$ and $3\times1\sigma$. 
Instead, the $98\%$ of the spaxels of the SF galaxy are in the pure star-forming region,  above the $1\sigma$ curve of the SDSS distribution and with log~U higher than $-3.2$. 
\subsubsection{The maps of the quenching regions}
\begin{figure}
\centering
\includegraphics{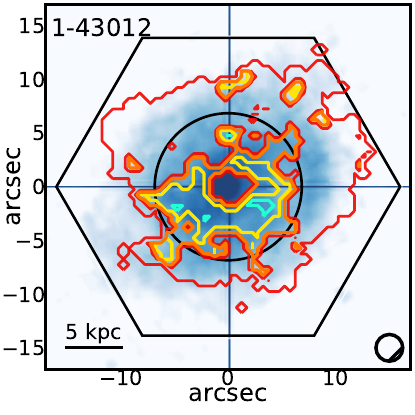} 
\!\!\includegraphics{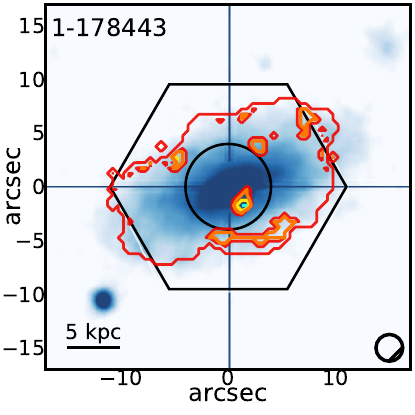} 
\captionof{figure}{The contours of the resolved \OH vs \NO diagram are superimposed to the G images (in false colours) of \und 1-43012 (\emph{left}) and  SF 1-178443 (\emph{right}). The contours colour code is based on the position of each spaxel on the diagram, and it is the same as in \autoref{fig:QGs_plane}. The cyan is representing quenching regions, followed by the yellow for the galactic regions that lie between $3\times1\sigma$ and $1\sigma$ of the diagram, orange for those between $1\sigma$ and the median and red for regions of pure star-formation that are above the median of the diagram. In the bottom-left corner is reported the scale of $5$ kpc.  Overlapped are the hexagonal shapes of the MaNGA IFU bundles, while the circles represent the R$_{50}$. The $2.5\arcsec$ circle in the bottom-right corner of the maps represent the PSF (FWHM) of MaNGA datacubes. }    
    \label{fig:map_quench}
\end{figure}

In \autoref{fig:map_quench} we show the contours of the resolved maps of the \OH vs \NO diagram for the two galaxies. 
For \und 1-43012 the quenching regions 
%LP% are mainly located in an irregular annulus around the centre of the galaxy, and they
cover an effective quenching area of $\sim7.1$ kpc$^2$, that becomes $\sim 67$ kpc$^2$ wide if we include also the spaxels lying between $1\sigma$ and $3\times1\sigma$ as regions in which the quenching could be started. 
Being contiguous to the proper quenching regions, it is likely that the quenching has started also in these regions.
%They are %LP% distributed around 
%contiguous to the proper quenching regions, and it is likely that the quenching is starting also in those regions. %LP% propagating in their direction. 
This extended quenching region is mainly located in an irregular annulus around the centre of the galaxy.
%%%%%%%%%%%%%%%%%%%%%%%%%%%%%%%%%%%%%%%%%%%%%%%%%%%%%%%%%%%%%%%%
%\begin{minipage}{\linewidth}
\begin{figure}
\centering
\includegraphics{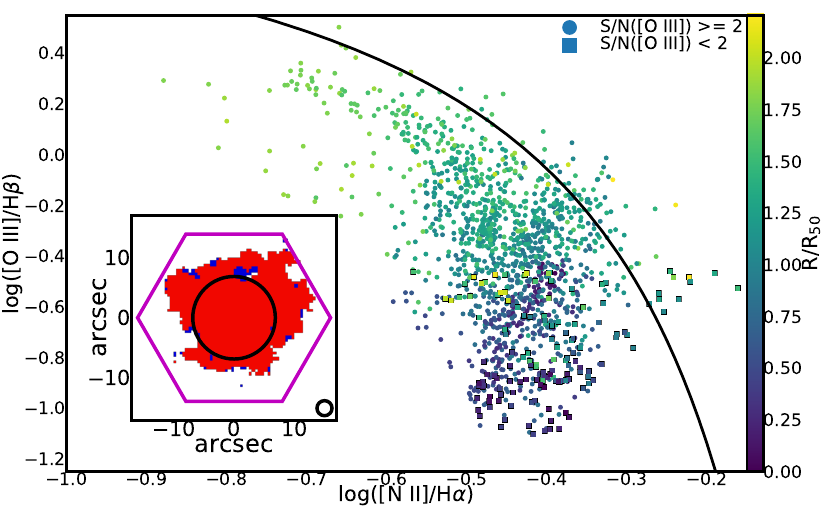} 
\includegraphics{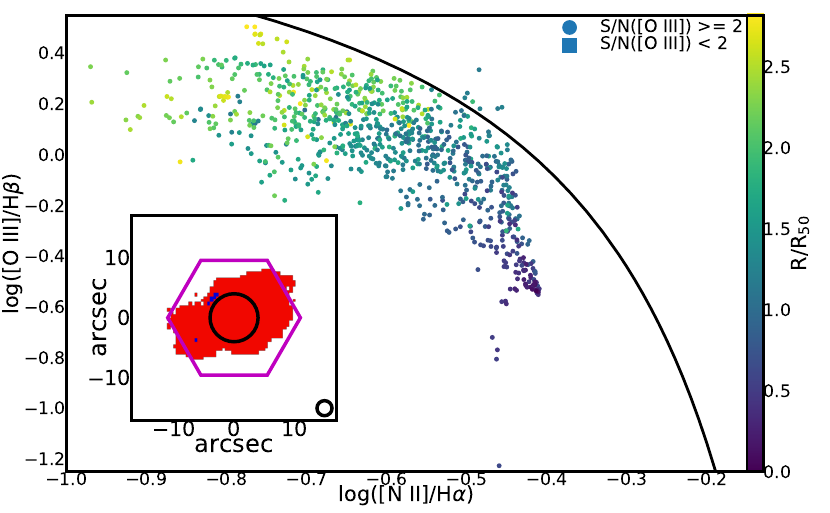} 
\captionof{figure}{The resolved BPT diagram of \und 1-43012 (\emph{top}) and of SF 1-178443 (\emph{bottom}). Each round dot represents a spaxel in which the S/N(\OIII) $\geq 2$, while the square dots represent spaxels in which the S/N(\OIII) $<2$ and their \OH values are upper-limits. The colours of the dots change according to the distance R/R$_{50}$ of the spaxels from the centre of the galaxy. The black curve is from \protect\cite{Kauffmann2003}.
The coloured maps represent the resolved BPT maps of the two galaxies. Spaxels with ionisation dominated by star formation (below the \protect\cite{Kauffmann2003} curve) are represented in red, while those whose the ionisation is dominated by AGN/LINERs radiation are represented in blue. Overlapped is the hexagonal shape of the MaNGA IFU bundle, while the circle represents the R$_{50}$. The $2.5\arcsec$ circle in the bottom-right corner of the maps represent the PSF (FWHM) of MaNGA datacubes.  }    
    \label{fig:BPT}
\end{figure}
%\end{minipage}
%%%%%%%%%%%%%%%%%%%%%%%%%%%%%%%%%%%%%%%%%%%%%%%%%%%%%%%%%%%%%%%%

\autoref{fig:BPT} shows the resolved diagnostic diagram of \citet[][herafter BPT]{Baldwin1981} for \und 1-43012 in which appears that the quenching regions are compatible with emission due to stellar ionisation, therefore, we can safely exclude the presence of an AGN. 
It should be noted that some spaxels, mostly located at the edge of the galaxy, lie above (but close) the BPT curve of \citet{Kauffmann2003} %that separates the region where the ionisation is due to star formation and that dominated by AGNs and LINERs systems. 
which distinguishes between galaxies where ionisation is due to star formation and the ones where it is due to AGN/LINER activity.
These spaxels are observed in almost all the  analysed galaxies (see the on-line material) and their behaviour is due to the uncertainties in \OH. 
The emission lines, indeed, become weaker towards the outskirts of galaxies, increasing the uncertainties of the emission line ratio measurements. 
For example, the typical S/N(\OH) within R$_{50}$ of \und 1-43012 is between $4$ and $25$, while it drops below $1.5$ above $\sim 1.6$ R$_{50}$.

%%%%%%%%%%%%%%%%%%%%%%%%%%%%%%%%%%%%%%%%%%%%%%%%%%%%%%%%%%%%%%%%
\subsection{Radial profiles}
%%%%%%%%%%%%%%%%%%%%%%%%%%%%%%%%%%%%%%%%%%%%%%%%%%%%%%%%%%%%%%%%
\begin{figure*}
%\begin{minipage}{\textwidth}
\centering
\includegraphics{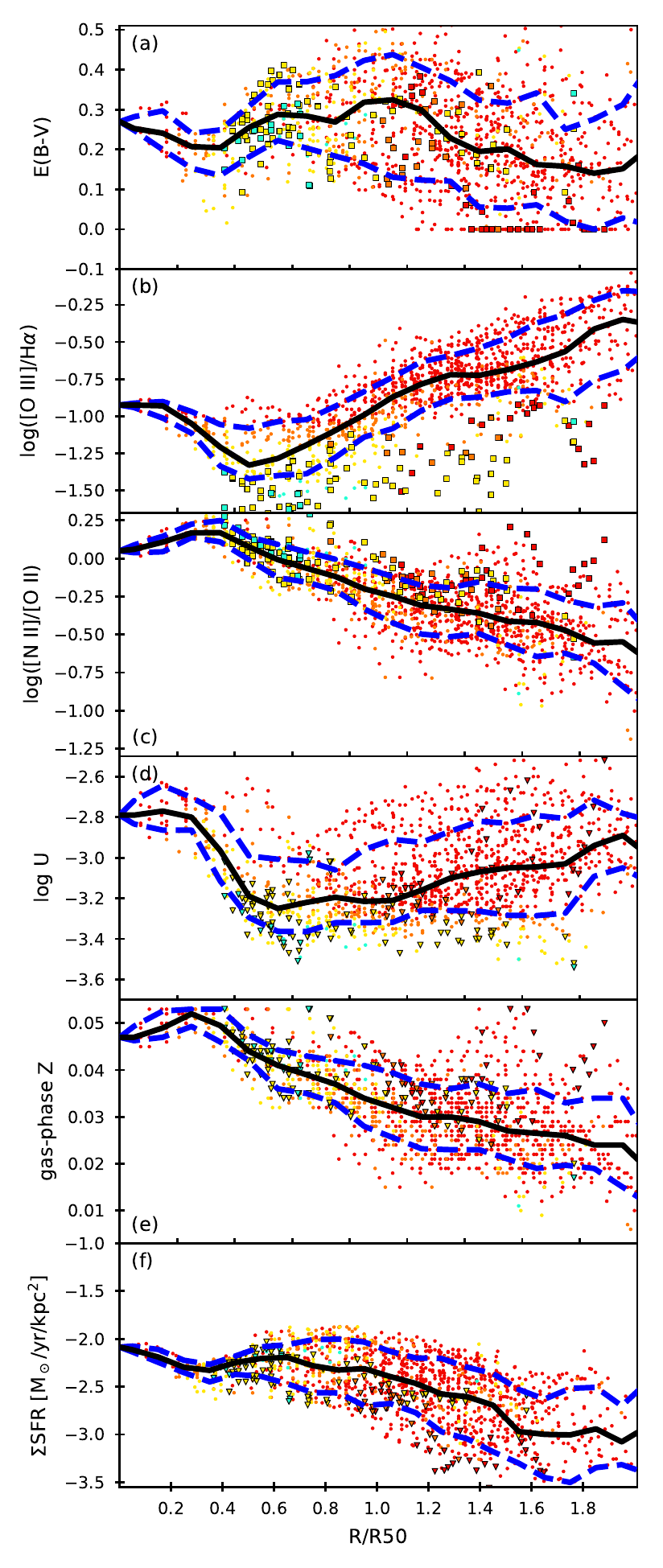}\includegraphics{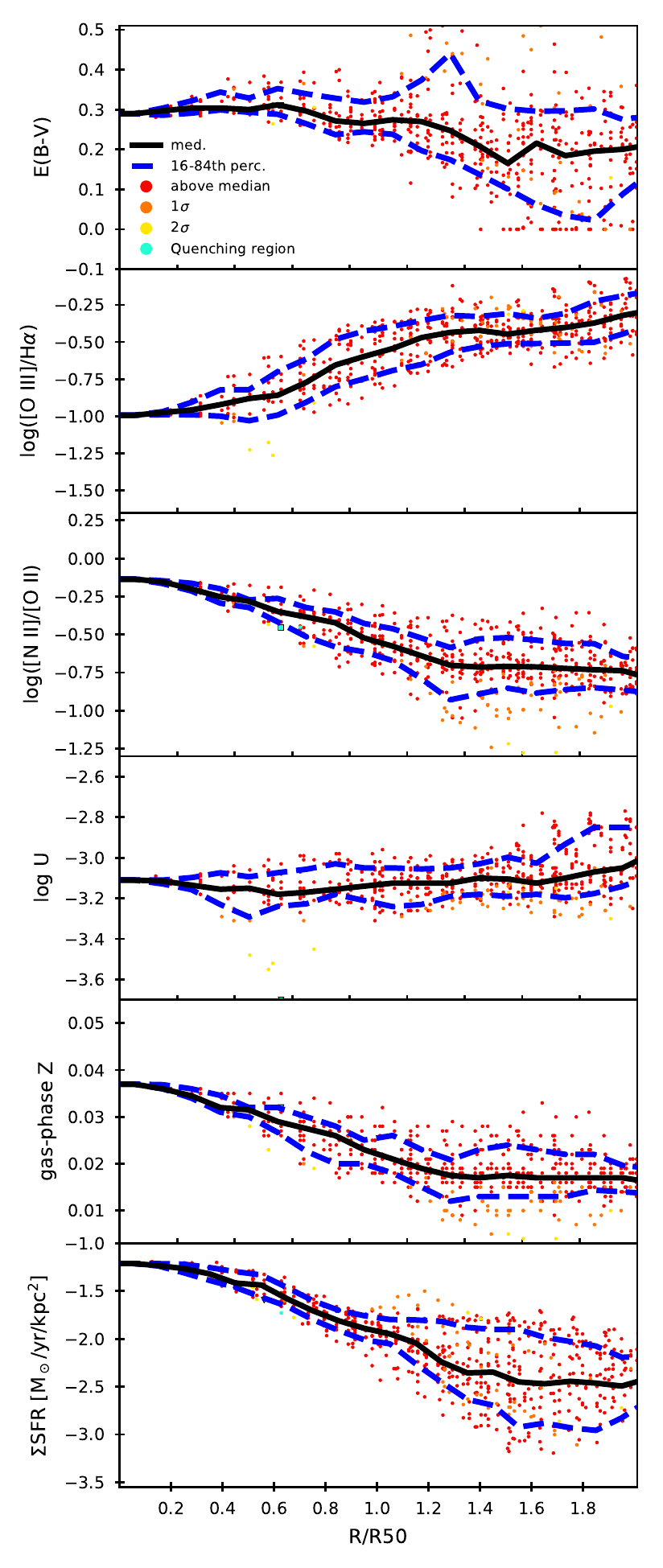}
 \captionof{figure}{Radial profiles of \und 1-43012 (\emph{left}) and of SF 1-178443 (\emph{right}): (a) E(B-V), (b) dust-corrected \OH, (c) dust-corrected \NO, (d) log~U, (e) gas-phase Z, (f) $\Sigma$SFR. 
  The black curves represent the median of the relations in bins of width $0.1$R/R$_{50}$ , while the blue ones represent the 16-84$^\text{th}$ percentile of the relations. Each round dot represents a spaxel in which the S/N(\OIII) $\geq 2$, while the square dots represent spaxels in which the S/N(\OIII) $<2$ and their \OH values are upper-limits. The dots colour code is the same as in \autoref{fig:map_quench}, and it is based on the position of each spaxel on the \OH vs \NO diagram (\autoref{fig:QGs_plane} and \autoref{fig:QGs_wDIST_Models}). The cyan is representing quenching regions, followed by the yellow for the galactic regions that lie between $3\times1\sigma$ and $1\sigma$ of the diagram, orange for those between $1\sigma$ and the median and red for regions of pure star-formation that are above the median of the diagram.}    
\label{fig:rad_prof_verNEW}
%\end{minipage}
\end{figure*}
In this section, we extend the analysis of the two galaxies by investigating the radial profiles of the main quantities used in this work.
%LP% We compute no projection of the galactic plane, and w
We normalise the distance to the elliptical R$_{50}$, that we consider as a circular radius.  
\autoref{fig:rad_prof_verNEW} shows the radial profiles of the
 colour excess E(B-V) and the observables \OH and \NO, the radial profiles of the parameters log~U and gas-phase Z and that of the star formation rate density. 
 Our findings can be briefly summarised as follows:
\begin{itemize}
\item The E(B-V) radial profile of QG 1-43012 is quite scattered between $0\leq$ E(B-V) $< 0.4$ at any radius. The spaxels
 marked as quenching regions show intermediate values of colour excess. 
The profile of SF 1-178443 is less scattered and it shows an almost flat median.
\item As mentioned in the previous sections, the \und 1-43012 \OH profile (i.e. ionisation level profile), shows a central peak, then it decreases down to a minimum log(\OH) $\sim -1.2 \pm 0.2$ between $\sim0.3 <$ R/R$_{50}$ $< \sim 0.75$, and steeply increases again towards larger radii.
 This minimum corresponds to the region in which 
almost all the spaxels compatible with quenching are concentrated 
% are concentrated almost all the spaxels compatible with quenching 
 (see also \autoref{fig:map_quench}).
Instead, SF 1-178443 shows a more homogeneous behaviour with a positive gradient in the \OH profile, which is steeper at small radii, while it grows slowly at larger radii. Moreover, the \OH values are higher than those of  \und 1-43012 at any radius.
 \item The profile of \NO for QGs follows a complementary pattern, with the off-centred peak  set near $0.25$ effective radius.
The relation is tight in the centre's proximity, with a $1\sigma$ scatter of about $0.1$ dex, but becomes higher than $0.2$ dex at radii larger than R$_{50}$. 
The SF galaxy shows lower \NO values, that suggests lower metallicity than \und 1-43012 at any radius, with a maximum value in the centre that decreases rapidly approximately at R$_{50}$, then becoming almost flat. 
\item The log~U profile of \und 1-43012 confirms the trend of the observable \OH, though with a larger spread.
It peaks at the centre of the galaxy, then decreasing down to a minimum
%It shows a peak in the centre of the galaxy that is followed by a decrement, which reaches the minimum 
log~U $= -3.2 \pm 0.1$ between $0.5$ and $0.75$ R$_{50}$ in correspondence of the \OH minimum. At larger radii the profile increases again, though the spaxels are scattered through all the available ionisation levels between log~U $-2.5$ and $-3.5$.
This findings suggests that the minimum in the observable \OH radial profile is due to a minimum in the ionisation level and not to the effect of the metallicity.
%In \autoref{sec:cases}, we find that the radial profiles of log~U (i.e. ionisation level) and $\Sigma$SFR  of \und 1-43012 have a minimum in an annular region around the centre of the galaxy and a corresponding maximum in the radial profile of the gas-phase metallicity. 
It is worth mentioning that other QGs show %\und 1-91760 shows a very similar pattern, with a vast annular region around the galactic centre with a low-ionisation level compatible with recent quenching.
%Other two QGs (i.e. 1-491193 and 1-392691) show 
these same features, though less clear (see online material and \autoref{sec:averageU})	. % than \und 1-43012 and 1-91760.}

Instead, the SF 1-178443 log~U radial profile is almost flat up to $1.8$ R$_{50}$, with log~U $\sim-3.1$,   suggesting that this star-forming galaxy is homogeneous in ionisation level and the increase of \OH is due the decreasing of \NO, i.e. metallicity. In general, the spaxels are less scattered than those of the \und and only a handful of them have log~U lower than $-3.2$. 
\item The gas-phase metallicity radial profile of \und 1-43012 can be adequately studied only at radii larger than $0.5$ R$_{50}$.
In fact, at closer distances, the metallicity estimate is linearly extrapolated from the \citetalias{Citro2017} models grid beyond its Z$=0.04$ limit, and up to Z $=0.054$. At such high metallicity, a secondary nucleosynthesis origin of the nitrogen could explain this behaviour.
In these circumstances, indeed, the \NO ratio (i.e. a tracer of the N/O ratio) overestimates the oxygen abundances (i.e. O/H ratio), leading to higher values of gas-phase metallicity. 
%For this reason, the plausible maximum in the metallicity profile, that can be inferred from the observable \NO radial profile, is hidden and it should be revealed using models which can reach higher metallicity.
The resulting gas-phase metallicity radial profile follows a trend with a peak near $0.3$R$_{50}$, that is similar to that of the observable \NO, and it is a typical metallicity profile found in galaxies with similar stellar mass \citep[e.g.,][]{Sanchez2014}.
% and as for the galaxy QG 1-91760 which shows a metallicity peak around 0.3 R$_{50}$ (see the online material).}
%LP% Also the galaxy \und 1-91760 (see the online materials) shows characteristics similar to those of \und 1-43012, though having, on average, a lower \NO (i.e. lower gas-phase metallicity). 
%LP%Its Z radial profile increases from the center of the galaxy and it reaches a maximum value around $0.3$ R$_{50}$ (see the online materials). 
%LP%\textcolor{blue}{This behaviour suggests that the true Z radial profile for \und 1-43012 should resemble to that of \und 1-43012} . At radii larger than $0.5$R$_{50}$ the Z profile decreases following the classical negative gradient, though showing a large spread.
% (e.g. at R$_{50}$ the metallicity is Z $=0.029 \pm 0.007$).
The Z radial profile of SF 1-178443 shows a negative gradient up to $1.25$ R$_{50}$; then it becomes almost flat.
In this galaxy, the gas-phase metallicity is lower than that of the \und one at any radius, and it shows a smaller spread.

\item The log~$\Sigma$SFR radial profiles of the two galaxies are different, with \und 1-43012 having $\Sigma$SFR values lower than those of SF 1-178443 at any radius.
Its radial profile shows a weak negative gradient. %between log~$\Sigma$SFR $\sim -2$ in the center to $\-3 \pm 0.4$ at $2$ R$_{50}$. 
A relative minimum is also visible around $0.3$ R$_{50}$, although less evident than that of the \OH  and log~U profiles.
%It can also be seen a relative minimum around $0.3$ R$_{50}$ but less evident than that of the \OH \textcolor{red}{\bf and log~U} profiles.
Instead, the $\Sigma$SFR radial profile of SF 1-178443 shows a negative gradient with a steeper slope until $\sim 1.25$ R$_{50}$, then the profile becomes almost flat. 
\end{itemize}

To summarise, the two galaxies show different characteristics, in terms of ionisation, gas-phase metallicity and distribution of star-formation rate surface density across the galactic plane. 
\section{Results II. Global comparison of QG and SF galaxies}
\label{sec:global_trends}
%%%%%%%%%%%%%%%%%%%%%%%%%%%%%%%%%%%%%%%%%%%%%%%%%%%%%%%%%%%%%%%%
In the previous section we gave details about the method we applied to each galaxy in our \und and SF samples. 
In this section, we present the general behaviour of the two samples. 
We stress that the two samples are in the same redshift range and they have same  average stellar mass and central gas-phase metallicity, therefore, we can directly compare their properties. We focus on the \OH vs \NO diagram, and on the average radial profiles of the quantities we showed in the previous section (i.e. E(B-V), \OH, \NO, log~U, gas-phase Z, $\Sigma$SFR). For each analysed  property we define the significance as the distance of the average differences in units of $\sigma$, where $\sigma$ is the error in the average.

\subsection{The average \OH vs \NO profile}
\begin{figure}
%\vspace{\baselineskip}
%\noindent
%\begin{minipage}{\linewidth}
\centering
\includegraphics[width=1\linewidth]
{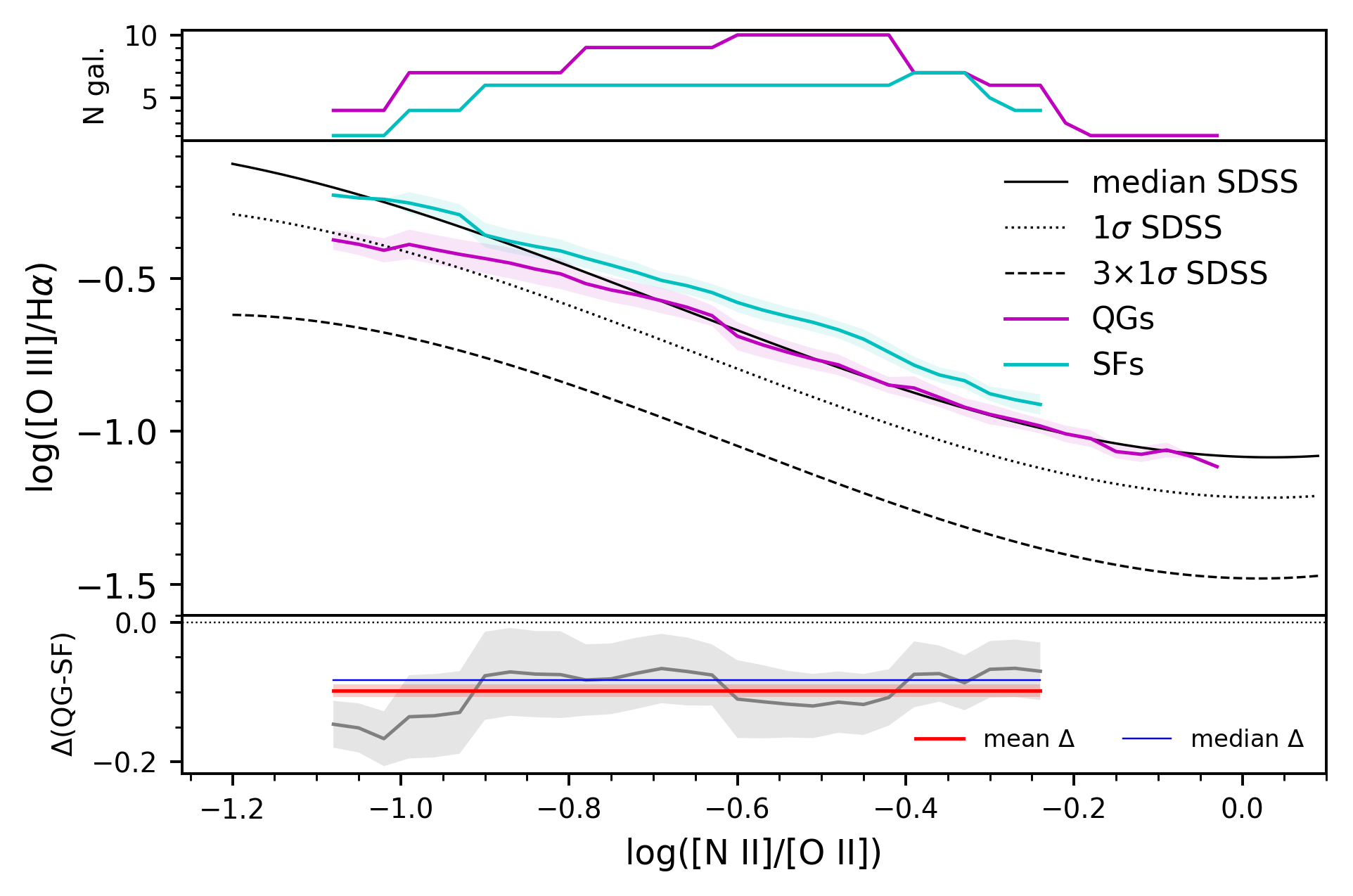} 
\captionof{figure}{The average dust-corretced \OH vs \NO diagram for the \undw and SF samples. 
{\emph Top:} the number N of galaxies that contribute to the average. 
%Gray dots represent the position on the diagram of all the spaxels of the galaxies in the sample. 
%The green curves represent the running mean of the galaxies in the \undw sample while the blue ones represent the running mean of the galaxies in the \tsf sample. %In both cases, the colour of the curves becomes darker with increasing metallicity (from \NO SDSS fiber measures). 
{\emph Centre:} The magenta curve represents the mean curve of the \und galaxies, that we obtain by averaging their means; the magenta shaded area shows the error of the average (i.e. $\sigma_\text{mean}/\sqrt{N}$). The cyan curve shows the mean curve of the \tsf sample.
The black curves represent median, $1 \sigma$ and $3\times1\sigma$ limit of the SDSS star-forming galaxies sample \citepalias[see][]{Quai2018}. 
{\emph Bottom:} the grey curve represents the differences in \OH, as a function of \NO, between QGs and SF galaxies and the grey shaded area shows the propagated errors. The red line and pink shaded area represent the weighted mean and error of these differences averaged over the \NO range (see \autoref{tab:averages}).}
\label{fig:trend_global_new_v}
\end{figure}

\autoref{fig:trend_global_new_v} shows the average curves of \undw and \tsf galaxies in the resolved \OH vs \NO diagram. 
%LP% The \undw curve is compatible with the trend of the median distribution of the \citetalias{Quai2018} SDSS sample, though at log(\NO) values lower than about $-0.6$  it deviates showing lower \OH values at fixed \NO.  
%This deviation in the slope from the average SDSS star-forming galaxies can also be seen in the average trend of the \tsf sample. 
The two samples share a very similar slope, compatible with the trend of the median distribution of the \citetalias{Quai2018} SDSS sample.
However, they differ in normalisation, being the average curve of \tsf sample above the \undw one %LP% of about $0.15$ dex 
at any \NO value.
The lower panel of \autoref{fig:trend_global_new_v} shows the difference in \OH, as a function of \NO, between QGs and SF galaxies. 
%LP% We average the differences and we define the significance to be the distance of this mean in units of $\sigma$, where $\sigma$ is the error in the average. 
We find  an average difference  of %$<\Delta$log(\OH)$>_\text{\NO} = 
$-0.12 \pm 0.01$ dex with a significance over 10$\sigma$ level (see \autoref{tab:averages}). The result does not change if we take the median in place of the average.  % weighted average.
%\end{minipage}
%%%%%%%%%%%%%%%%%%%%%%%%%%%%%%%%%%%%%%%%%%%%%%%%%%%%%%%%%%%%%%%%

\subsection{The average E(B-V) radial profile}
\begin{figure}
%\begin{minipage}{\linewidth}
\centering
\includegraphics[width=1\linewidth]
{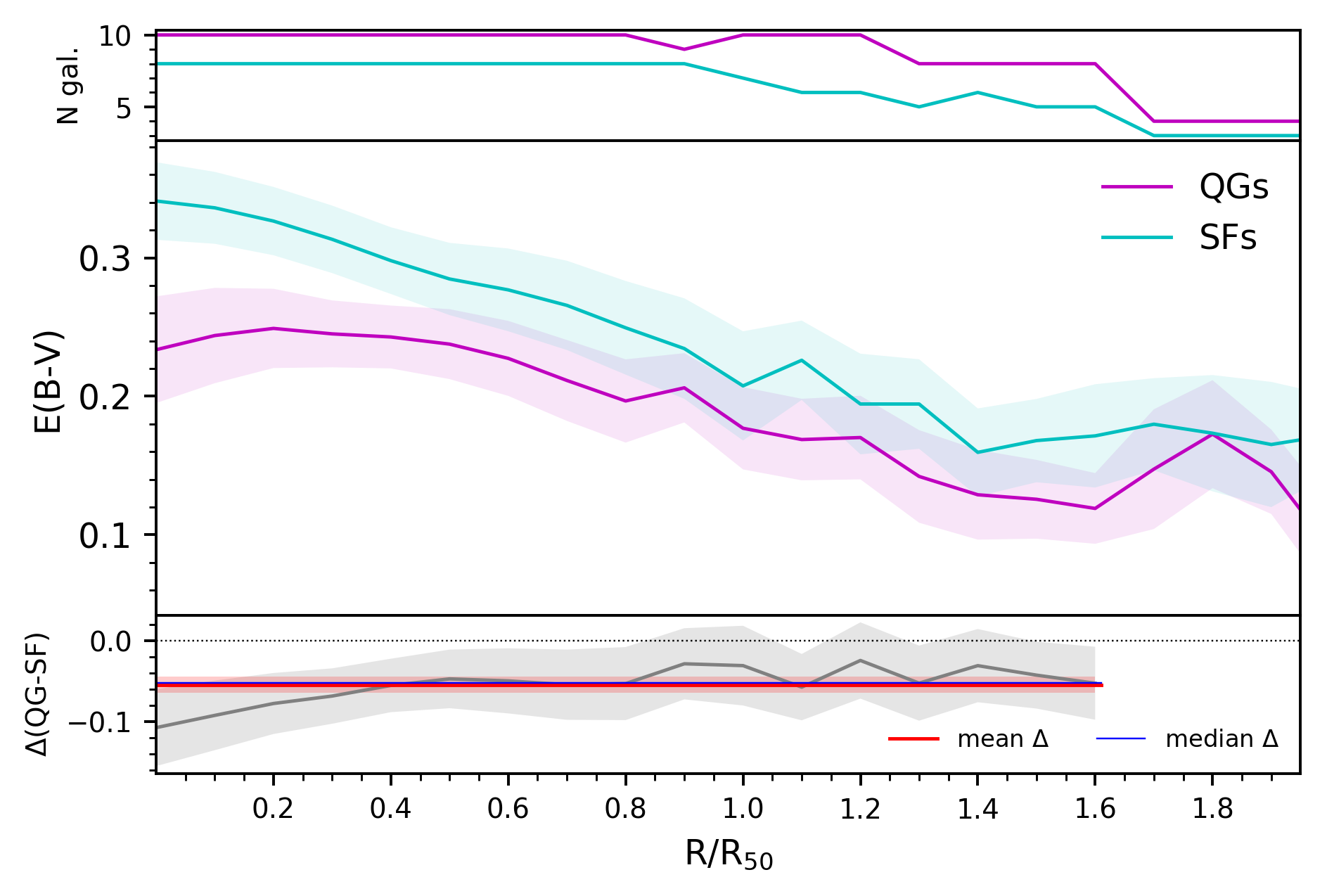} 
\captionof{figure}{The average radial profiles of E(B-V)  for \undw  and \tsf  samples. The colour code is the same as in \autoref{fig:trend_global_new_v}. The upper panels of the two plots show the number N of galaxies that contribute to the average. The grey curve in the lower panel of the two plots represents the differences in log~U and Z, respectively, as a function of R/R$_{50}$, between QGs and SF galaxies. The red line and pink shaded area represent the mean and error of these differences averaged over R/R$_{50}$  (see \autoref{tab:averages}). We exclude from the significance analysis the trends at galactocentric distances larger than 1.6 R$_{50}$ due to the exiguous number of objects in our samples that extend at this radii. }
\label{fig:EBV_global}
%\end{minipage}
\end{figure}

\autoref{fig:EBV_global} shows the average radial profiles of the colour excess E(B-V) of the two samples. 
The SFs have higher extinction at any radius, with increasing values toward the centre. 
\autoref{fig:EBV_global} shows also the difference in E(B-V) as a function of R/R$_{50}$, between QG and SF galaxies. 
%We average the differences and we estimate the significance as the distance of the mean in units of $\sigma$, where $\sigma$ is the error in the average. 
We find  that the average difference between QGs and SFs is about $0.05$ and is confirmed at a significance of about $5\sigma$ level (see \autoref{tab:averages}). 
This result does not change by using the median in place of the 
average. 
 
%%%%%%%%%%%%%%%%%%%%%%%%%%%%%%%%%%%%%%%%%%%%%%%%%%%%%%%%%%%%%%%%
%\vspace{\baselineskip}
%\noindent
%%%%%%%%%%%%%%%%%%%%%%%%%%%%%%%%%%%%%%%%%%%%%%%%%%%%%%%%%%%%%%%%

% ZZZ

\subsection{The average radial profiles of \OH and \NO ratios}
\begin{figure}
%\begin{minipage}{\linewidth}
\centering
\includegraphics[width=1\linewidth]
{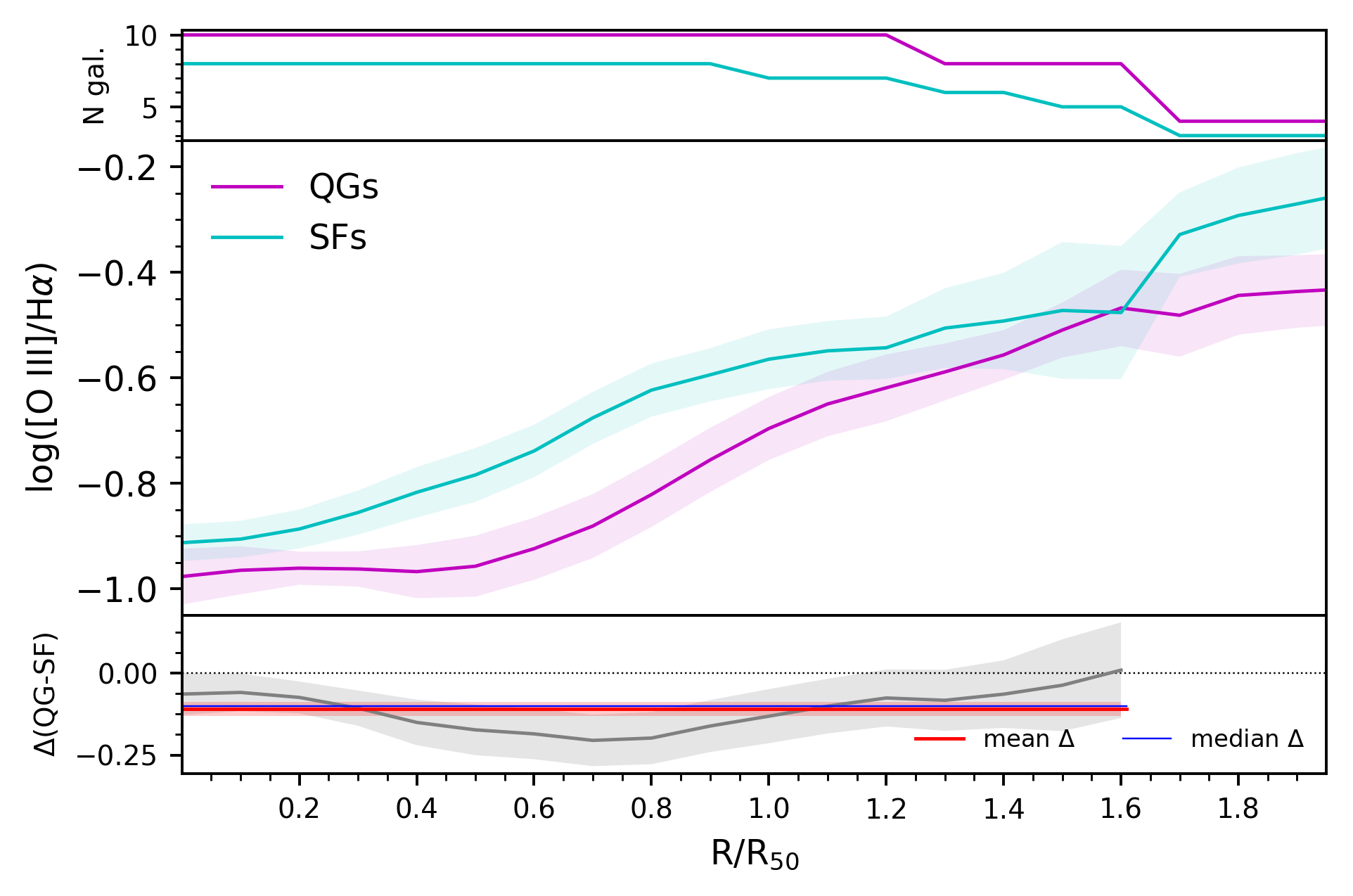}  \\
\includegraphics[width=1\linewidth]
{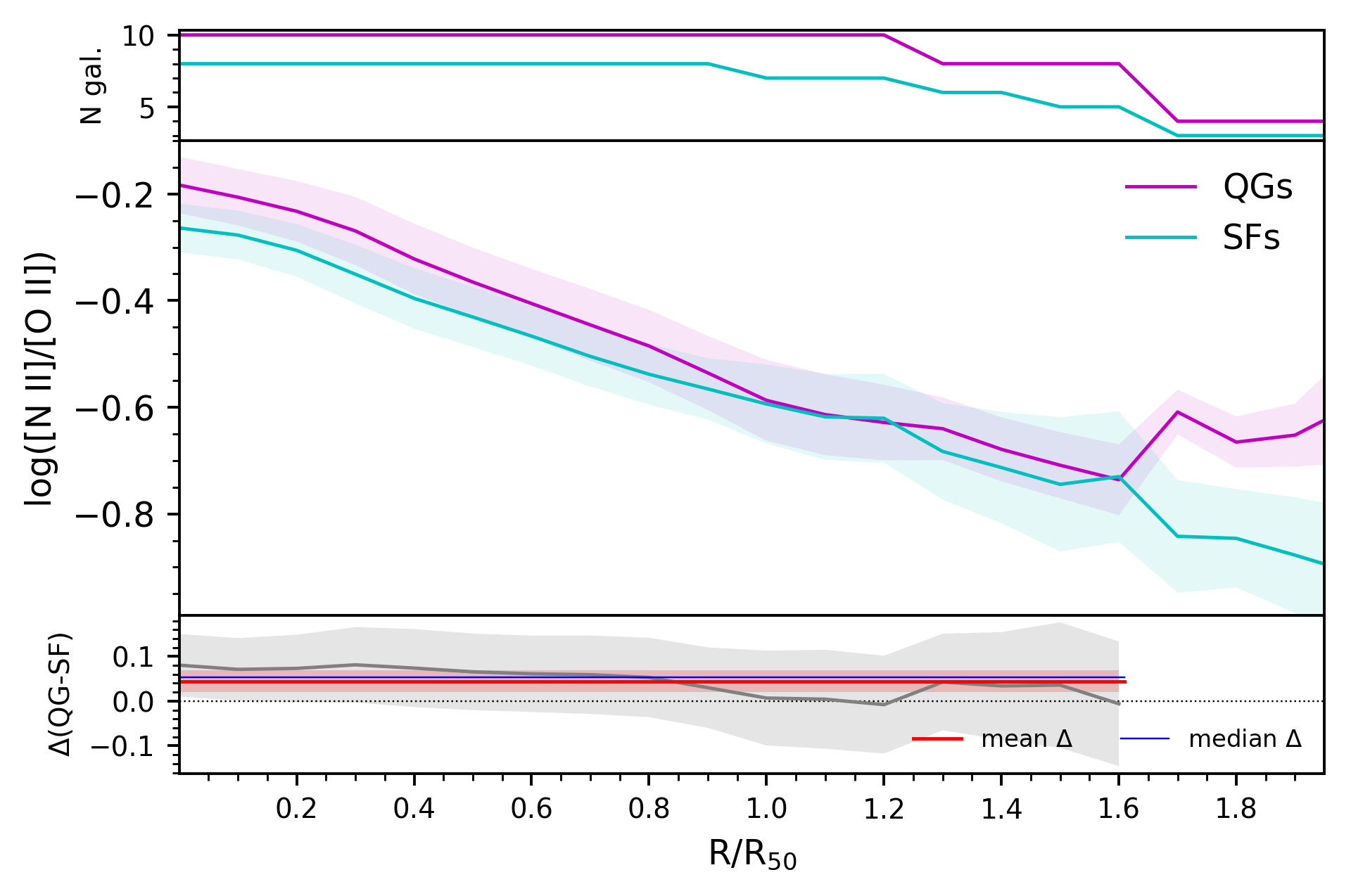} 
\captionof{figure}{The average radial profiles of dust-corrected \OH (\emph{top}) and \NO (\emph{bottom}) for \undw and \tsf samples. The layout of the figure is the same as in \autoref{fig:trend_global_new_v}.}
\label{fig:trend_global_ratios_revis}
%\end{minipage}
\end{figure}

In \autoref{fig:trend_global_ratios_revis} we show the average radial profiles of \OH and \NO ratios.
% and log~U  of the two new samples.
%The trend of these observables confirms differences between the two samples at any radius, showing again similar slopes and different normalisations. 
At any radius, the \tsf sample shows, on average, higher \OH and a slightly lower \NO values than the \undw population, though this one has larger errors. 
%In the same way as in \OH as a function of \NO, we study the significance of the differences in \OH and \NO as a function of R/R$_{50}$, between QGs and SFs. 
We find a difference in \OH at a significance level of about $5\sigma$ and a weak difference in \NO at a significance level slightly lower than $<3\sigma$ (see \autoref{tab:averages}). This result does not change if we use the median in place of the 
average.  %weighted average. 

%%%%%%%%%%%%%%%%%%%%%%%%%%%%%%%%%%%%%%%%%%%%%%%%%%%%%%%%%%%%%%%%
%\vspace{\baselineskip}
%\noindent
%%%%%%%%%%%%%%%%%%%%%%%%%%%%%%%%%%%%%%%%%%%%%%%%%%%%%%%%%%%%%%%%

\subsection{The average log~U radial profile}
\label{sec:averageU}
\begin{figure}
%\begin{minipage}{\linewidth}
\centering
\includegraphics[width=1\linewidth]
{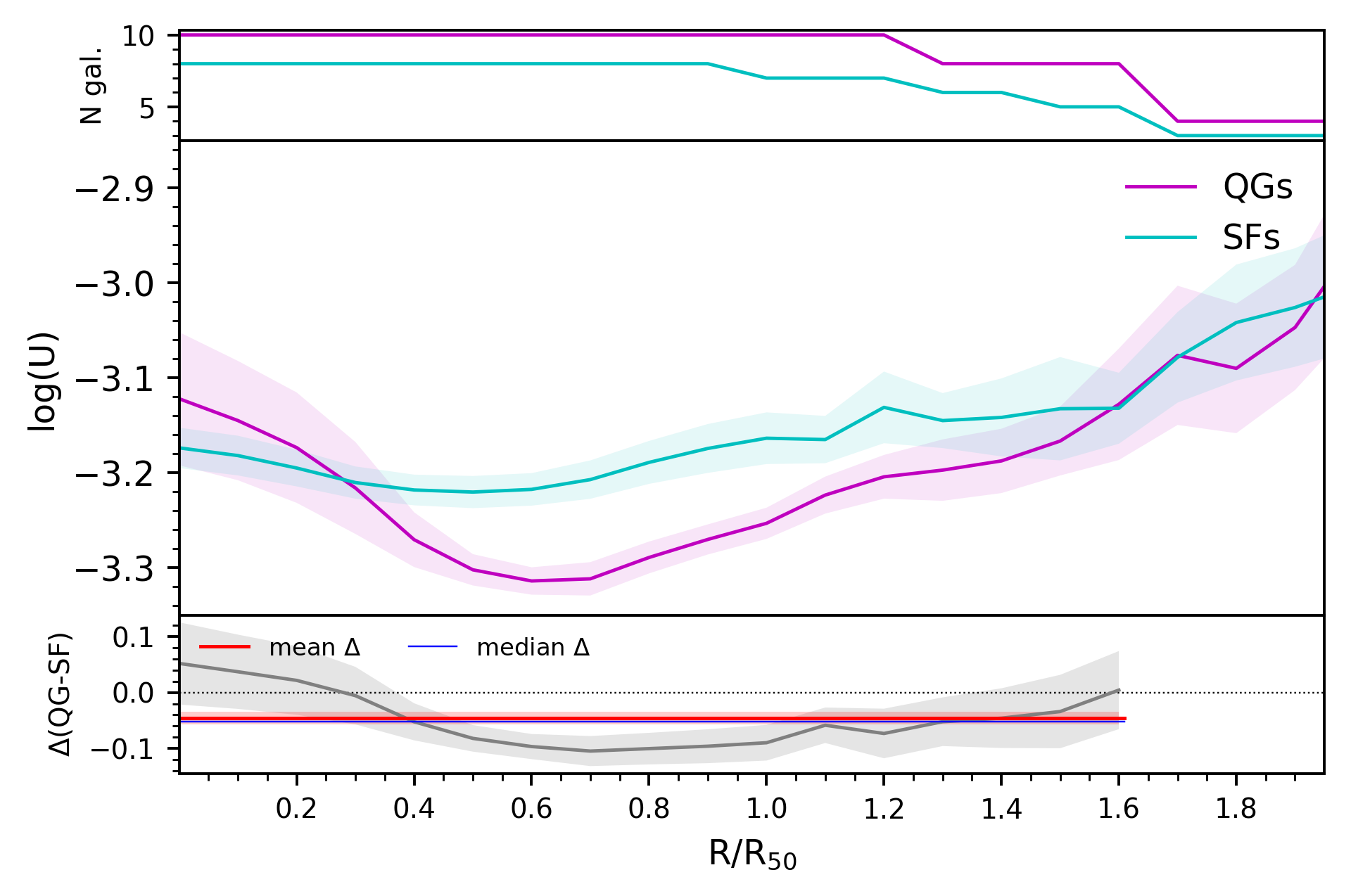} 
\captionof{figure}{The average radial profiles of log~U  for \undw  and \tsf  samples. The layout of the figure is the same as in \autoref{fig:trend_global_new_v}.}
\label{fig:U_global}
%\end{minipage}
\end{figure}

\autoref{fig:U_global} log~U radial profiles for the two samples. 
The average log~U radial profile of \tsf galaxies increases very slowly from log~U $\sim -3.2$ in the centre, toward $\sim -3.1$ in the outskirts.
%, with a behaviour similar to that of SF 1-178443 (see \autoref{fig:trends_U_Z}). 
The QG galaxies log~U profile, instead, has a maximum of log~U $\sim -3.1$ in the centre, then it decreases to a minimum log~U $\sim -3.3$ around $0.5$ R$_{50}$ before rising again to log~U $\sim -3.1$ towards the outskirts. 
We note that $8$ out to $10$ QGs show such shape in their ionisation radial profile, while only $1$ SF galaxy shows a similar trend (see on-line material). 
On the contrary, only $1$ QGs has a minimum in log~U in the center (QG 1-491193).
\autoref{fig:U_global} shows the difference in log~U as a function of R/R$_{50}$, between QGs and SF galaxies.
In the inner region there is no evidence of difference between the two samples, while they strongly differ between $0.3$ and $1.2$ R/R$_{50}$. Therefore, the average difference is confirmed at a high significance of about $3.95\sigma$ level (see \autoref{tab:averages}). 

%%%%%%%%%%%%%%%%%%%%%%%%%%%%%%%%%%%%%%%%%%%%%%%%%%%%%%%%%%%%%%%%
%\vspace{\baselineskip}
%\noindent
%%%%%%%%%%%%%%%%%%%%%%%%%%%%%%%%%%%%%%%%%%%%%%%%%%%%%%%%%%%%%%%%

\subsection{The average gas-phase metallicity radial profile}
\begin{figure}
%\begin{minipage}{\linewidth}
\centering
\includegraphics[width=1\linewidth]
{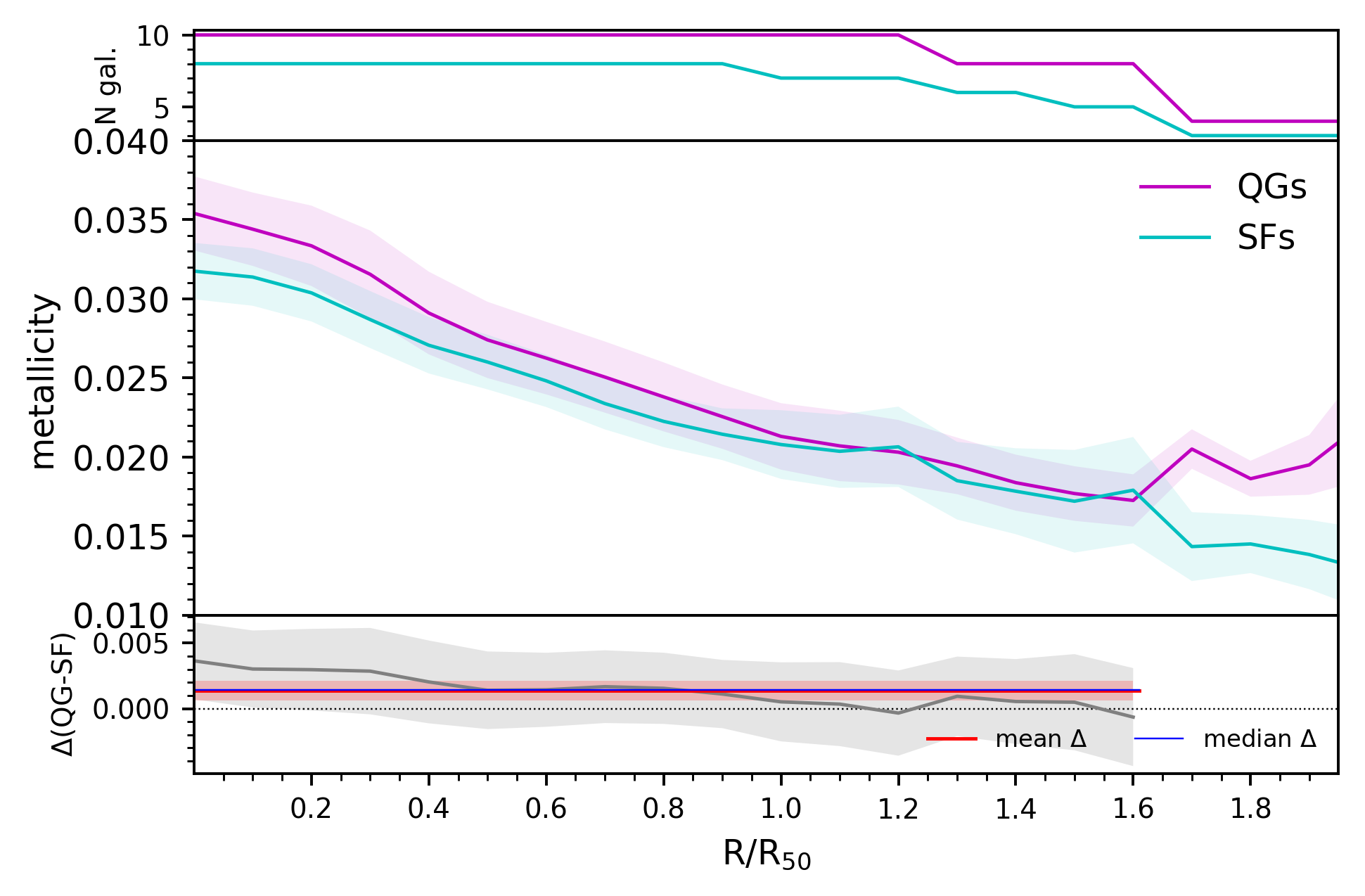} 
\captionof{figure}{The average radial profiles of the gas-phase metallicity Z  for \undw  and \tsf  samples. The layout of the figure is the same as in \autoref{fig:trend_global_new_v}. 
}
%The two dotted lines \textcolor{red}{\bf DOVE SONO??} represent the linear fit of the radial profiles between $0.5$ and $1.6$ R$_{50}$. }
\label{fig:Z_global}
%\end{minipage}
\end{figure}

The gas-phase metallicity radial profiles (see \autoref{fig:Z_global}) of the two samples have a similar slope and normalisation. 
We fit with a straight line to the average radial profile at galactocentric distances larger than $0.5$ R$_{50}$ \citep[as suggested in][to avoid smearing effects due to the MaNGA PSF in the inner part of galaxies]{Belfiore2017b} and smaller than $1.6$ R$_{50}$. 
We obtain a slope of $\sim -0.01/$R$_{50}$ for QGs \citep[consistent with typical slopes of star-forming galaxies with similar stellar mass, e.g.][]{Sanchez2014, Belfiore2017b}, though the average profile of the \undw galaxies shows a larger error than that of \tsf ones\footnote{If we convert the gas-phase metallicity in terms of 12+log(OH) (i.e. 12+log(OH) $=$ log(Z/Z$_\odot$) + $8.69$, with Z$_\odot = 0.02$), we will obtain a slope of $\sim -0.14$dex/R$_{50}$ for QGs.}. 
In this case, the difference between the sample is rejected (significance level $<3\sigma$). 
This result confirms that the two samples have a similar gas phase metallicity. 

%%%%%%%%%%%%%%%%%%%%%%%%%%%%%%%%%%%%%%%%%%%%%%%%%%%%%%%%%%%%%%%%
%\vspace{\baselineskip}
%\noindent
%%%%%%%%%%%%%%%%%%%%%%%%%%%%%%%%%%%%%%%%%%%%%%%%%%%%%%%%%%%%%%%%

%ZZZ

\subsection{The average $\Sigma$SFR radial profile}
\begin{figure}
\centering
\includegraphics[width=1\linewidth]
{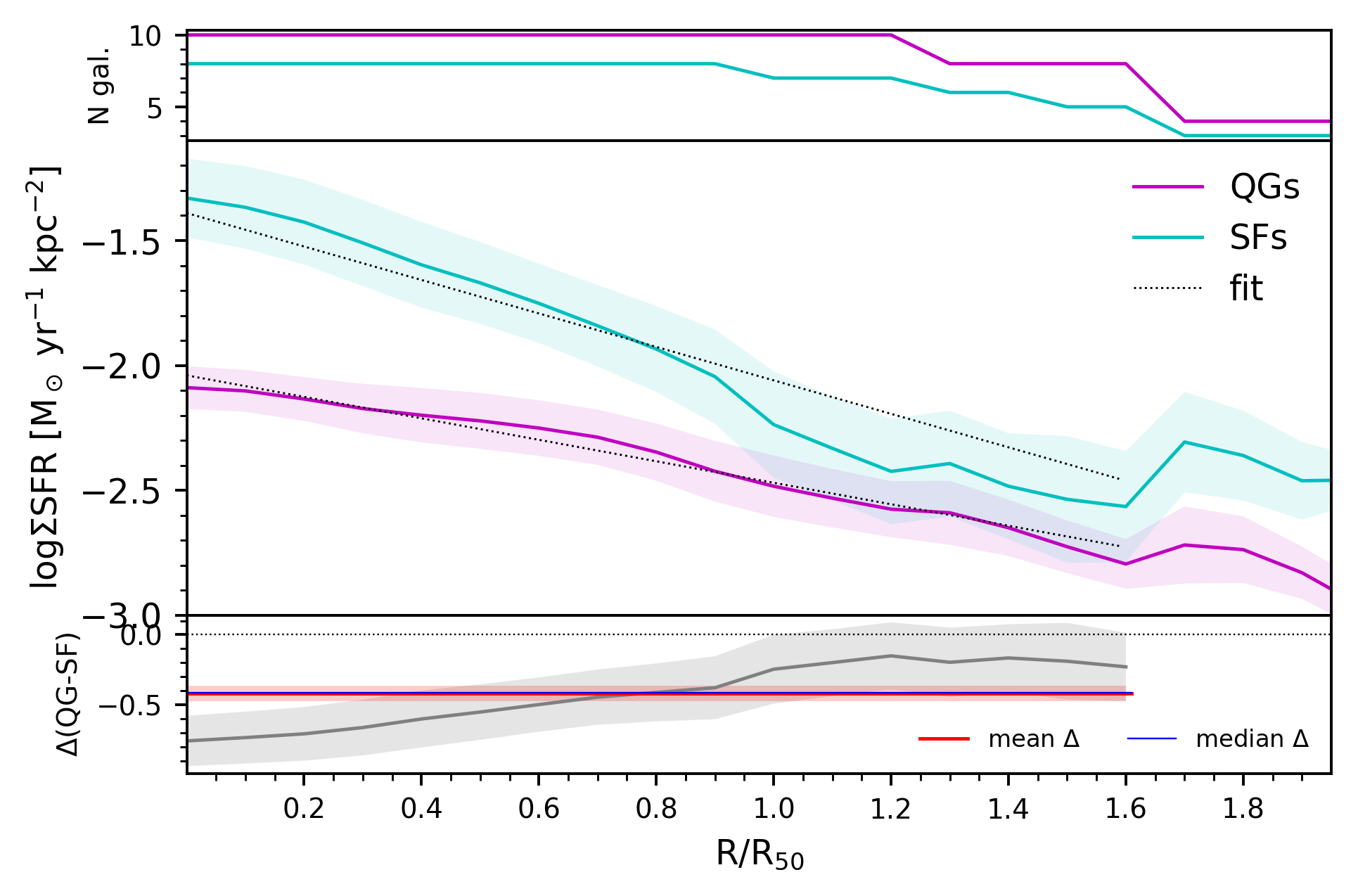} 
\captionof{figure}{The SFR surface density (log~$\Sigma$SFR) radial profile for \undw and \tsf samples. The dotted black lines represent the linear fit to the radial profiles. The layout of the figure is the same as in \autoref{fig:trend_global_new_v}. }
\label{fig:trend_global_SFRD_revis}
%\end{minipage}
\end{figure}

From the analysis of the average radial profiles of the SFR surface density (log~$\Sigma$SFR) of the two samples (see \autoref{fig:trend_global_SFRD_revis}) it turns out that, on average, in the central regions the \tsf galaxies show higher SFR density than \undw, with a difference up to $0.6$ dex. 
The average profile of \und galaxies shows a slow decline towards large radii.
As shown in the central panel of \autoref{fig:trend_global_SFRD_revis}, the average SFR surface density of the SFs has an exponential profile, and its linear fit shows a slope of $-0.67$ \citep[consistent with][]{Spindler2018}. %LP% and an intercept of $-1.39$. 
Instead, the average trend of QGs shows suppression of SFR with respect to the exponential trend of SFs, with a linear fit characterised by a slope of $-0.43$.  % and an intercept of $-2.04$.
The mean behaviour of the SFR in QGs is validated by the analysis of the trends of individual galaxies (see on-line material), with $9$ out to $10$ QGs showing an SFR suppressed at any radius with respect to the average trend of the SFs.

% up to log~$\Sigma$SFR $\sim -2.5$ at $1.5$R$_{50}$. 
The lower panel of \autoref{fig:trend_global_SFRD_revis} shows the difference in log~$\Sigma$SFR as a function of R/R$_{50}$, between QGs and SF galaxies.  As expected, the difference is larger at small galactocentric distances and it becomes negligible in the outskirts. %LP% In the same way, as for the other parameters, we average the differences, and we define the significance as the distance of this mean in units of $\sigma$, where $\sigma$ is the error in the average.
We find a strong differences in the median $\Sigma$SFR between the two samples, of $\sim 0.4$ dex, with a significance at about $8\sigma$ level  (see \autoref{tab:averages}). 
%However, considering the scatter of the \tsf population, there is no appreciable difference in the global behaviour of $\Sigma$SFR between the two samples, at least up to $1.5$ R$_{50}$ but even at larger radii, though there is less statistic.
%We note that the average behaviour of the \tsf population follows 

It is interesting to discuss why the average star formation surface density radial profile of our QGs does not show a minimum similar to that of the ionisation parameter. The different behaviour of these two parameters can be due to the fact that the SFR, derived from the \halpha luminosity, is sensitive to the overall presence of O and B stars. Hence, its suppression represents an evidence of the general decrease of massive stars. Instead, as argued in the previous sections and in \citetalias{Citro2017, Quai2018}, the ionisation parameter promptly reacts to the disappearance of short-lived O stars, therefore, it is a better tracer of the actual distribution of quenching regions within galaxies. 
To confirm this, however, we need to discuss (see next section) the possibility that the absence of O stars can be connected with a stochasticity on the IMF in low SFR regime. 
%the fact that the SFR estimate reacts slower to the quenching than the \OH ratio, because the \halpha luminosity starts to decrease only after the disappearance of late-B stars, following the drop in the soft-UV radiation. 

%Our findings appear to disagree with the results of s
Several studies \citep[e.g.][]{Tacchella2015, Belfiore2018, Ellison2018, Morselli2018, Lin2019} interpret the suppression of the inner SFR in galaxies that lie below the star forming main sequence as an evidence of an inside-out quenching. But in that case, %LP%In this scenario, 
the quenching starts as the effect of a mechanism acting in the centre of the galaxies (e.g. AGN feedback) and then it propagates towards the outskirts. 
However, our method is not sensitive to AGN feedback \citep[e.g.][]{DeLucia2006, Fabian2012, Cimatti2013, Cicone2014} as quenching mechanism, due to our a priori exclusion of AGNs. 
Also in our QGs we find a SFR density flatter than SF galaxies, suggesting a suppression of the SFR in the inner part of our galaxies. However, the ionisation parameter (log~U) has its minimum off-center, suggesting a quenching scenario more complex than an inside-out one.

%%%%%%%%%%%%%%%%%%%%%%%%%%%%%%%%%%%%%%%%%%%%%%%%%%%%%%%%%%%%%%%%
%\vspace{\baselineskip}
%\noindent
%\begin{minipage}{\linewidth}
%%%%%%%%%%%%%%%%%%%%%%%%%%%%%%%%%%%%%%%%%%%%%%%%%%%%%%%%%%%%%%%%

%%%%%%%%%%%%%%%%%%%%%%%%%%%%%%%%%%%%%%%%%%%%%%%%%%%%%%%%%%%%%%%%
\begin{table}
%\begin{minipage}{\linewidth}
\centering
\captionof{table}{Median, mean, mean error and significance (expressed in units of $\sigma$)  of the differences in the listed quantities between QGs and SF galaxies.}
\label{tab:averages}
\begin{tabular}{lcccc} 
 & median & mean & error & \#$\sigma$ \\
\hline
 <E(B-V)>$_\text{R/R$_{50}$}$									& -0.05 	& -0.05 	& 0.01 		& 5.2 \\							
 <$\Delta$log(\OH)>$_\text{\NO}$     					& -0.08 		& -0.1 		& 0.01 		& 10.7 \\		
 <$\Delta$log(\OH)>$_\text{R/R$_{50}$}$ 					& -0.1 		& -0.1 		& 0.02 		& 5.2 \\			
 <$\Delta$log(\NO)>$_\text{R/R$_{50}$}$ 					& 0.05 	& 0.04 	& 0.02		& 1.8 \\		
 <$\Delta$log(U)>$_\text{R/R$_{50}$}$     					& -0.05 	& -0.05 	& 0.01 		& 3.95 \\
 <$\Delta$ Z>$_\text{R/R$_{50}$}$            					& 0.001 & 0.001 &0.0007 & 1.8  \\
 <$\Delta$log($\Sigma$SFR)>$_\text{R/R$_{50}$}$ & -0.4 	& -0.4 	& 0.05 	& 7.9 \\
\hline
\end{tabular}		 
%\end{minipage}
\end{table}
%%%%%%%%%%%%%%%%%%%%%%%%%%%%%%%%%%%%%%%%%%%%%%%%%%%%%%%%%%%%%%%%

%This analysis of the global trends suggests that both the samples behave statistically as star forming populations, however, they differ because the \undw galaxies have, on average, a lower ionisation levels and higher gas-phase metallicity than the \tsf ones. 

%%%%%%%%%%%%%%%%%%%%%%%%%%%%%%%%%%%%%%%%%%%%%%%%%%%%%%%%%%%%%%%%
\subsubsection{Stochasticity on the initial mass function}
%
%In the last decades emerged that the SFRs estimated from \halpha luminosity is inconsistent with those obtained from UV continuum in galaxies with low SFRs such those measured in dwarf galaxies \citep[e.g.][]{Sullivan2000, Lee2009, Meurer2009, Fumagalli2011, daSilva2014}. 
%Since the UV continuum light is primarily produced in the photospheres of long-lived stars more massive than $3$ \smass, the plausible explanations of this trend are mainly $2$: lower massive galaxies are predominantly bursty and/or the massive end of their IMF is different from that expected from a fully populated IMF. 
\cite{Lee2009} showed that assuming a Salpeter IMF, a conservative level of SFR of $1.4 \times 10^{-3}$ \smass yr$^{-1}$ (i.e. log~SFR = $-2.8$) is required to sustain the ability of robustly populate the entire IMF.
%there are only $10$ stars more massive than $18$ \smass for a total stellar mass of $4.3 \times 10^3$ \smass and that 
Otherwise, the massive end of the IMF could result depleted by a certain amount, with the number of the most massive stars regulated by stochasticity.
Therefore, for values of log~SFR $\leq-3$ and fixed metallicity emerges a degeneracy between low \OIII flux due to incomplete sampling of the massive end of the IMF and the quenching of the star-formation. 
It is important to note that these kinds of studies regard the total SFR through the galaxy, especially dwarf ones. 
In our sample, the lowest total log~SFR from SDSS data is $-0.79$ \smass yr$^{-1}$ (i.e. MaNGA 1-352114), however, with MaNGA we measure SFR on small scales (i.e. about a squared kpc per spaxel) and we can study the impact of stochasticity on the spaxels of our galaxies. 
In \und 1-43012 (i.e. the galaxy analysed in previews sections) the vast majority of the spaxels show log~$\Sigma$SFR $< -2$ (see \autoref{fig:rad_prof_verNEW}) but only a small amount of $8\%$ have log~$\Sigma$SFR $< -3$ M$_\odot$ yr$^{-1}$ kpc$^{-2}$. 
We observe a similar situation in the other \und galaxies (see \autoref{fig:trend_global_SFRD_revis} and on-line materials). 

\cite{Paalvast2017} widely studied the impact of the stochastic sampling of the mass function on the production of lines requiring high energetic photons (i.e. \OIII) relative to that of the Balmer ones. 
All of their stochastic models predict a significant increase of the scatter of \OHb ratios with decreasing of the SFR with respect to the typical BPT values, while at higher SFR the models well reproduce the BPT locus of the SDSS star-forming galaxies. 
For log~SFR $<-3$, the lack of massive stars extends the scatter of \OHb for solar metallicity from the BPT locus to values of log(\OHb) lower than $-4$ (i.e. $\sim -4.5$ using the \OH ratio, alternatively), but also for log~SFR $\sim-2$ the scatter is considerably larger than that expected for a fully populated IMF. 
Moreover, the effect becomes more relevant with increasing metallicity.

In order to evaluate the impact of the stochasticity on our sample, we study galactic regions within our galaxies showing the lowest SFR values and super-solar gas phase metallicity. This combination of parameters maximises the effect in \citet{Paalvast2017} models.
To this aim, we gathered all the spaxels with $-3 \leq$ log~$\Sigma$SFR $< -2.5$ and super-solar metallicity (i.e. $0 \leq $\NO$<0.2$)  and we analyse the distribution of their \OH. 
By definition, due to the stochasticity, we should obtain a wide distribution that covers the scatter due to the poor IMF sampling.
We analyse an extreme conservative case in which all the spaxels showing an upper limit in \OIII (i.e. spaxels with S/N(\OIII)$<2 $, about $22\%$ of the spaxels in the bin) are considered with log \OH $=-4$, as if all of them are the lowest outcome of the stochasticity models of \cite{Paalvast2017}. 
%%%%%%%%%%%%%%%%%%%%%%%%%%%%%%%%%%%%%%%%%%%%%%%%%%%%%%%%%%%%%%%%
\begin{figure}
\centering
\includegraphics[width=0.8\linewidth]
{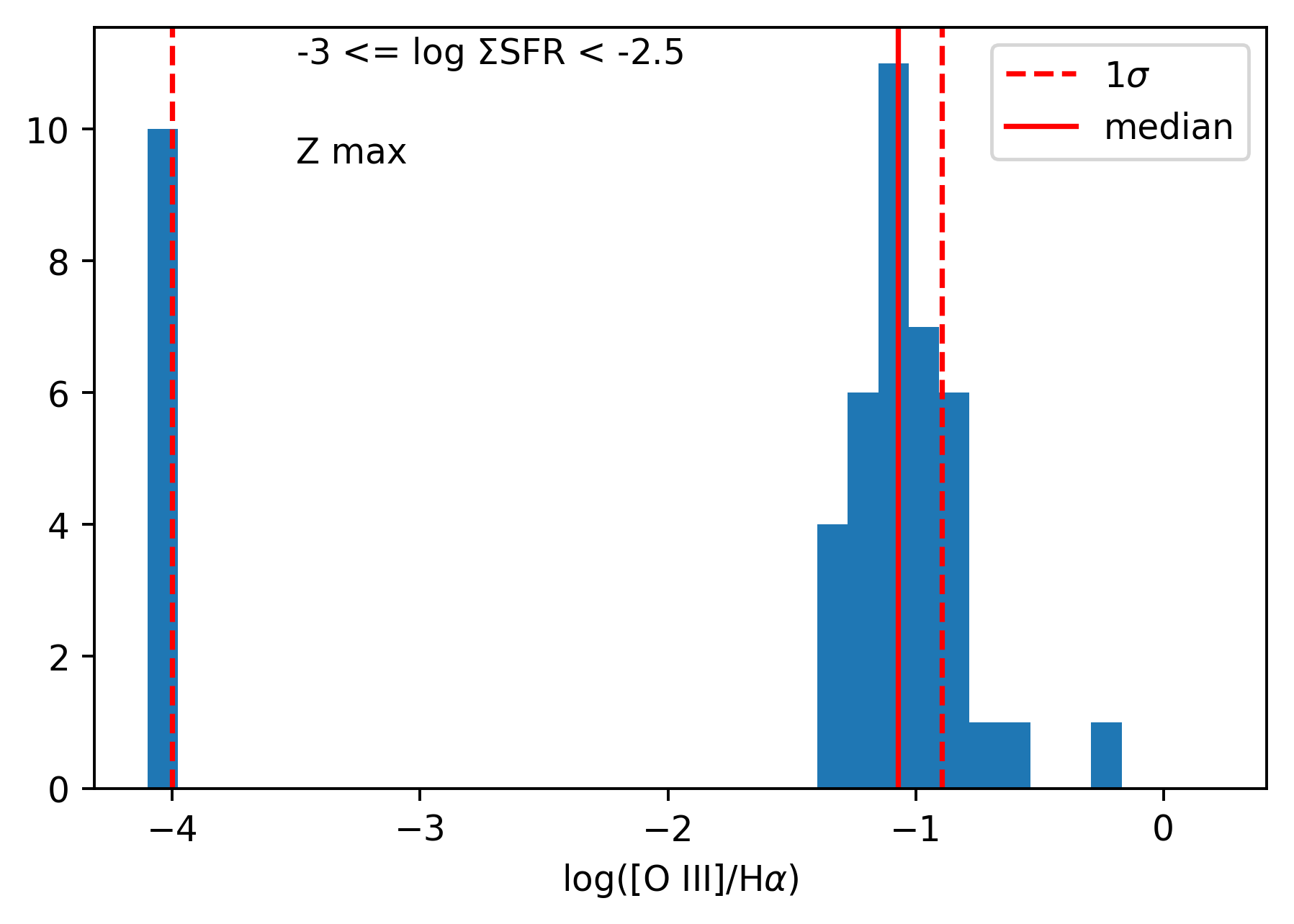} 
\includegraphics[width=0.8\linewidth]
{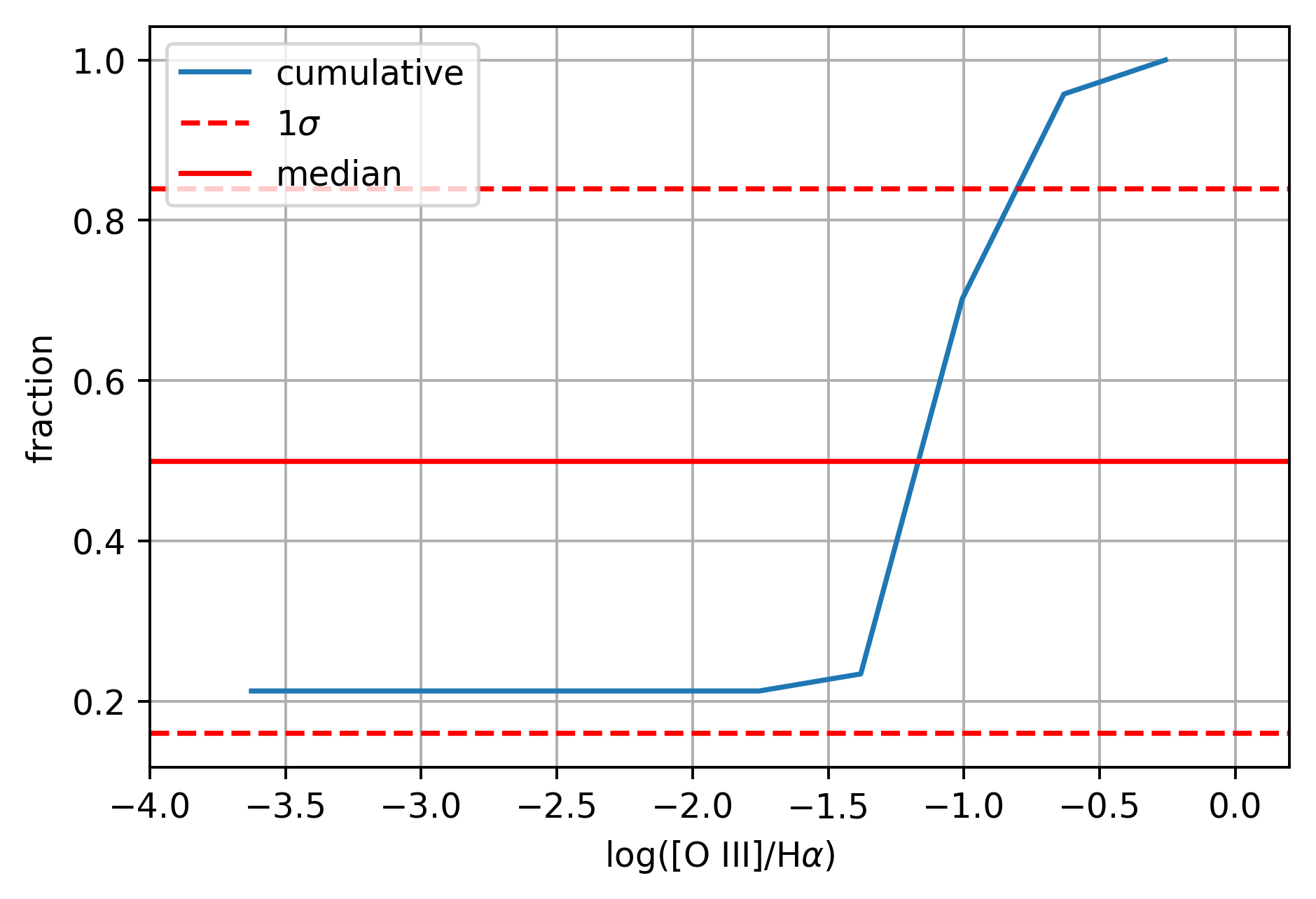} 
\caption{(\emph{Top:}) the dust-corrected \OH distribution for the spaxels with $-3 \leq$ log~$\Sigma$SFR $< -2.5$ and $0 \leq $\NO$<0.2$. The peak at \OH $= -4$ represents the limit point that we choose for spaxels showing S/N(\OIII)$<2$. (\emph{Bottom:}) the cumulative curve of the distribution.}
\label{fig:stochast}
\end{figure}
%%%%%%%%%%%%%%%%%%%%%%%%%%%%%%%%%%%%%%%%%%%%%%%%%%%%%%%%%%%%%%%%
In \autoref{fig:stochast} we show the \OH distribution together with its cumulative curve. 
Even in this conservative situation, the median of the distribution is at log(\OH)$\sim-1.1$, with a small scatter. 
Therefore, we can exclude the stochasticity on the IMF as the primary cause of low ionisation values observed in our galaxies. 
However, we cannot exclude that %Moreover, 
in a galaxy with recent quenching the low ionisation could be due both to stochasticity on the IMF sampling and the death of the most massive stars.

% could be due 
%is likely that low \OH values are due 
%both to stochasticity on the IMF sampling and to the death of the massive stars. %LP% and the degeneracy between the two effects is less relevant.

%%%%%%%%%%%%%%%%%%%%%%%%%%%%%%%%%%%%%%%%%%%%%%%%%%%%%%%%%%%%%%%%%
%\end{document}
\subsection{The spatial distribution of the quenching}
In this section, we focus on the \undw galaxies with the aim to analyse the spatial distribution and extension of their quenching regions.  
None of the QGs shows extended regions compatible with our quenching criteria, instead they have groups of small regions ($2-5$ regions each), with extension between $2 $ and $4$ kpc$^2$, that are smaller than the MaNGA PSF (i.e. $2.5\arcsec$ of FWHM, or about $5$ kpc$^2$ at these redshifts that corresponds to a percentage of the entire galaxies between $\sim 1\%$ and $\sim 8 \%$). 
However, these groups of quenching regions are always interconnected in more extended areas that are characterized by slightly higher ionisation levels, and they lie in the region between $1\sigma$ and $3\times1\sigma$ of the SDSS data within the \OH vs \NO diagram (see \autoref{fig:QGs_plane}, and, for example the yellow area in \autoref{fig:map_quench}).
Therefore, for effect of the PSF, it is likely that the actual size of the quenching regions is broader than that observed and it is mixed with the adjacent areas.
%%%%%%%%%%%%%%%%%%%%%%%%%%%%%%%%%%%%%%%%%%%%%%%%%%%%%%%%%%%%%%%%
\begin{figure}
\centering
\includegraphics[width=0.7\linewidth]
{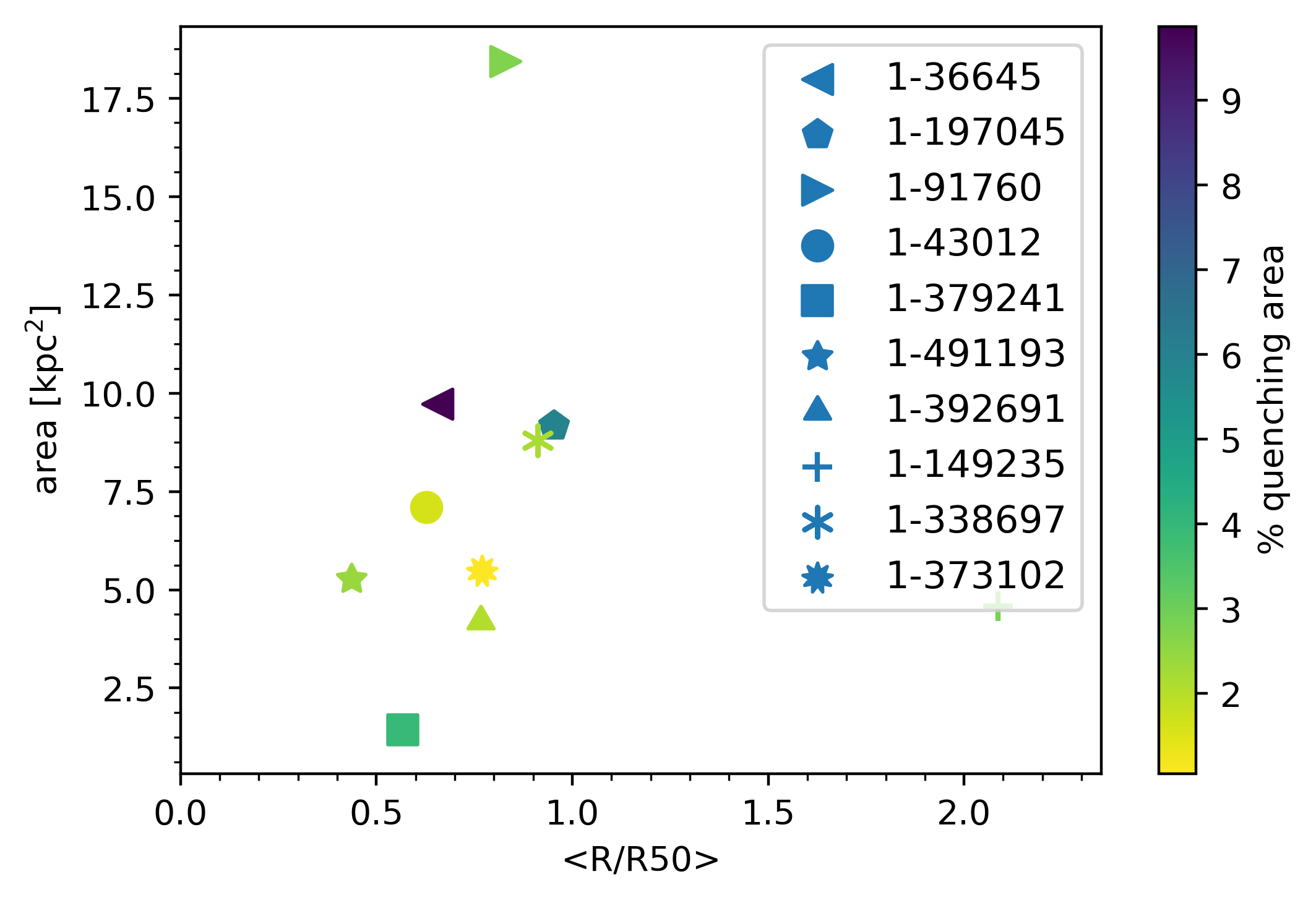} 
\includegraphics[width=0.7\linewidth]
{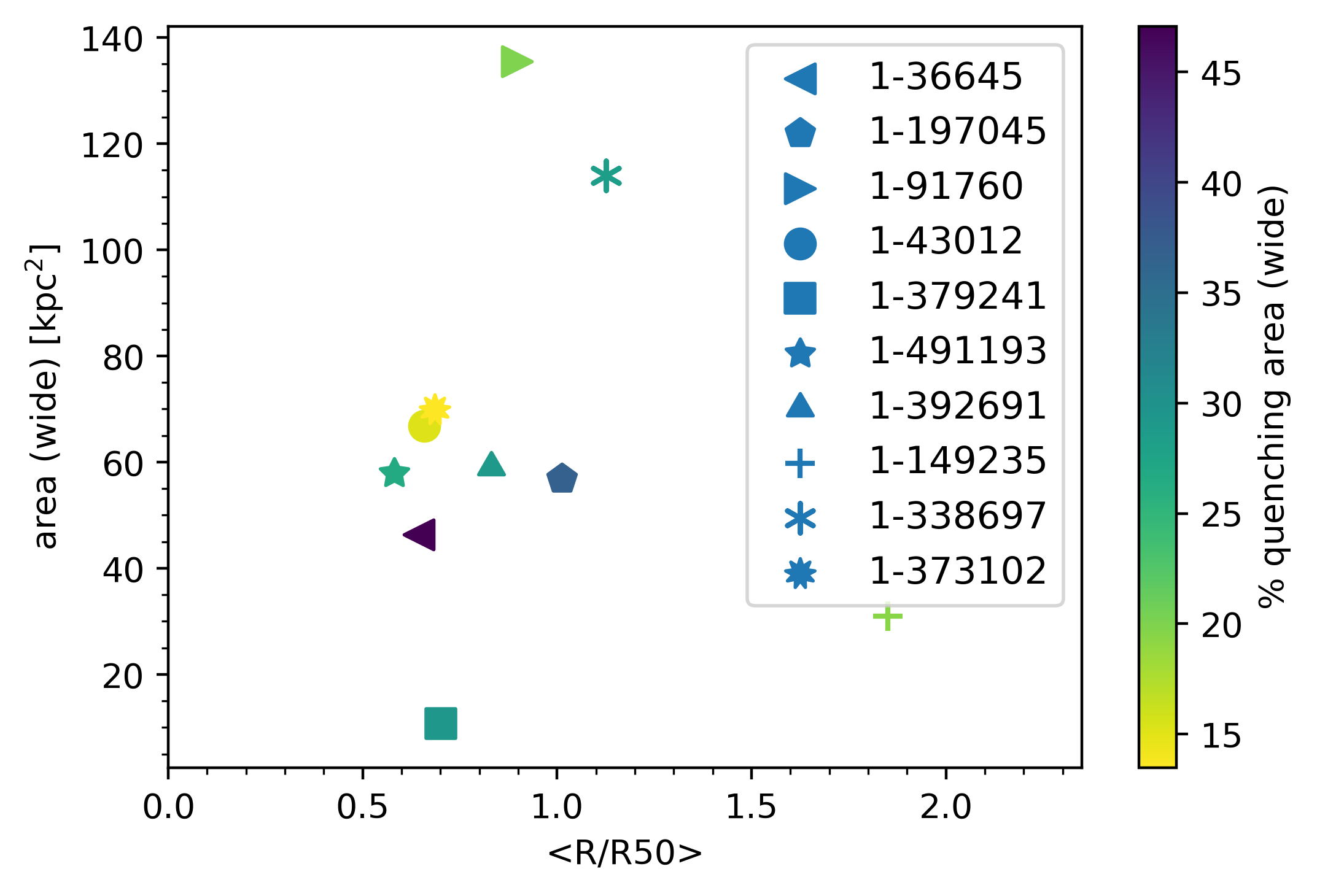}
\caption{The total size of the likely quenching regions as a function of their average distance (R/R$_{50}$) from the galactic centre of the \undw galaxies.  In the \emph{top} panel we consider only the regions that satisfied our quenching criteria, while in the \emph{bottom} panel we consider the area comprising the plausible quenching regions and the less extreme adjacent areas. 
The symbols represent different galaxies, while the colour represents the percentage of spaxels in the area with respect to the whole spaxels of the galaxy. }
\label{fig:size_vs_dist}
\end{figure}
%%%%%%%%%%%%%%%%%%%%%%%%%%%%%%%%%%%%%%%%%%%%%%%%%%%%%%%%%%%%%%%%

%\end{document}
\autoref{fig:size_vs_dist} shows the total size of the likely quenching regions (i.e. the sum of the size of these regions) as a function of their average distance (R/R$_{50}$) from the galactic centre of the \undw galaxies. 
%Only $2$ out of $10$ galaxies have significant quenching areas. They are \undw 1-38802, with an area of about $47$ kpc$^2$ and \undw 1-91760, with an area of about $20$ kpc$^2$. \textcolor{red}{\bf QUANTO IN PERCENTUALE ???????}.
Only $2$ out of $10$ galaxies have significant quenching areas. They are \undw 1-36645 and \undw 1-197045 which have quenching regions covering an area of about $10\%$ and $6\%$ of the entire galaxies, respectively (i.e. corresponding to an area of about $10$ and $9$ kpc$^2$, respectively).
The other $8$ galaxies have quenching regions which cover between $\sim1.1\%$ and $\sim6\%$ in percentage of the entire galaxies (i.e. corresponding to an area between about $1.5$ and $20$ kpc$^2$).
Moreover, all of them are located, on average, between $0.5$ and $1$ R$_{50}$ and only one of the galaxies have quenching regions in the outskirts at radii larger than $2$ R$_{50}$ (i.e. QG 1-149235). 

It is interesting to consider, as an upper limit of the total size of the quenching distribution in these galaxies, the broader areas obtained by summing the dimension of quenching regions together with that of their adjacent less extreme areas. 
The result is shown in \autoref{fig:size_vs_dist}. 
In this case, the quenching areas in QGs cover percentages between $13.5\%$ and $47\%$ of all the spaxels, to which correspond sizes larger than $30$ kpc$^2$ in $9$ out of $10$ QGs.
Even when a broader area is considered, the average distance from the centre of the galaxies remains between about $0.5$ and  $1.1$ R$_{50}$, confirming that only one of our galaxies have quenching regions in their center. 
Finally, we note that in $4$ QGs these extended quenching regions shape an annulus, even if irregular and incomplete, around the galactic centre (see  galaxies 1-392691, 1-373102, 1-43012 and 1-91750 in \autoref{fig:OIII_info_1}). 

%\textcolor{blue}{Our findings appear to disagree with the results of several studies \citep[e.g.][]{Tacchella2015, Belfiore2018, Ellison2018, Morselli2018, Lin2019} which interpret the suppression of the inner SFR in galaxies that lie below the star forming main sequence as an evidence of an inside-out quenching. In this scenario, the quenching starts as the effect of a mechanism acting in the centre of the galaxies (e.g. AGN feedback) and then it propagates towards the outskirts. 
%However, our method is not sensitive to AGN feedback \citep[e.g.][]{DeLucia2006, Fabian2012, Cimatti2013, Cicone2014} as quenching mechanism, due to our a priori exclusion of AGNs. 
%}

How can a quenching mechanism be compatible with our findings of a off-centre start of the quenching?
Several simulations \citep[e.g.][]{Prendergast1983, Sellwood1993,  Regan2003} show that in the inner regions of barred galaxies the gas flows toward the centre and it is often concentrated near the Lindblad resonance because of the dynamical interaction with the potential of the bar. There, the gas feeds intense episodes of star formation. 
However, we see neither bars nor rings of stars in the optical images of our galaxies. Nevertheless, even if the star formation is mostly in the ring, hardly ever it leads to form a ring of long-lived stars \citep[e.g.][]{Kormendy2004}. 
Near-infrared images have shown that bars are hidden in approximately two-thirds of spiral galaxies, despite their appearance in optical wavelengths \citep[e.g.][]{Block1991, Eskridge2002, Block2001, Laurikainen2002, Kormendy2004}.
%The star formation is supported by a continuous replenishment of fresh gas from the corona. Uninterrupted radial flows of gas towards the centre end up in the annular region for effect of the bar where the gas is converted into stars at high rates. 
%The metallicity peak we observe in the annular regions around the galactic centre of some QGs is compatible with high SFR typically associated with rings. However, we find that there is a depletion of the \halpha luminosity and a minimum of the parameter of ionisation. 
It is necessary to highlight that it is challenging to recognise bars and other galaxy substructures on the SDSS optical images of our QGs. 
However, since about $2/3$ of star-forming galaxies show bars when observed at infrared wavelengths \citep[see][]{Kormendy2004}, it results therefore likely that a large fraction of our QGs should be composed by barred galaxies. 

If there was a mechanism able to interrupt the inflow of gas toward the centre or the replenishment of gas from the surrounding hot halo, the star formation would continue by burning the remaining gas in the galaxy's reservoir. 
\citet{Tacconi2013} found that this reservoir can sustain the star formation for less than $1$ Gyr before a typical star-forming galaxy runs out of gas. 
In these circumstances, at high star formation rates, the region of the ring should be the first to consume the fuel and to interrupt the star formation. 
Therefore, it is plausible that we are witnessing the very early phase of the quenching in our QGs. 
%LP%Then, the quenching would propagate from the annular region, reaching soon the close inner part of the galaxies.  
A possible interpretation of the results regarding the QGs, is that in these galaxies a quenching wave is presumably propagating from the inner region towards the outskirts because of the entire consumption of the available gas in intense episodes of star formation and we are witnessing the quenching phase on an annular region around the galactic centre.

It is interesting to report the recent results by \citet[][]{Chen2019}. In the MaNGA survey, they found a population of galaxies with ring-like post-starburst regions (RPSB). These regions are spatially distributed as the quenching regions in our QGs, and since the post-starburst phase traces timescales between $\sim0.3-1$Gyr after the quenching (i.e. stellar population dominated by A-type stars), the RPSB can represent a population of QG-like galaxies in a subsequent quenching phase. 
If we could confirm any affinity between RPSB and QG populations, we would prove that (i) there is a population of galaxies that experience a sharp and off-centre interruption of the star formation,  and (ii) in these galaxies the quenching phase lasts $300 - 1$Gyr (i.e. the post-starburst phase)

However, the exiguous number of galaxies in our sample, does not allow to establish whether our QGs really are progenitors of the RPSBs. Extending this study to the whole MaNGA population, also including a study of morphology, stellar and gas kinematic, would be critical to address this question.

\section{Summary and conclusion}
\label{sec:Summary}
In this paper, we present a spatially resolved study of $18$ MaNGA galaxies \citep[extracted from SDSS-IV MaNGA DR14,][]{Bundy2015, Abolfathi2018}, that is aimed at deriving spatial information about the quenching process within galaxies. 
For each galaxy, we obtain maps and radial profiles for SFR surface density (from the \halpha luminosity corrected for dust extinction), E(B-V) (from Balmer decrement),  \OH and \NO emission lines ratios (corrected for dust extinction), and for ionisation parameters and gas-phase metallicity (derived from photoionisation models by \citetalias{Citro2017}).
We classify the galaxies according to the spaxels distribution in the \OH vs \NO diagram, %\citepalias[
extending the method devised by \citetalias{Quai2018} to IFU data, and we obtain two samples: 
%i.e. an extension to IFU data of the method devised in
%][to find quenching galaxies in the SDSS survey]{

\begin{itemize}
\item QGs sample: $10$ galaxies which show regions compatible with a recent quenching of the star formation. These quenching regions are those satisfying the criteria devised in \citet{Quai2018}, showing \OH ratios that are too low to be explained by metallicity effects. 
\item SFs sample: a control sample of $10$ galaxies which are compatible with ongoing star-formation. 
\end{itemize}
The galaxies in the two samples are in the same mass and redshift ranges. However, in our analysis we exclude $2$ low-metallicity SFs to preserve also the same gas-phase metallicity range (see \autoref{tab:sample_prop} for a list of the main properties of the two samples). 
%The aim of this work is to derive spatial information about the quenching process within galaxies. 

We discuss the general characteristics of \und and SF galaxies concerning gas-phase metallicity, ionisation status, color excess E(B-V) and SFR. Our main results are summarised as follows:
\begin{itemize}
\item The average gas-phase metallicity radial profile of QG galaxies is slightly higher than that of SF ones at any radius. However, the difference between the two profiles is not significant. This result confirms that the two samples have a statistically similar gas-phase metallicity. This entails that the following results cannot be ascribed to metallicity effects.

\item The average ionisation parameter log~U radial profile of the SF sample shows a slow increase of log~U towards the outskirts of the galaxies. Despite QGs show central log~U values similar to those of SFs, their log~U profile drops at radii $\sim0.2$ R/R$_{50}$ and reaches a minimum at effective radii between $0.5$ and $0.8$. 
This trend reveals a lack of O stars in the region surrounding the galactic center of the analysed QGs, suggesting that the quenching could be originated off-centre. 
We confirm the difference between the QGs and the SFs at a high significance level of about $5.5\sigma$.

\item The average radial profile of the star formation rate surface density of QGs is lower than that of the SFs, at any radii, suggesting an overall suppression of the star-formation rate. The difference is larger approaching small radii. As expected, this trend is similar to that of the colour excess E(B-V). We confirm the difference between the QGs and the SFs at a high significance level of about $8\sigma$. 
%Moreover, despite a different approach this result is perfectly consistent with other studies of star formation rate surface density radial profiles of galaxies selected in the SFR - stellar mass diagram \citep[e.g.][]{Belfiore2018, Ellison2018, Morselli2018, Lin2019}. 

\item The quenching regions within our QGs are located between $0.5$ and $1.1$ effective radii from the centre and occupy a total area between $\sim15\%$ and $\sim 45\%$ (i.e. between $\sim 10$ and $\sim 140$ kpc$^2$ of the total galactic area\footnote{When we refer to the total galactic area, we mean the total area of all the spaxels covering a galaxy.}).
 It is interesting to note that none of these quenching regions is found in the inner part of our QGs, despite the low level of the  measured SFRs. 
The recent findings by \citet{Chen2019} of a population of MaNGA galaxies showing post-starburst phase in a ring-like region spatially distributed as the quenching regions in our QGs, encourage us to analyse affinities between the two populations, and to study the QGs as progenitors of post-starburst galaxies. 

\end{itemize}

We stress that we do not expect to find galaxies in an advanced quenching phase among the few analysed QGs, since they are not as extreme as the quenching candidates found by \citetalias{Quai2018}, which should provide decisive clues on the early phase of the quenching. 
However, it can be expected that the quenching will propagate from the off-centre regions over all the galaxy. %LP% annular region reaching soon the close galactic centre of the QGs and at the same time it will start to spread towards the galactic outskirts. 
We interpret the off-centre distribution of quenching regions in QGs as an early phase of the quenching. %LP% occurred as the consequence of the interruption of cold gas inflow. 
In the case of no %LP%Without 
replenishment of new gas, the galactic regions with low gas depletion rapidly run out of fuel \citep[e.g.][]{Tacconi2013}.
Therefore, the quenching distribution in regions around the galactic centre of our QGs suggests that a shortage of cold gas started recently in the proximity of Lindblad in barred galaxies. There, the depletion time is shorter because of the accumulation of gas gathered by the effect of the gravitational potential of the bar, that induces high star formation rates \citep[e.g.][]{Kennicutt2012}. 
Although %LP%It is necessary to highlight that 
it is challenging to recognise bars and other galaxy substructures in the SDSS optical images of our QGs,
%However, 
it turns out that about $2/3$ of star-forming galaxies show bars when observed at infrared wavelengths \citep{Kormendy2004}. Therefore, it is likely that a large fraction of our QGs should be composed by barred galaxies.

A critical question that needs to be addressed is whether the QGs are actually %LP% quenching galaxies (i.e. they are 
starting the permanent quenching phase, or if the quenching regions they host are indicative of a minimum in the star formation history of the galaxy due to the interruption of inflows of fresh gas \citep[see][for a discussion]{Wang2019}. 
In a future perspective, an extension of our method to the whole sample of galaxies in the SDSS IV MaNGA data release 15 \citep{Aguado2019}, together with a multi-wavelength sample including maps of the distribution of the cold gas would help to disentangle the two possibilities and shed lights on the mechanism driving the quenching.

\section*{Acknowledgements}
The authors thank the anonymous referee for helpful suggestions and constructive comments.
We are grateful to Filippo Mannucci and Roberto Maiolino for useful suggestions and discussions. The authors also acknowledge the grants 
PRIN MIUR 2015, ASI n.I/023/12/0 and ASI n.2018-23-HH.0. 
JB acknowledges support by Funda{\c c}{\~a}o para a Ci{\^e}ncia e a
Tecnologia (FCT) through national funds (UID/FIS/04434/2013) and Investigador FCT
contract IF/01654/2014/CP1215/CT0003., and by FEDER through COMPETE2020 (POCI-01-0145-FEDER-007672).
Funding for the Sloan Digital Sky Survey IV has been provided by the Alfred P. Sloan Foundation, the U.S. Department of Energy Office of Science, and the Participating Institutions. SDSS-IV acknowledges
support and resources from the Center for High-Performance Computing at
the University of Utah. The SDSS web site is www.sdss.org.
SDSS-IV is managed by the Astrophysical Research Consortium for the 
Participating Institutions of the SDSS Collaboration including the 
Brazilian Participation Group, the Carnegie Institution for Science, 
Carnegie Mellon University, the Chilean Participation Group, the French Participation Group, Harvard-Smithsonian Center for Astrophysics, 
Instituto de Astrof\'isica de Canarias, The Johns Hopkins University, Kavli Institute for the Physics and Mathematics of the Universe (IPMU) / 
University of Tokyo, the Korean Participation Group, Lawrence Berkeley National Laboratory, 
Leibniz Institut f\"ur Astrophysik Potsdam (AIP),  
Max-Planck-Institut f\"ur Astronomie (MPIA Heidelberg), 
Max-Planck-Institut f\"ur Astrophysik (MPA Garching), 
Max-Planck-Institut f\"ur Extraterrestrische Physik (MPE), 
National Astronomical Observatories of China, New Mexico State University, 
New York University, University of Notre Dame, 
Observat\'ario Nacional / MCTI, The Ohio State University, 
Pennsylvania State University, Shanghai Astronomical Observatory, 
United Kingdom Participation Group,
Universidad Nacional Aut\'onoma de M\'exico, University of Arizona, 
University of Colorado Boulder, University of Oxford, University of Portsmouth, 
University of Utah, University of Virginia, University of Washington, University of Wisconsin, 
Vanderbilt University, and Yale University.

%colleagues, acknowledge funding agencies, telescopes and facilities used etc.
%Try to keep it short.
%
%%%%%%%%%%%%%%%%%%%%%%%%%%%%%%%%%%%%%%%%%%%%%%%%%%%
%
%%%%%%%%%%%%%%%%%%%%% REFERENCES %%%%%%%%%%%%%%%%%%
%
%% The best way to enter references is to use BibTeX:
%

\bibliographystyle{mnras}
\bibliography{biblio_Quench_MANGA_v23Nov} 
%%%%%%%%%%%%%%%%%%%%%%%%%%%%%%%%%%%%%%%%%%%%%%%%%%

%%%%%%%%%%%%%%%%% APPENDICES %%%%%%%%%%%%%%%%%%%%%

%

\include{mnras_QS_QUENCH_MANGA_ARXIV_VERSION__ONLINE_MATERIAL}
\end{document}

%% file: mnras_QS_QUENCH_MANGA_ARXIV_VERSION__ONLINE_MATERIAL.tex
\appendix
\begin{figure*}
\vspace{0.1in}
\begin{flushleft}
\section{Main parameters radial profiles of individual galaxies}
\label{app:radial_prof}
In this section we show the radial profiles for each galaxy in the QGs and SFs samples of the main parameters analysed in this paper (i.e. \OH, \NO, ionisation parameter log(U), gas-phase metallicity Z and the SFR surface density $\Sigma$SFR). 
\end{flushleft}
\vspace{0.1in}
%\includegraphics[width=0.495\linewidth, valign=t]
%{plot_QGs_EBV_orig_vs_DIST_v01R50_SingleProfiles_v2_0.pdf}\includegraphics[width=0.495\linewidth, valign=t]
%{plot_SFs_EBV_orig_vs_DIST_v01R50_SingleProfiles_v2_0.pdf}
%\caption{Radial profiles of E(B-V) of each galaxy belonging to \und sample (left in red) and to SF sample (right in blue). The magenta and cyan curves represent the average radial profiles of \und and SFs populations (as in \autoref{fig:EBV_global}).}
%\label{fig:EBVradProf}
%%\end{minipage}
%\end{figure*}
%
%\begin{figure*}
\includegraphics[width=0.495\linewidth, valign=t]
{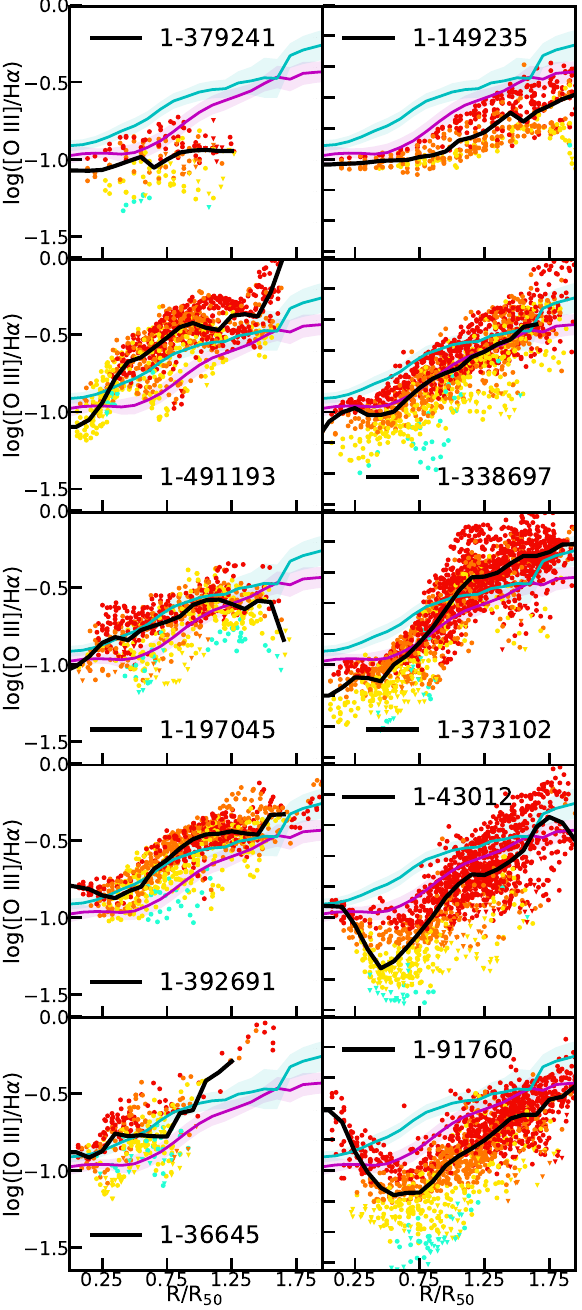}\includegraphics[width=0.495\linewidth, valign=t]
{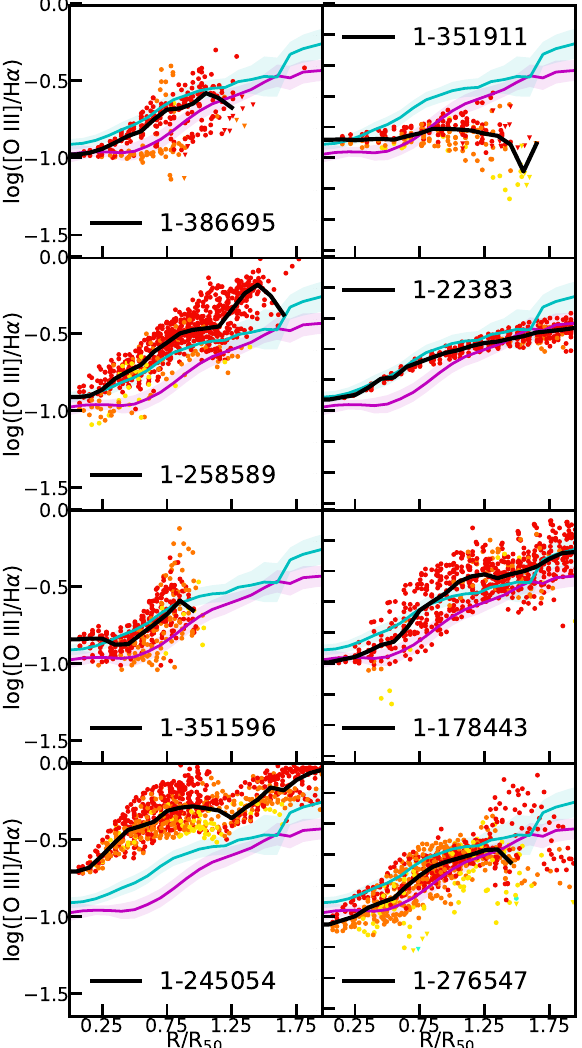}
\caption{Radial profiles of\OH ratio of each galaxy belonging to \und sample (left) and to SF sample (right). 
The black curves represent the median of the relations in bins of width $0.1$R/R$_{50}$ , while the blue ones represent the 16-84$^\text{th}$ percentile of the relations. Each round dot represents a spaxel in which the S/N(\OIII) $\geq 2$, while the square dots represent spaxels in which the S/N(\OIII) $<2$ and their \OH values are upper-limits. The dots colour code is the same as in \autoref{fig:map_quench}, and it is based on the position of each spaxel on the \OH vs \NO diagram (\autoref{fig:QGs_plane} and \autoref{fig:QGs_wDIST_Models}). The cyan is representing quenching regions, followed by the yellow for the galactic regions that lie between $3\times1\sigma$ and $1\sigma$ of the diagram, orange for those between $1\sigma$ and the median and red for regions of pure star-formation that are above the median of the diagram.
The magenta and cyan curves represent the average radial profiles of \und and SFs populations (as in \autoref{fig:EBV_global}).}
\label{fig:O3HaradProf}
%\end{minipage}
\end{figure*}

\begin{figure*}
\includegraphics[width=0.495\linewidth, valign=t]
{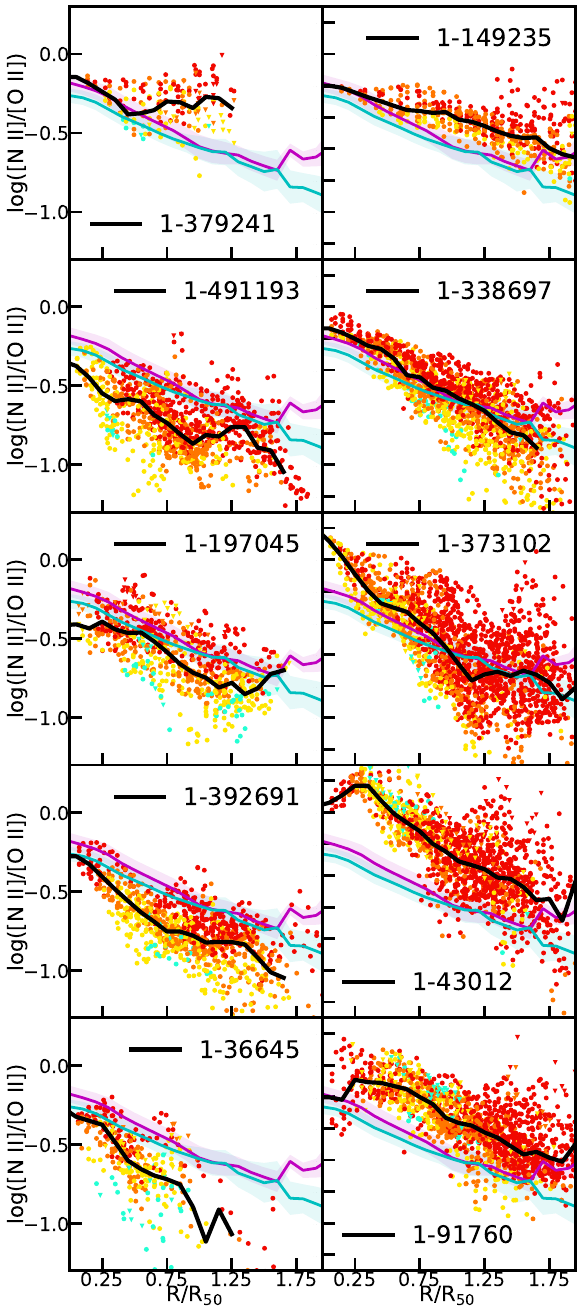}\includegraphics[width=0.495\linewidth, valign=t]
{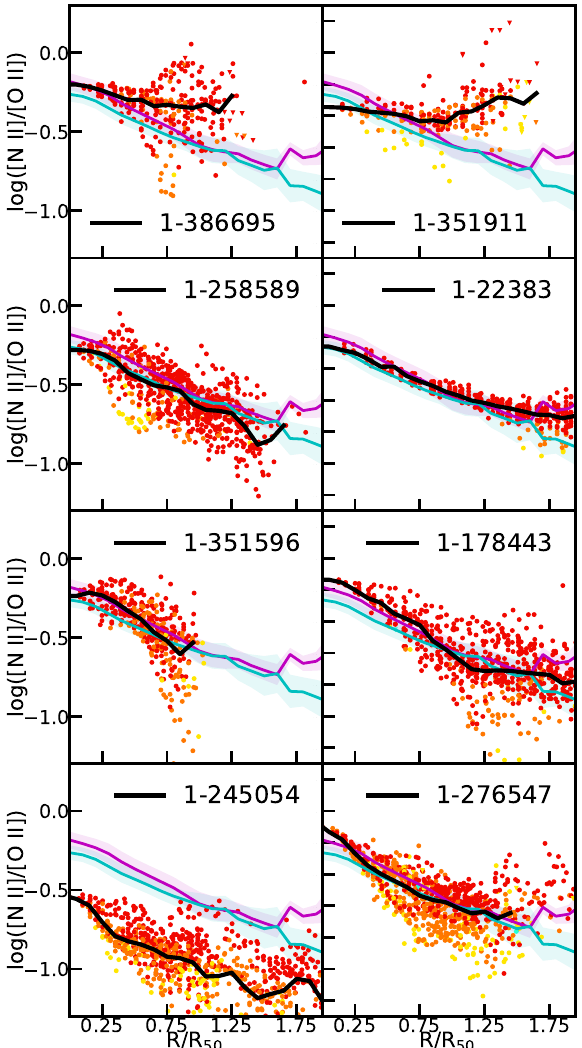}
\caption{Radial profiles of \NO of each galaxy belonging to \und sample (left) and to SF sample (right). See \autoref{fig:O3HaradProf} for panels description.}
\label{fig:N2O2radProf}
%\end{minipage}
\end{figure*}

\begin{figure*}
\includegraphics[width=0.495\linewidth, valign=t]
{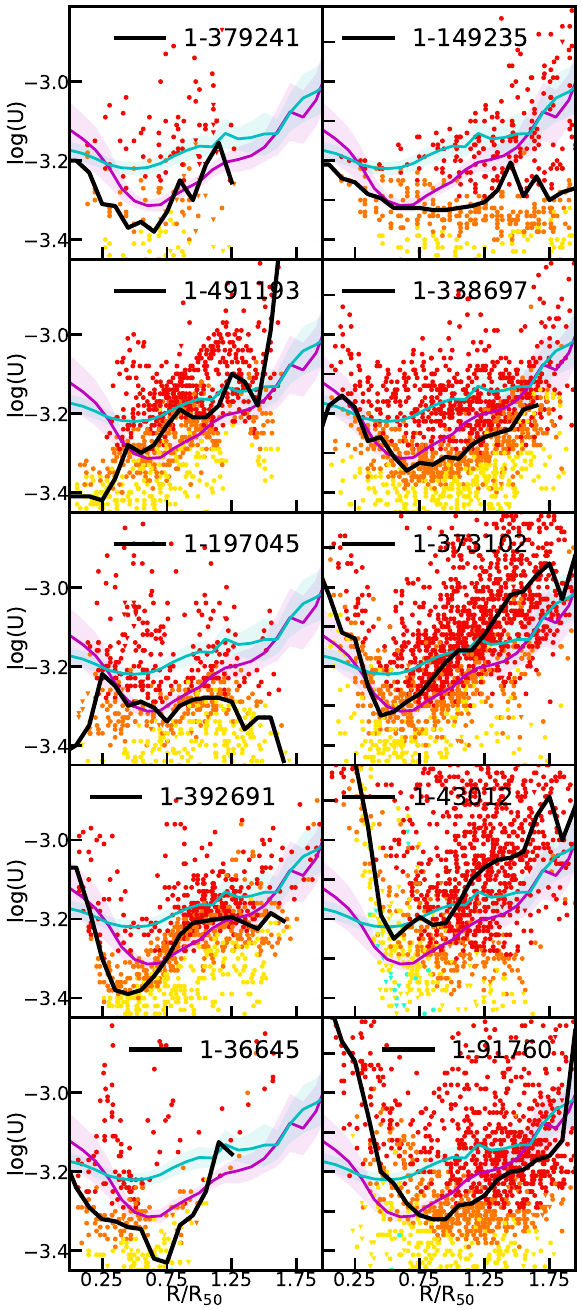}\includegraphics[width=0.495\linewidth, valign=t]
{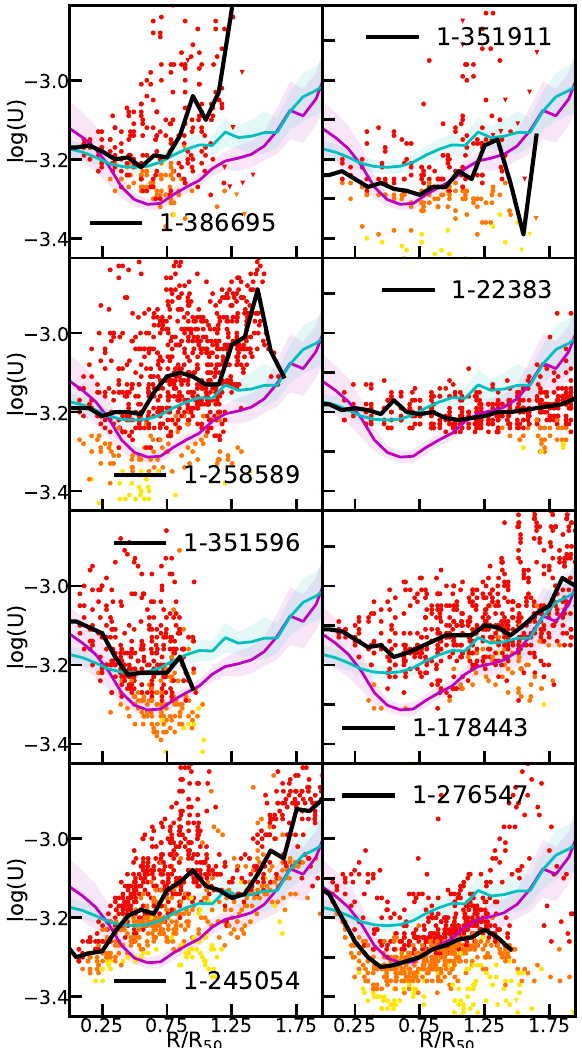}
\caption{Radial profiles of ionisation parameter log(U) of each galaxy belonging to \und sample (left) and to SF sample (right). See \autoref{fig:O3HaradProf} for panels description.}
\label{fig:UradProf}
%\end{minipage}
\end{figure*}

\begin{figure*}
\includegraphics[width=0.495\linewidth, valign=t]
{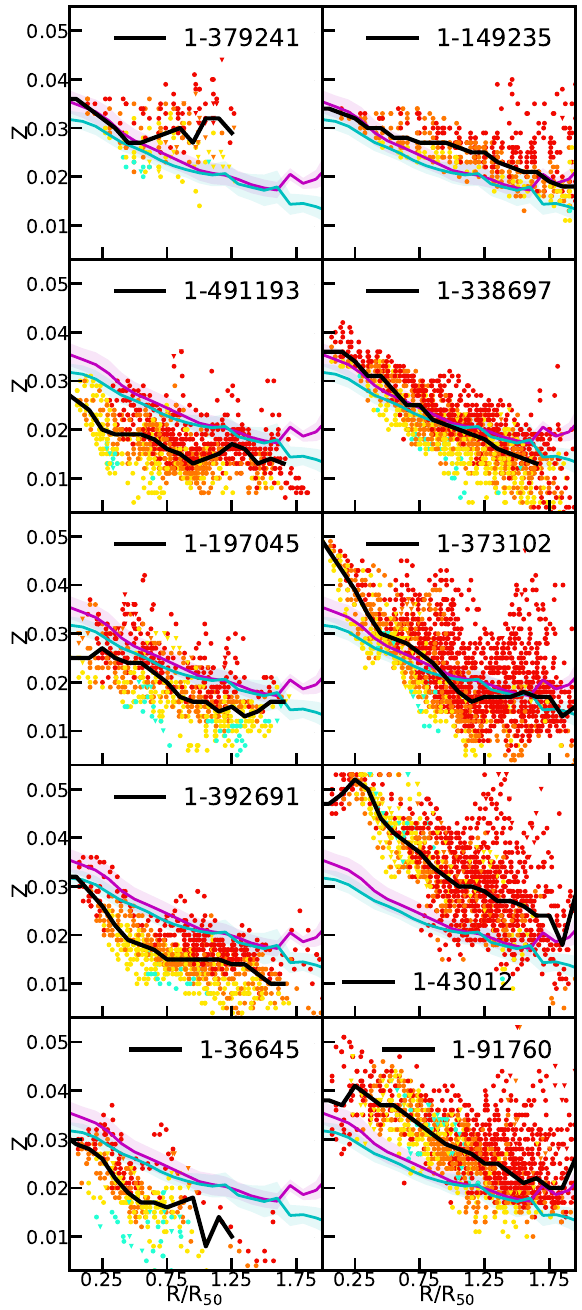}\includegraphics[width=0.495\linewidth, valign=t]
{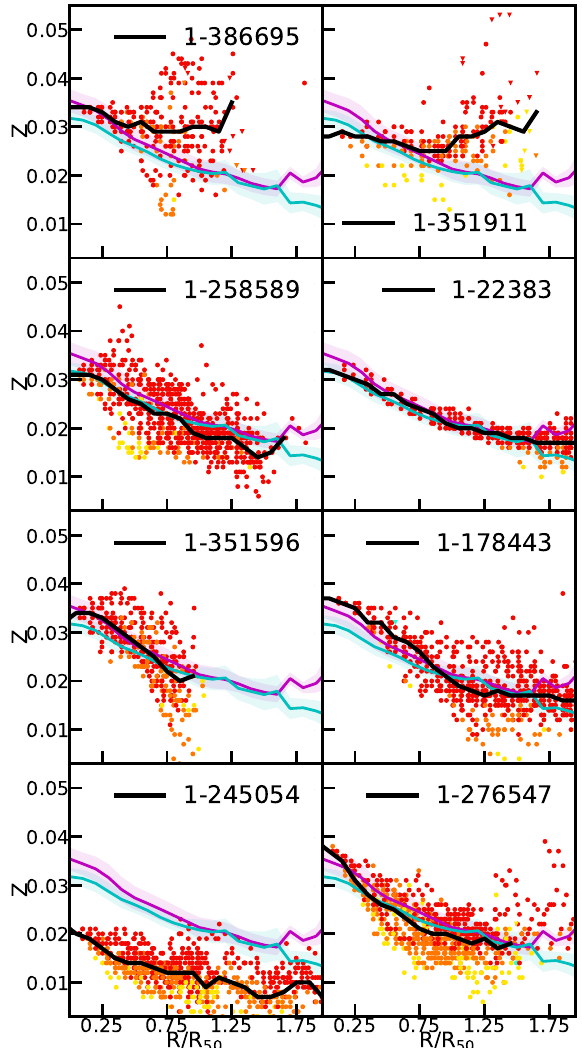}
\caption{Radial profiles of gas phase metallicity Z of each galaxy belonging to \und sample (left) and to SF sample (right). See \autoref{fig:O3HaradProf} for panels description.}
\label{fig:ZradProf}
%\end{minipage}
\end{figure*}

\begin{figure*}
\includegraphics[width=0.495\linewidth, valign=t]
{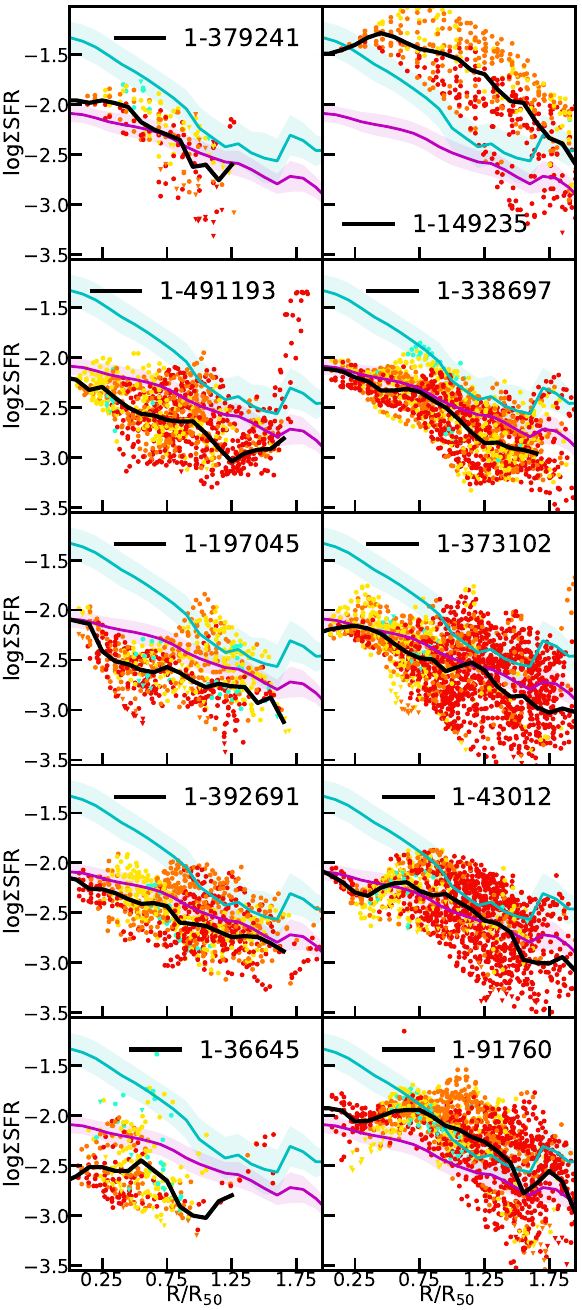}\includegraphics[width=0.495\linewidth, valign=t]
{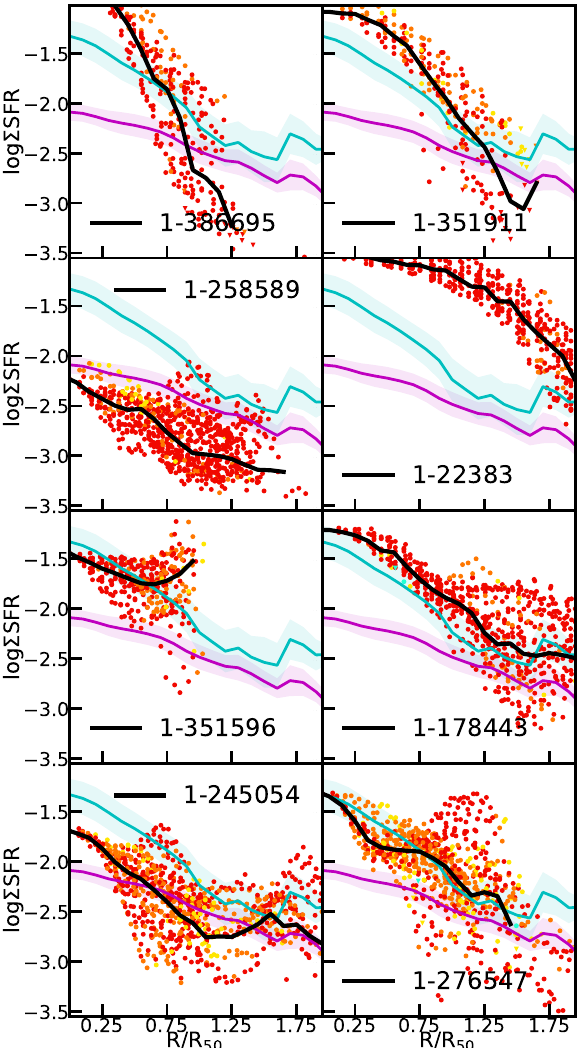}
\caption{Radial profiles of the star formation rate surface density $\Sigma$SFR of each galaxy belonging to \und sample (left) and to SF sample (right). See \autoref{fig:O3HaradProf} for panels description.}
\label{fig:UradProf}
%\end{minipage}
\end{figure*}